\documentclass{aa}
\usepackage{graphicx}
\usepackage{graphicx}
\usepackage{times}
\usepackage{epsfig}
\usepackage{longtable}
\usepackage{lscape}
\usepackage{amssymb}
\usepackage{latexsym}

\usepackage{txfonts}
\def\kms{$\rm km\;s^{-1}$}

\def\pa{\theta}
\def\ha{H$\alpha$}

\def\niip{[N~{\small II}]$\,\lambda6548$} 
\def\niig{[N~{\small II}]$\,\lambda6583$} 
\def\niipg{[N~{\small II}]$\,\lambda\lambda6548,6583$}

\begin{document}
\title{The search for inner polar disks with integral field
  spectroscopy: the case of NGC 2855 and NGC 7049\thanks{Based on
  observations carried out at the European Southern Observatory (ESO
  73.B-0803).}}

\author{L. Coccato \inst{1}
   \and E. M. Corsini$^{2,3}$
   \and A. Pizzella \inst{2}
   \and F. Bertola$^{2}$}

\offprints{L. Coccato, \\ e-mail: {\tt coccato@astro.rug.nl}}

\institute{Kapteyn Astronomical Institute, University of Groningen, Postbus 800, 
         9700 AV Groningen, The Netherlands\\
  \and Dipartimento di Astronomia, Universit\`a di Padova, vicolo
         dell'Osservatorio 3, I-35122 Padova, Italy\\
  \and Scuola Galileiana di Studi Superiori, via VIII Febbraio 2, I-35122
         Padova, Italy}

\date{\today}

\abstract
%
{The presence of non-circular and off-plane gas motions is frequently
observed in the inner regions of disk galaxies.}
%
{We have measured with integral-field spectroscopy the
surface-brightness distribution and kinematics of the ionized gas in
NGC~2855 and NGC~7049. These two early-type spiral galaxies were
selected to possibly host a kinematically-decoupled gaseous component
in orthogonal rotation with respect to the galaxy disk.}
%
{We have modeled the ionized-gas kinematics and distribution of both
galaxies assuming that the gaseous component is distributed either on
two orthogonally-rotating disks or in a single and strongly
warped disk.}
%
{In both galaxies the velocity field and distribution of the inner gas
are consistent with the presence of an inner polar disk. In NGC~2855
it correponds to the innermost and strongly warped portion of the main
disk. In NGC~7049 it is a central and geometrically-decoupled disk,
which is nested in the main disk.}
%
{}
\keywords{galaxies: kinematics and dynamics -- galaxies: spirals --
galaxies: structure -- galaxies: individual: NGC 2855, NGC 7049}

\titlerunning{Inner polar disks in NGC 2855 and NGC 7049}

\authorrunning{Coccato et al.}  

\maketitle

\section{Introduction}
\label{sec:intro}

In a recent paper we pointed that about $50\%$ of bright unbarred
galaxies show a remarkable gas velocity gradient along the optical
minor axis (Coccato et al. 2004). This phenomenon is observed all
along the sequence of disk galaxies from S0's to Sm's. Since
minor-axis velocity gradients are unexpected if the gas is moving onto
circular orbits in a disk coplanar to the stellar one, we concluded
that non-circular and off-plane gas motions are not rare in the inner
regions of disk galaxies. Further support to this picture is given by
the analysis of the velocity fields of large samples of lenticular
galaxies (Sarzi et al. 2006), early (Falc\'on-Barroso et al. 2006) and
late-type spirals (Ganda et al. 2006) observed with integral-field
spectroscopy.

The presence of a velocity gradient along the minor axis is
characteristic of galaxies hosting an inner polar disk (IPD
hereafter). IPDs are small disks of gas and/or stars ($R\approx300$
pc), which are located in the center of lenticular and spiral galaxies
and are rotating in a plane perpendicular to that of the main disk of
their host. We call them IPDs instead of inner polar rings (Sil'chenko
2006) since there is no clearcut evidence they have an anular
structure.  Most of these orthogonally-rotating disks have been
discovered in last few years (Corsini et al. 2003; Sil'chenko \&
Afanasiev 2004; Shalyapina et al. 2004; Sil'chenko \& Moiseev 2006;
Coccato et al. 2005).
The acquisition of external gas via merging or accretion on nearly
polar orbits by a pre-existing galaxy, and the transfer of gas onto
highly-inclined anomalous orbits of a triaxial bulge or a bar, which
is tumbling about its short axis, are both viable mechanism to build
a orthogonally-rotating disk.
According to these different scenarios, it can be either a
geometrically-decoupled structure or the inner portion of a strongly
warped and larger gaseous disk, respectively. Therefore, constraining
the structural properties of a sample of IPDs will give the clues to
understand the processes driving their formation (see Sil'chenko 2006
for a review).

\begin{table*}
\centering
\caption{Instrumental setup}
\begin{tabular}{l c c c c}
\hline
\hline
\noalign{\smallskip}
Parameters & \multicolumn{4}{c}{Channels} \\ 
           & 1 & 2 & 3 & 4 \\
\noalign{\smallskip}
\hline
\noalign{\smallskip}
Grism                                    & HR\_red & HR\_red & HR\_red & HR\_orange \\
Order sorting filter                     & GG475   & GG475   & GG475   & GG435      \\
Spatial resolution ($''$ fiber$^{-1}$)   & 0.67    & 0.67    & 0.67    & 0.67       \\ 
Reciprocal dispersion (\AA\ pixel$^{-1}$)& 0.58    & 0.58    & 0.58    & 0.62       \\
Readout noise (e$^-$)                    & 4.5     & 3.2     & 3.2     & 3.7        \\
Gain  (e$^-$ ADU$^{-1}$)                 & 1.9     & 1.8     & 1.7     & 2.0        \\
Wavelength range (\AA)                   & 6300 -- 8700 & 6300 -- 8700 & 6300 -- 8700 & 5200 -- 7600  \\
Instrumental FWHM at \ha\ (\kms )        & 68      & 68      & 68      & 70  \\ 
\noalign{\smallskip}
\hline
\noalign{\smallskip}
\end{tabular}
\label{tab:setup}
\end{table*}

Over the course of the last few years, we have undertaken a program
aimed at detecting IPDs in disk galaxies using long-slit spectroscopic
observations (Bertola et al. 1999; Sarzi et al. 2000; Corsini et
al. 2002, 2003; Coccato et al. 2004, 2005). In long-slit spectra the
kinematic signature of an IPD is the presence of a central velocity
gradient along the disk minor axis and a zero-velocity plateau along
the disk major axis, respectively. On the contrary, the presence of a
gas velocity gradient along both major and minor axis is
characteristic of gas moving onto elliptical orbits in the disk plane
of a triaxial bulge (de Zeeuw \& Franx 1989; Corsini et al. 2003).
Our goal was to identify galaxies, which are good candidates to host a
IPD, to be followed up with integral-field spectroscopy at high
spatial resolution in order to determine the size and orientation of
such a kinematically-decoupled component.

In this paper we present the analysis of the bidimensional kinematics
and distribution of the ionized-gas component of the two most
promising objects we found in our investigation. They are the nearby
early-type spirals NGC~2855 and NGC~7049. For both galaxies the
presence of a major-axis velocity plateau together with a minor-axis
velocity gradient is suggestive of an IPD as discussed by Corsini et
al. (2002, 2003). The reader is refereed to them for a summary of the
main properties of the two galaxies.
The paper is organized as follows. The integral-field spectroscopic
observations and data reduction are described in
Sect. \ref{sec:observations}. We present and model the ionized-gas
kinematics and distribution in Sect. \ref{sec:ionized} and
\ref{sec:model}, respectively. Our conclusions are discussed in
Sect. \ref{sec:conclusions}.

\section{Observations and data reduction}
\label{sec:observations}

The integral-field spectroscopic observations of NGC~2855 and NGC~7049
were carried out with the Very Large Telescope (VLT) at the European
Southern Observatory (ESO) in Paranal (Chile) on February 26--27, 2004
and May 13--14, 2004. The Unit Telescope 3 (Melipal) mounted the
Visible Multi Object Spectrograph (VIMOS) in the Integral Field Unit
(IFU) configuration.
The field of view of the four VIMOS channels was $27'' \times 27''$
and it was projected onto a microlenses array. This was coupled to
optical fibers which were rearranged on a linear set of microlenses to
produce an entrance pseudoslit to the spectrograph. The pseudoslit was
$0\farcs95$ wide and generated a total of 1600 spectra covering the
field of view with a spatial resolution of $0\farcs67$ per fiber. Each
channel was equipped with either the HR\_red or HR\_orange high
resolution grism and a thinned and back illuminated EEV44 CCD with
$2048\times4096$ pixels of $15\times15$ $\mu$m$^2$. The details of the
instrumental setup are given in Table \ref{tab:setup}.

For each galaxy we obtained $4\times19$-min exposures. They were taken
in service mode by executing two different observing blocks of two
exposures each, as listed in Table \ref{tab:obserlog}. Different quarz
and arc lamp spectra were taken after every object exposure to ensure
accurate flatfield correction and wavelength calibration,
respectively. During observations seeing FWHM ranged between
$0\farcs8$ and $1\farcs0$ as measured by the ESO Differential Image
Meteo Monitor.

\begin{table}
\centering
\caption{Observing log}
\begin{tabular}{l c c c}
\hline
\hline
\noalign{\smallskip}
Galaxy & Date  & Observing block & Exposure time      \\
\noalign{\smallskip}
\hline
\noalign{\smallskip}
NGC 2855 & 27 Feb 2004 & 154289 &$2\times19$ min \\
NGC 2855 & 27 Feb 2004 & 154292 &$2\times19$ min \\
NGC 7049 & 13 May 2004 & 154293 &$2\times19$ min \\
NGC 7049 & 14 May 2004 & 154296 &$2\times19$ min \\
\noalign{\smallskip}
\hline
\noalign{\smallskip}
\end{tabular}
\label{tab:obserlog}
\begin{minipage}{8.2cm}
Notes -- During the execution of the observing block 154293 (13 May
2004) the data of channel 4 were lost because the Grism Exchange Unit
of channel 4 was out of order.
\end{minipage}
\end{table}

For each VIMOS channel all the spectra were traced, identified, bias
subracted, flatfield corrected, corrected for relative fiber
transmission, and wavelength calibrated using the routines of the ESO
Recipe Execution pipeline\footnote{{\tt ESOREX} and {\tt MIDAS} are
developed and maintained by the European Southern Observatory}.
Cosmic rays and bad pixels were identified and cleaned using standard
MIDAS$\;^1$ routines.
We checked that the wavelength rebinning was done properly by
measuring the difference between the measured and predicted wavelength
for the brightest night-sky emission lines in the observed spectral
ranges (Osterbrock et al. 1996). The resulting accuracy in the
wavelength calibration is better than 5 \kms .
The intensity of the night-sky emission lines was used to correct for
the different relative transmission of the VIMOS channels.
The contribution of the night-sky emission lines was determined from a
number of spectra by fitting a Gaussian to each line and a
straight line to the adjacent continuum. We selected spectra where the
night-sky emission lines did not overlap with the \niipg\ and \ha\
emission lines of the galaxy. Then night-sky emission lines were
subtracted from all the available spectra. It was not possible to
subtract the contribution of the night-sky continuum alone since
both galaxies covered the entire field of view of the integral field
unit.
The processed spectra were organized in a datacube using the tabulated
correspondence between each fiber and its position in the field of
view.
{\bf From} the two exposures of each observing block we built a single
datacube.  In fact, the two spectra taken by the same fiber were
coadded after checking that there was no pointing offset between the
two exposures. This was done by comparing the position of the
intensity peaks of the two reconstructed images obtained by collapsing
the datacubes along the wavelength direction.
Finally, for each galaxy we coadded the two available datacubes
by using the intensity peaks of the flux maps as a reference for the
alignment. In this way we produced a single datacube to be analyzed in
order to derive the surface brightness and kinematic maps.

\section{Kinematics and distribution of the ionized gas}
\label{sec:ionized}

\subsection{Analysis}
\label{sec:analysis}

In the coadded spectrum of the galaxy datacube we measured the
position, full width at half maximum (FWHM), and uncalibrated flux of
the \ha\ and \niig\ emission lines by fitting a Gaussian to each line
and a straight line to the adjacent continuum. This includes the
contributions of both galaxy and night sky. No double-peaked emission
line was observed.  We neglected the \niip\ line because it was not
clearly detected in all the spectra, and we averaged the spectra of
adjacent fibers in the outer regions where the intensity of the \ha\
and \niig\ lines was low. We performed a $2\times2$ binning in order
to reach a minimum signal-to-noise ratio $S/N=10$. To assess the
data quality we show some examples of the profile of the
\niig\ emission line and its Gaussian fit in different regions of the
two galaxies. They are plotted in Figs. \ref{fig:example2855} and
\ref{fig:example7049} for NGC~2855 and NGC~7049, respectively. The
central wavelength of the fitting Gaussian was converted into
line-of-sight velocity in the optical convention. The standard
heliocentric correction was applied. The Gaussian FWHM was
corrected for the instrumental FWHM and then converted into the
intrisic line-of-sight velocity dispersion. The flux of the fitting
Gaussian was converted into surface brightness according the fiber
area.  No flux calibration was performed.
%

\begin{figure}
\centering
\hbox{
  \psfig{file=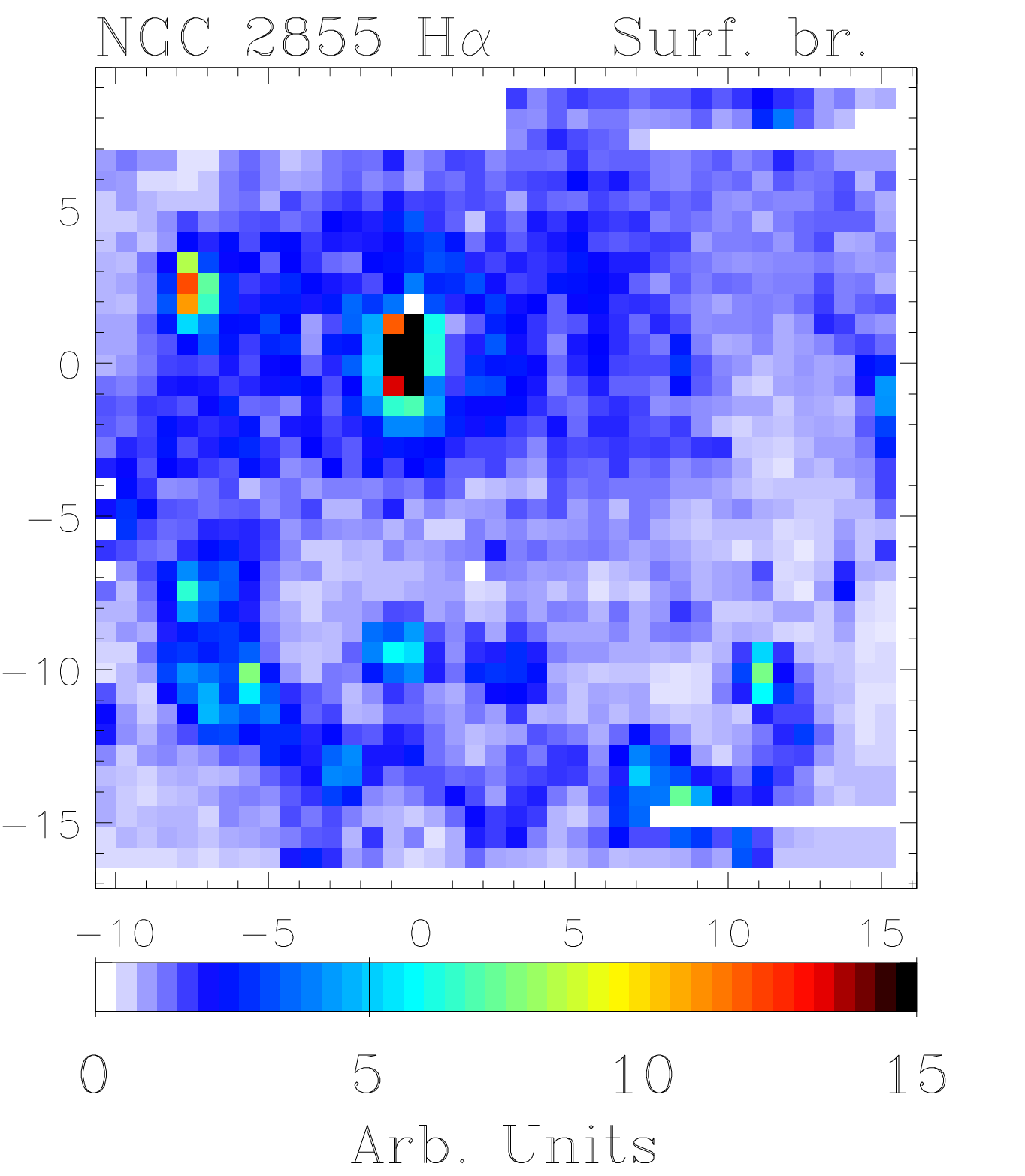,width=2.93cm,clip=}
  \psfig{file=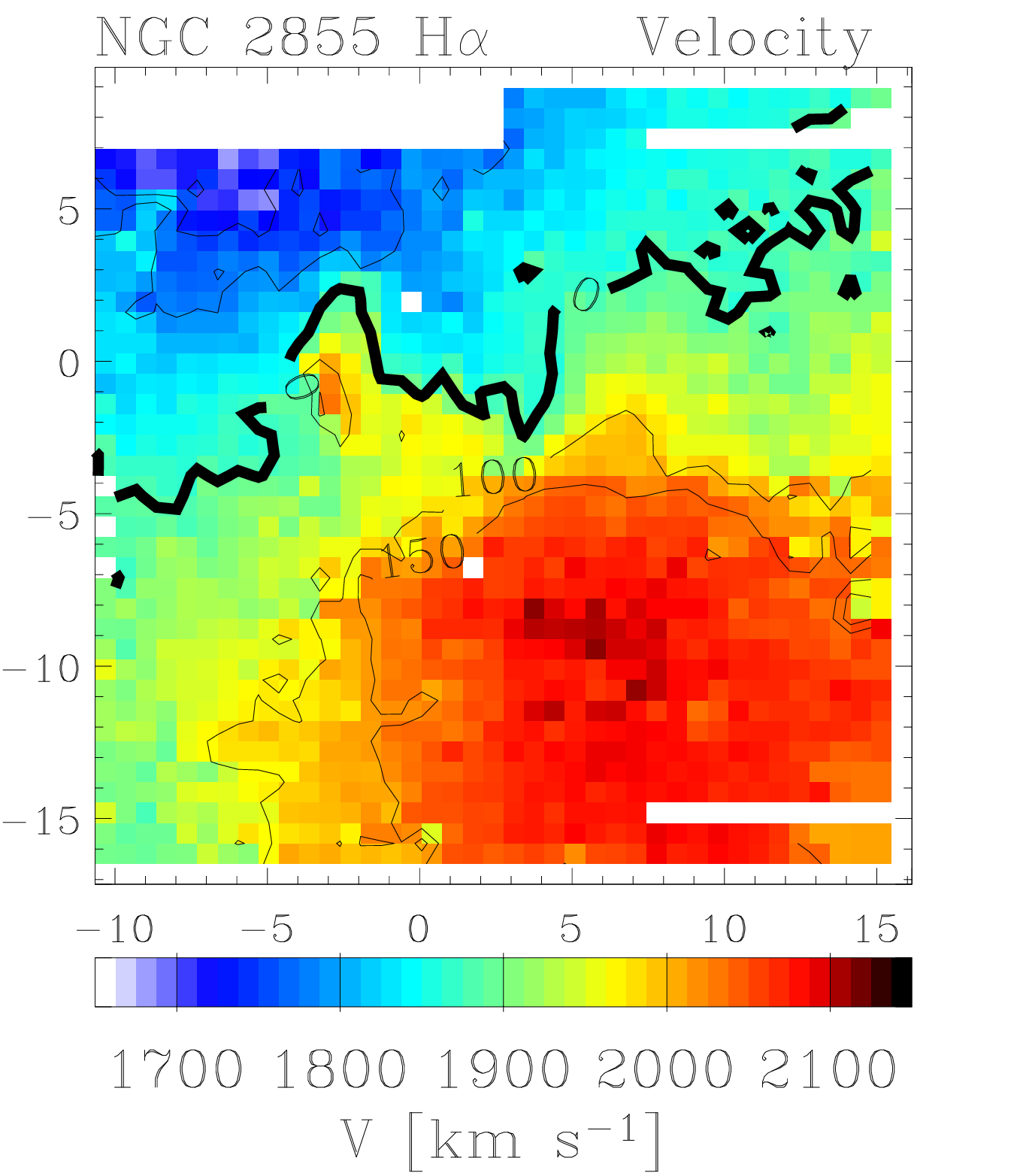,width=2.93cm,clip=}
  \psfig{file=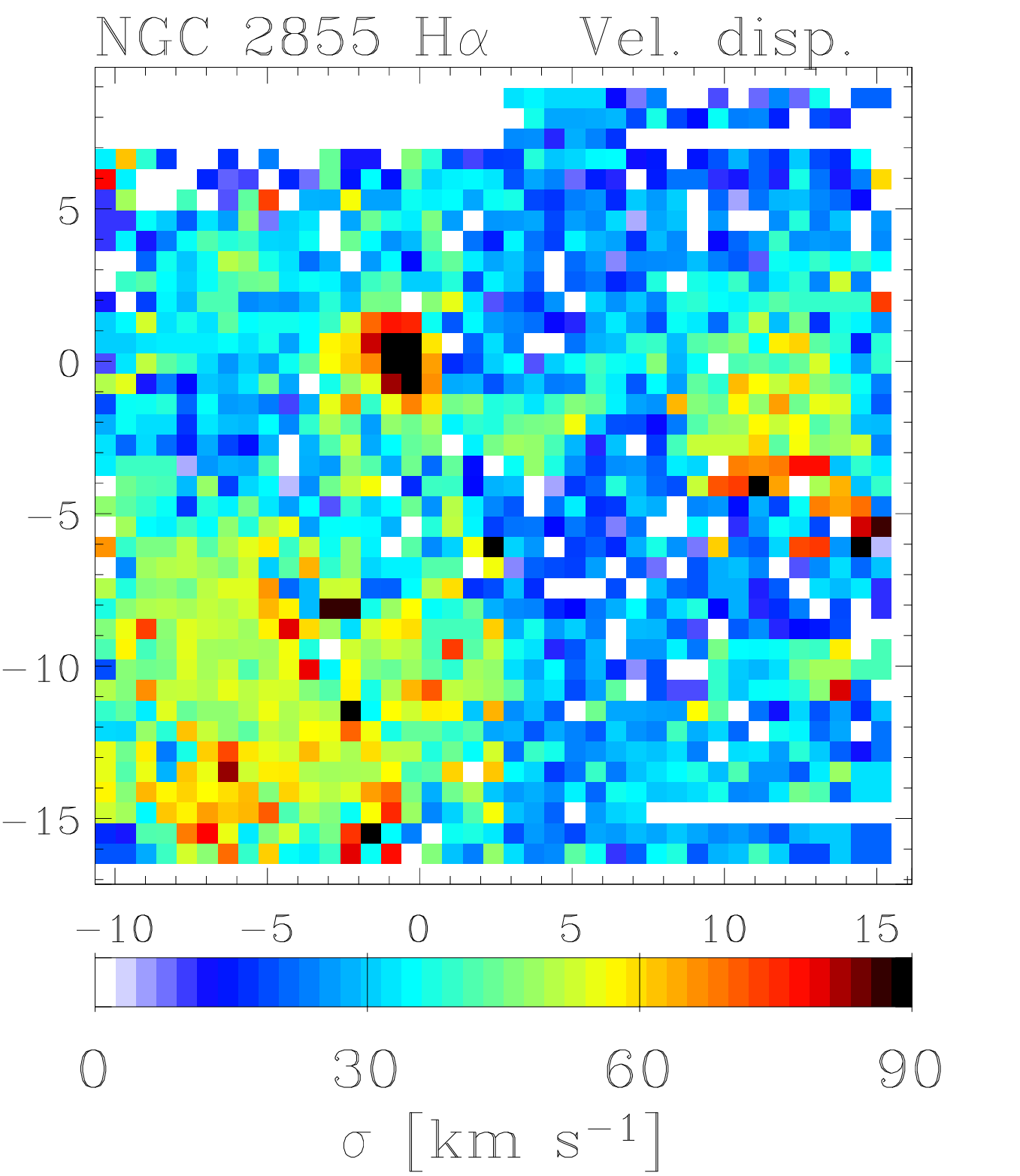,width=2.93cm,clip=}}
\hbox{
  \psfig{file=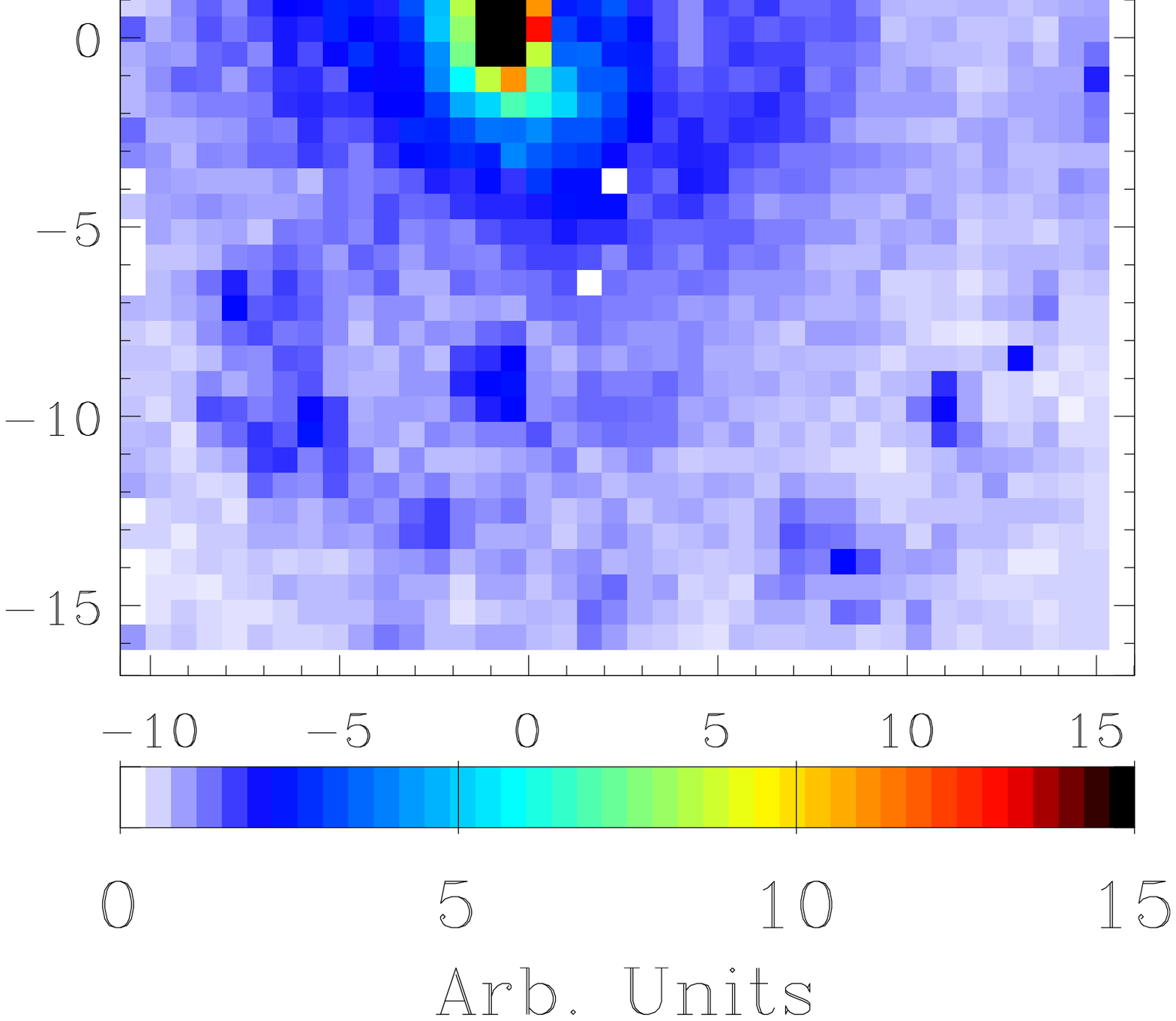,width=2.93cm,clip=}
  \psfig{file=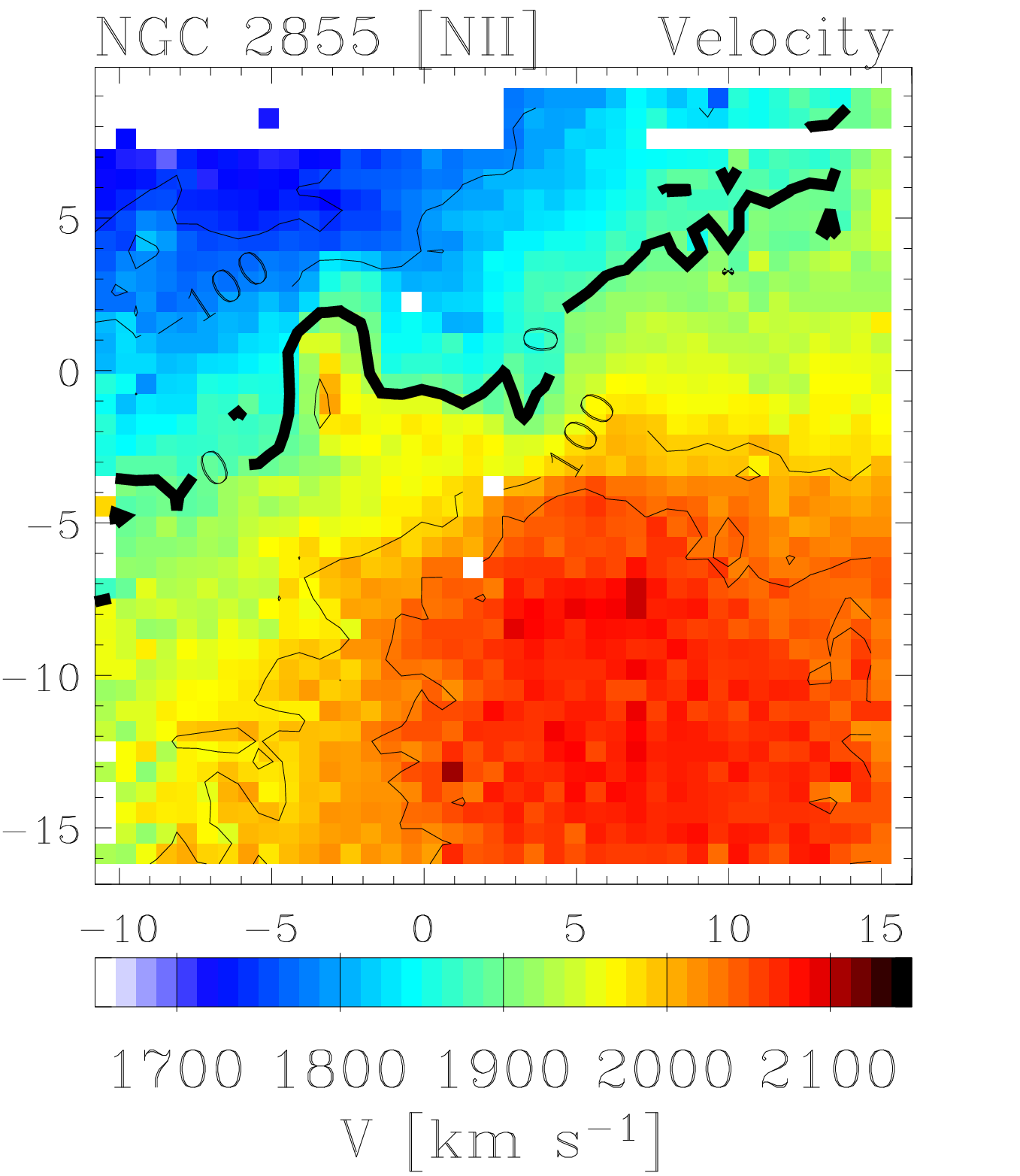,width=2.93cm,clip=}
  \psfig{file=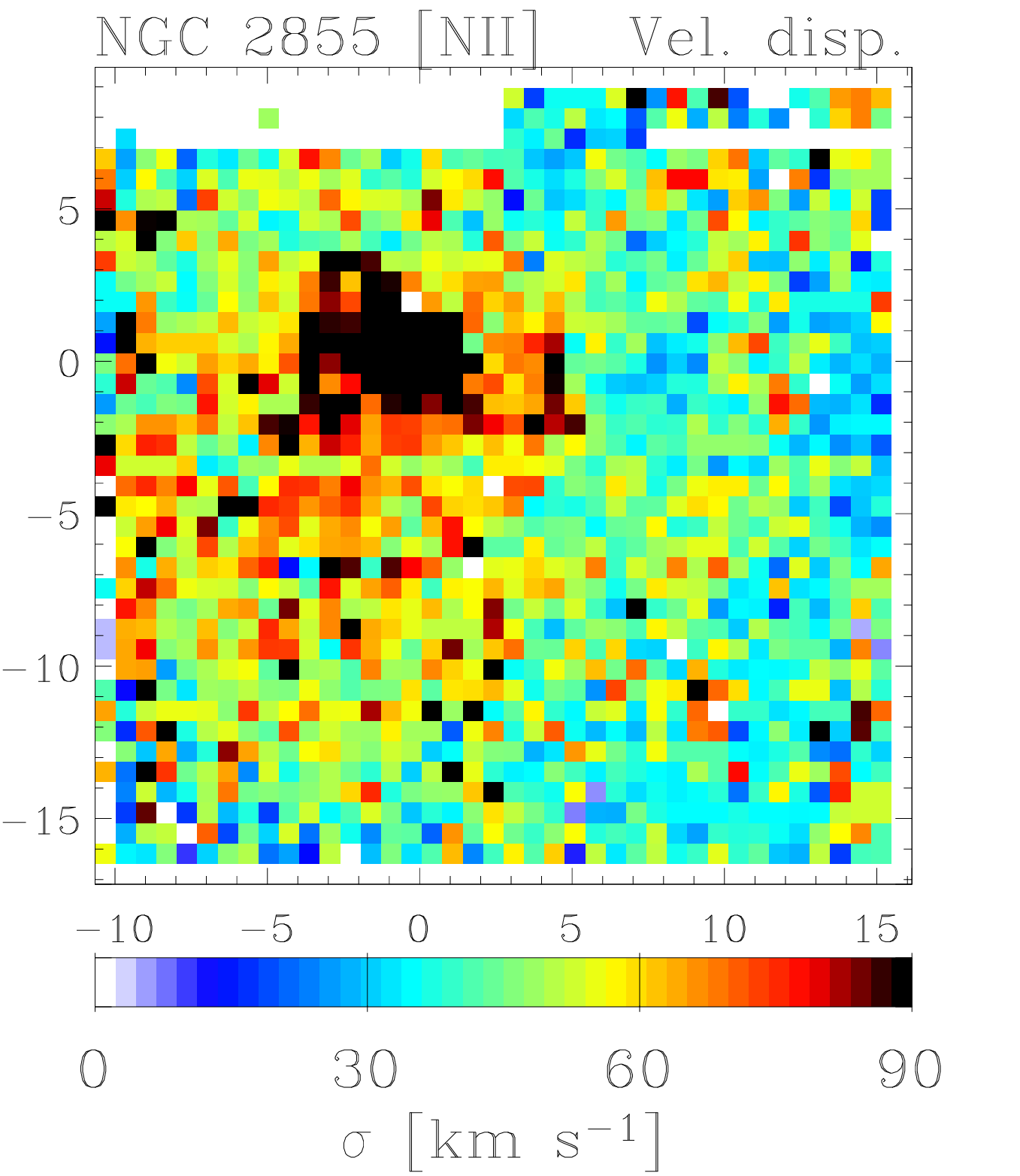,width=2.93cm,clip=}}
\caption{Maps of the ionized-gas kinematics of NGC~2855 derived from
the \ha\ ({\em top panels}) and \niig\ ({\em bottom panels}) emission
lines. The field of view of each panel is $27''\times27''$. In
each panel the spatial coordinates with respect to the galaxy center
are given in arcsec. East is up and North is right. The ranges are
indicated at the bottom of each panel. The white color is used for
bins where no measurement was obtained on account of a signal-to-noise
ratio below the selected thresold due to either faint emission line or
poor fiber transmission. {\em Left panel:} Surface brightness in
arbitrary linear units. {\em Central panel:} Heliocentric
line-of-sight velocity in \kms\ without applying any correction for
galaxy inclination. Isovelocity contours correspond to velocities
after the subtraction of the systemic velocity. The zero-velocity
contour is highlighted with a thicker solid line. {\em Right panel:}
Line-of-sight velocity dispersion in \kms\ corrected for
instrumental velocity dispersion.}
\label{fig:n2855_vfield}
\end{figure}

\begin{figure}
\centering
\hbox{
  \psfig{file=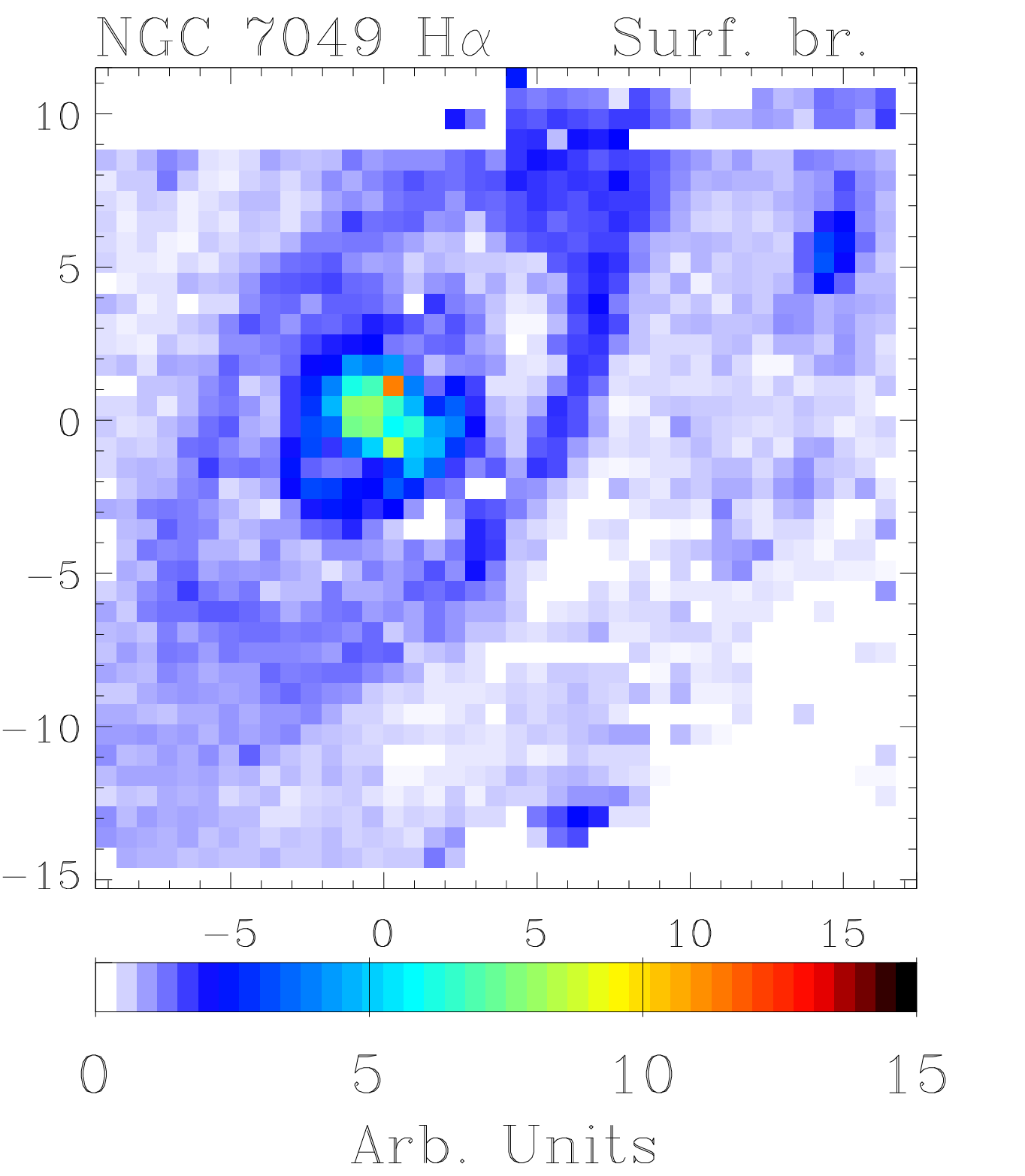,width=2.93cm,clip=}
  \psfig{file=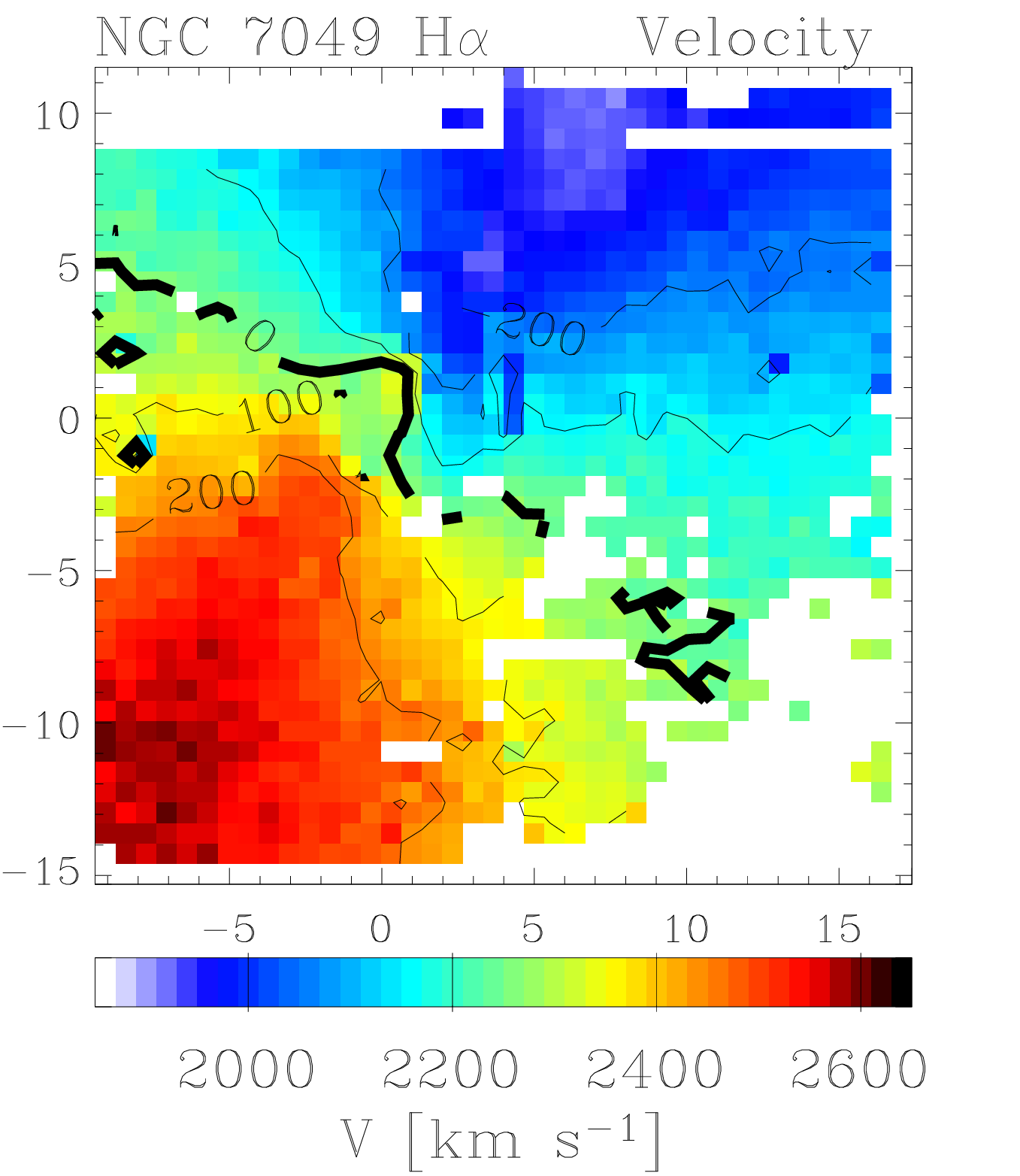,width=2.93cm,clip=}
  \psfig{file=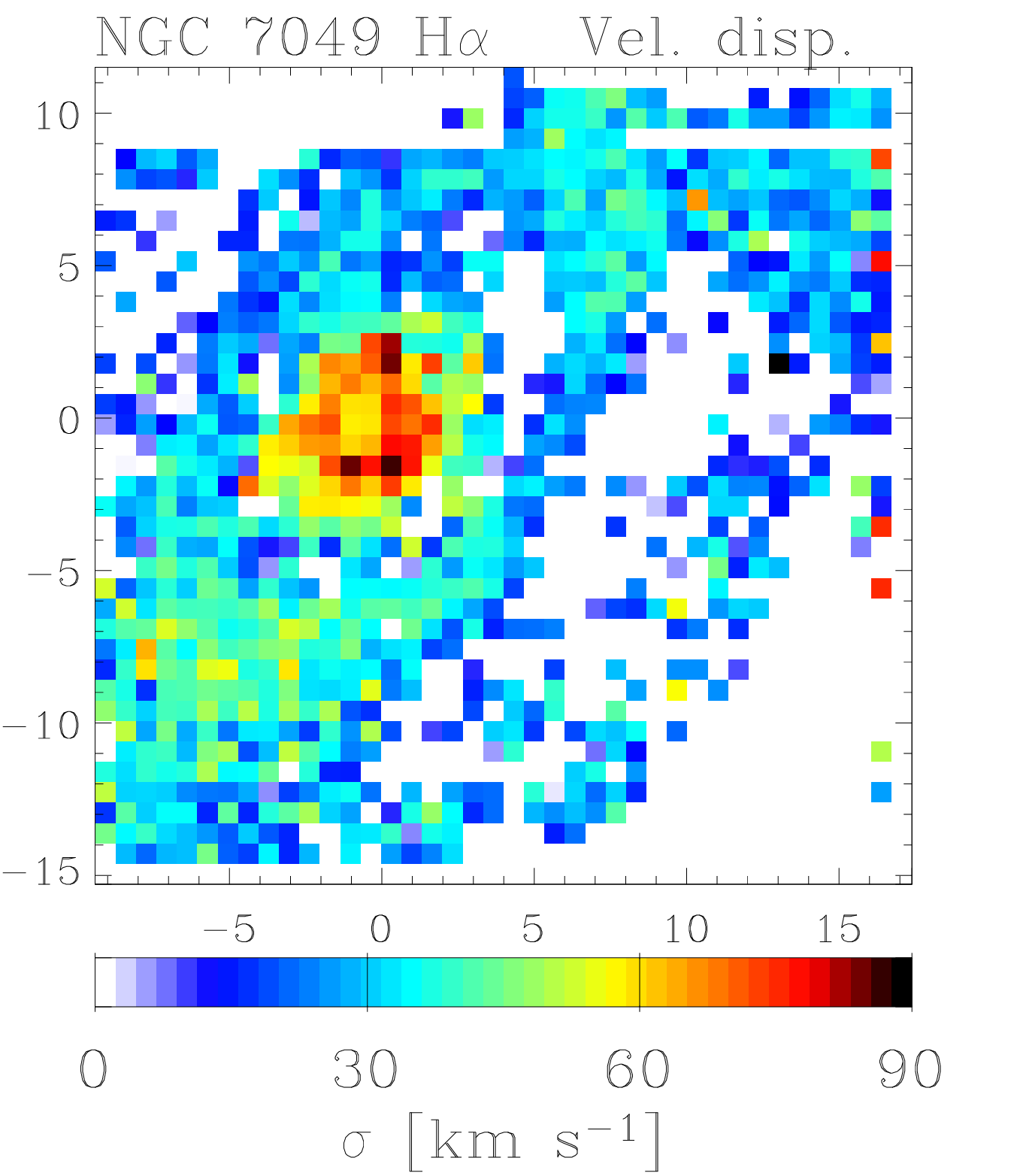,width=2.93cm,clip=}}
\hbox{
  \psfig{file=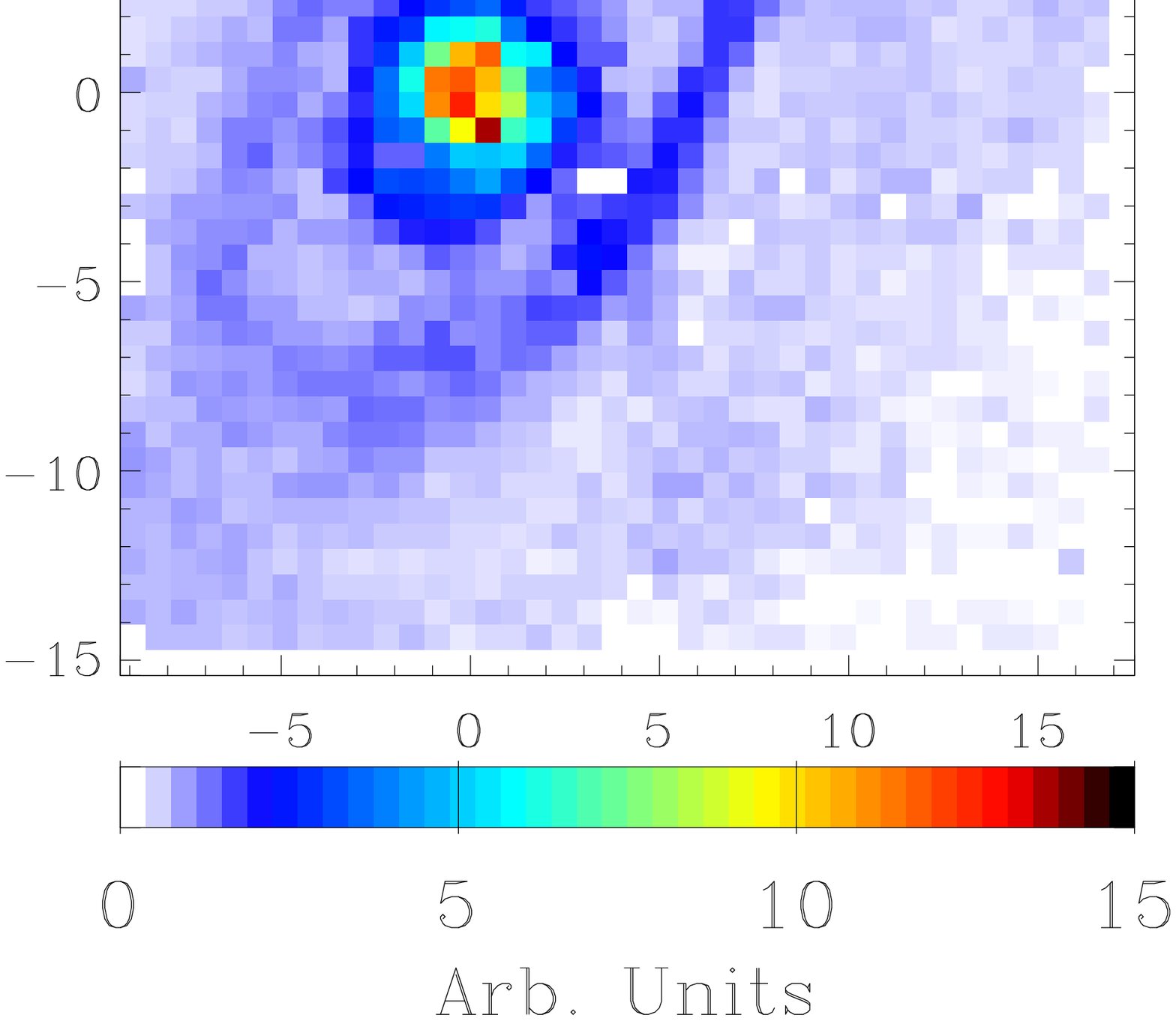,width=2.93cm,clip=}
  \psfig{file=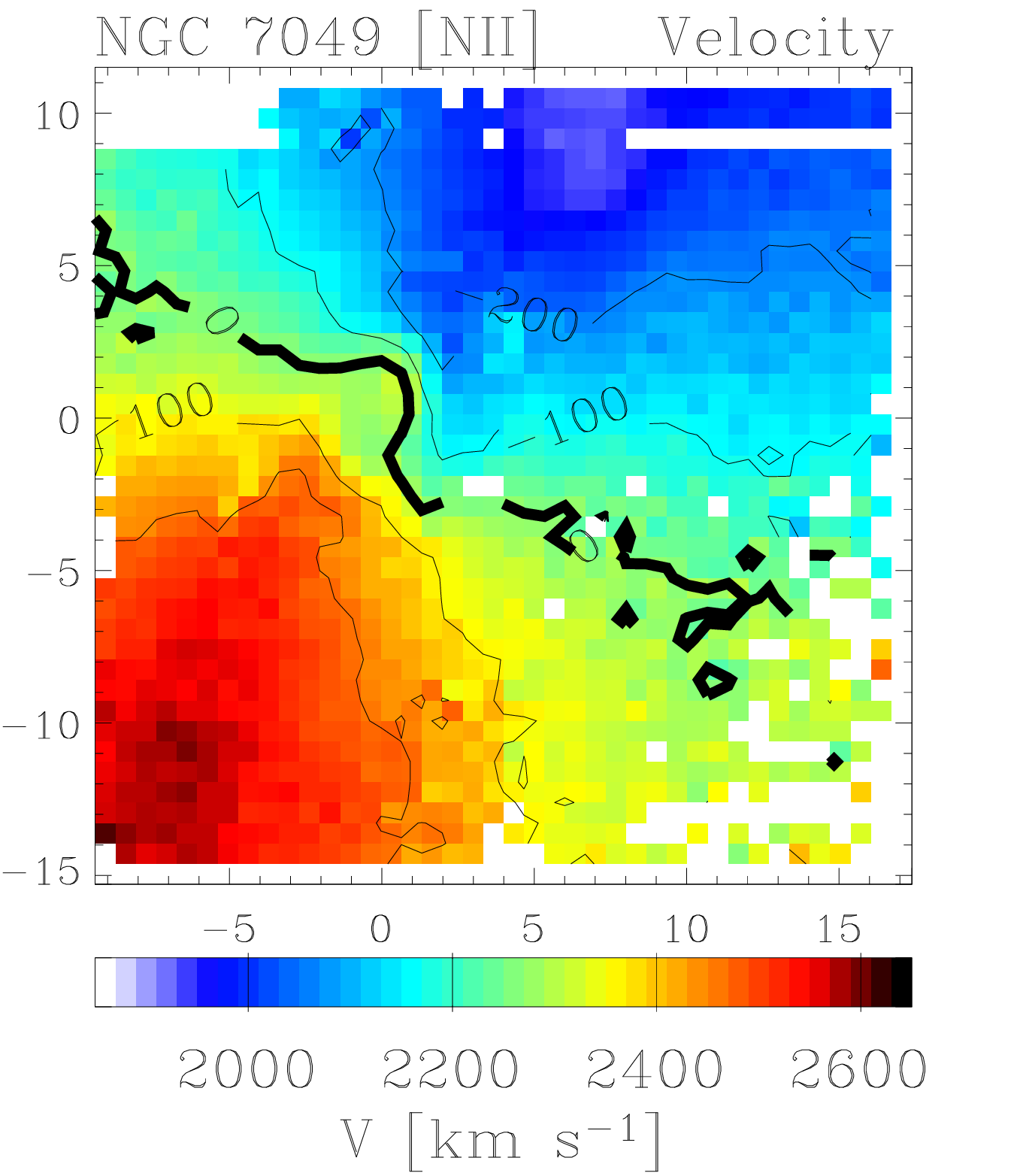,width=2.93cm,clip=}
  \psfig{file=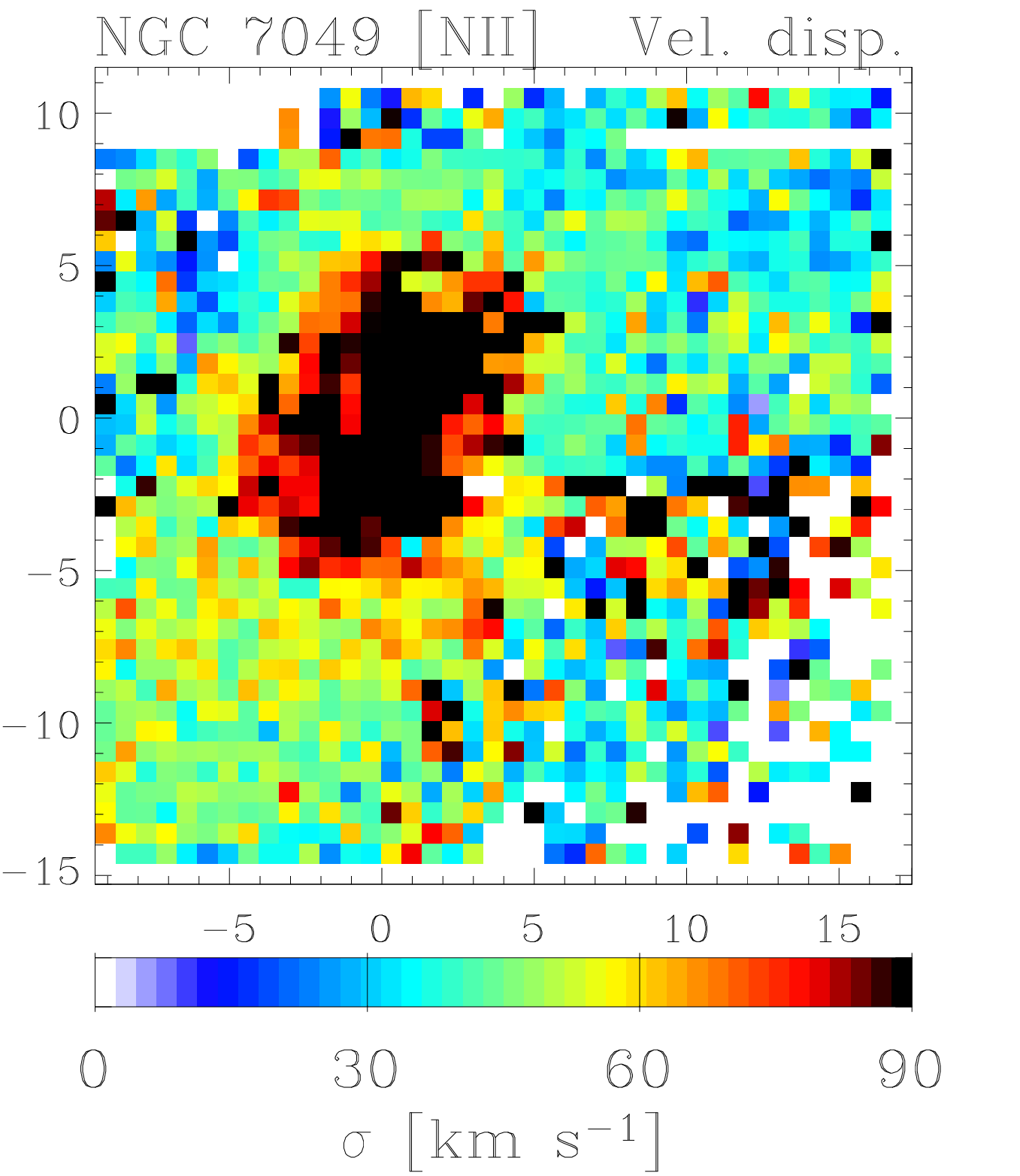,width=2.93cm,clip=}}
 \caption{Same as in Fig. \ref{fig:n2855_vfield} but for NGC 7049.}
\label{fig:n7049_vfield}
\end{figure}

\begin{figure*}
\centering
\psfig{file=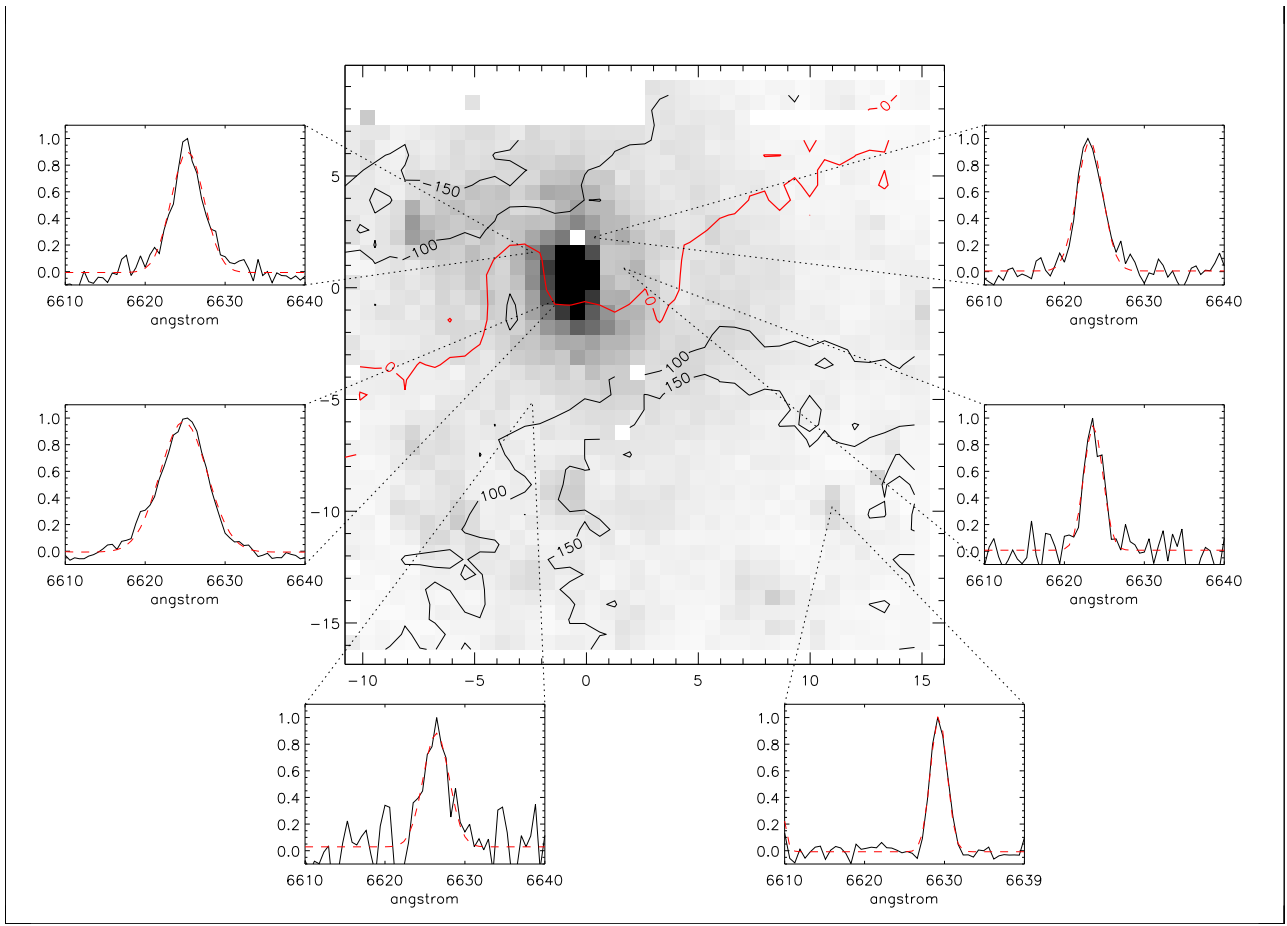,width=17.0cm,clip=}\\
\caption{ {\em Central panel}: Maps of the surface brightness
and line-of-sight velocity field derived from the \niig\ emission line
for NGC~2855. The field of view, orientation, ranges and isovelocity
contours are as in Fig. \ref{fig:n2855_vfield}. {\em Outer panels:}
The \niig\ emission line profile (solid line) and its Gaussian fit
(dashed line) in different positions of the field of view.  Line
profiles are normalized to the peak intensity to allow comparison.}
\label{fig:example2855}
\end{figure*}

\begin{figure*}
\centering
\psfig{file=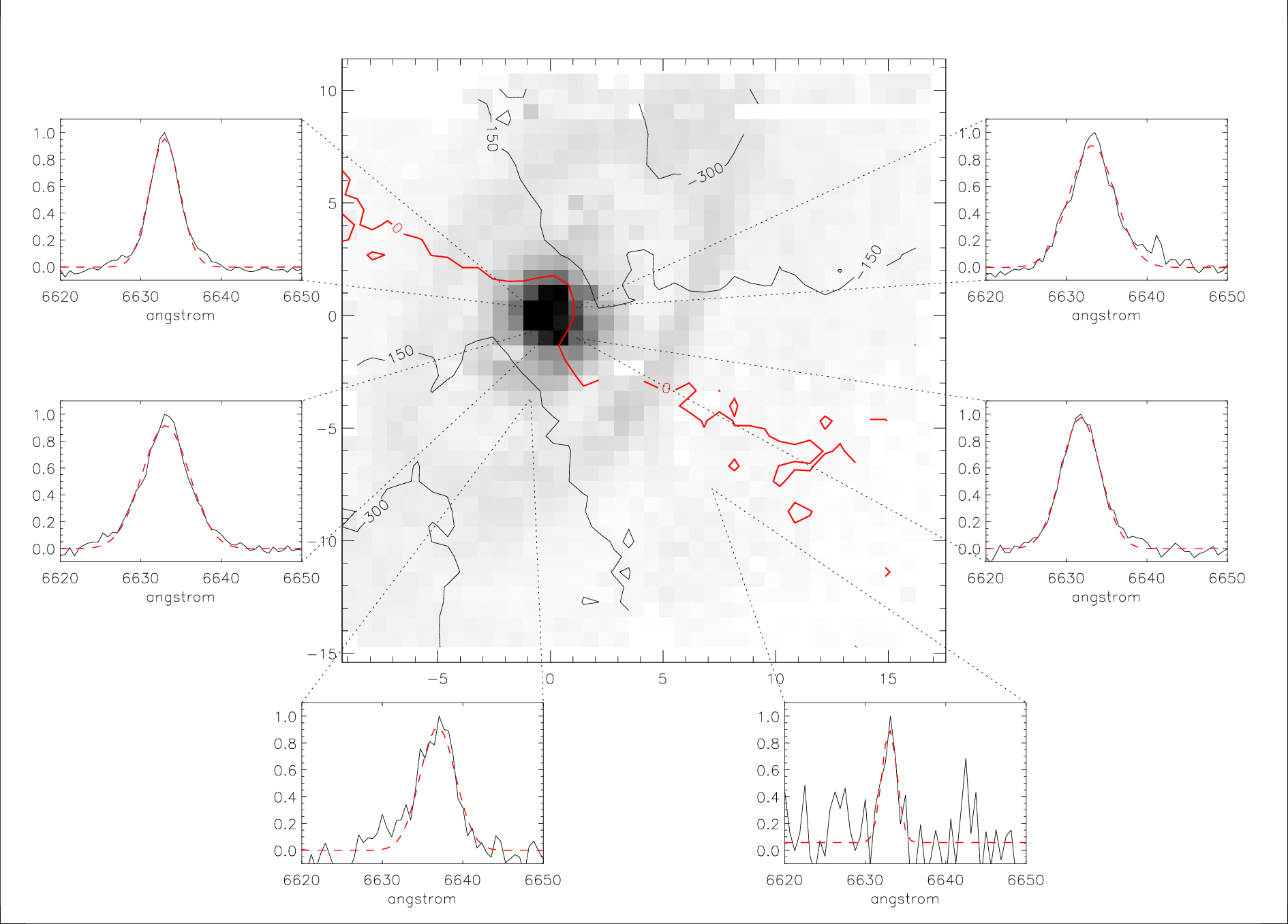,width=17.0cm,clip=}\\
\caption{Same as Fig. \ref{fig:example2855} but for NGC 7049.}
\label{fig:example7049}
\end{figure*}

The result is shown in Figs. \ref{fig:n2855_vfield} and
\ref{fig:n7049_vfield} for NGC~2855 and NGC~7049, respectively.

\begin{figure}
\centering
 \psfig{file=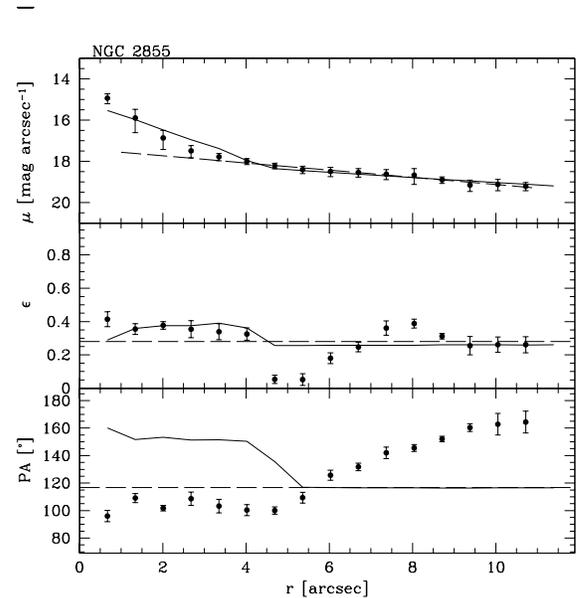,width=8.0cm,clip=}\\
\caption{Radial profiles obtained from the isophotal analysis of the
\niig\ surface-brightness map measured for NGC 2855. {\em Top panel:}
Surface brightness calculated adopting an arbitrary zero point of 20
mag arcsec$^{-1}$. {\em Middle panel:} Ellipticity. {\em Bottom
panel:} Position angle. In each panel the solid line corresponds to
the prediction of the model which assumes that the ionized-gas
component is distributed onto two orthogonal disks (see
Sect. \ref{sec:orthodisk}), while the dashed line corresponds to
the prediction for the single disk model (see
Sect. \ref{sec:maindisk}).}
\label{fig:n2855_ellipse}
\end{figure}

\begin{figure}
\centering
 \psfig{file=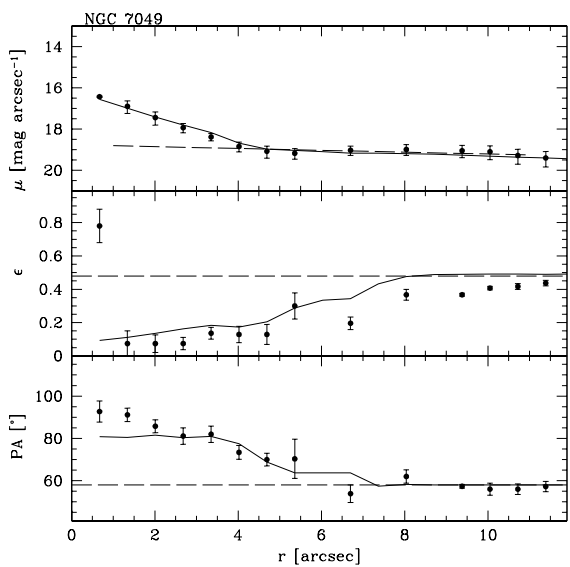,width=8.0cm,clip=}
\caption{Same as in Fig. \ref{fig:n2855_ellipse} but for NGC 7049.}
\label{fig:n7049_ellipse}
\end{figure}

To study the distribution of the ionized gas we fitted elliptical
isophotes to the \niig\ surface brightness map by means of the {\tt
IRAF}\footnote{{\tt IRAF} is distributed by National Optical Astronomy
Observatories, which is operated by Association of Universities for
Research in Astronomy, Inc. under cooperative agreement with the
National Science Foundation.} taks {\tt ELLIPSE}. We first fitted
ellipses allowing the center to vary to test the regularity of the gas
distribution. We found no evidence of a varying center with the fits
for NGC 2885. The ellipse fits were then repeated with the ellipse
center fixed. This was not the case of NGC 7049. The position of
center was constant within the errors only for $r\lesssim4''$, while
it changed by $\approx2''$ at larger radii.  The radial profiles of
the surface brightness, ellipticity and position angle derived from
the \niig\ surface-brightness map are shown in
Figs. \ref{fig:n2855_ellipse} and \ref{fig:n7049_ellipse} for NGC~2855
and NGC~7049, respectively.

\subsection{Results}
\label{sec:results}

In NGC 2855 the ionized gas is concentrated in the nucleus and its
distribution at larger radii follows the spiral pattern of the galaxy
(Fig. \ref{fig:n2855_vfield}).
The abrupt change of ellipticity (from 0.05 to 0.45) and position
angle (from $105^\circ$ to $155^\circ$) of the ellipses fitting the
\niig\ surface-brightness map (Fig. \ref{fig:n2855_ellipse}) are the
photometric signatures of such a distribution of the gaseous
component.
The gas velocity field is complex. It is characterized by a S-shaped
zero velocity line in the nuclear region. In the outer region it shows
a regular rotation around the galaxy minor axis.
The velocity dispersion increases from the outer parts towards the
center. It peaks to $\approx$120 \kms\ in \ha\ and to $\approx150$
\kms\ in \niig\ and is systematically smaller for
\ha\ with respect to \niig\ (Fig. \ref{fig:n2855_vfield}).

In NGC 7049 the gas is mostly concentrated in the nucleus and in a
ring-like structure (Fig. \ref{fig:n7049_vfield}). This corresponds to
the prominent dust lane, which is visible in the optical images of the
galaxy (Sandage \& Bedke 1994, Panel 74). The photometric profiles
derived from the \niig\ surface-brightness map
(Fig. \ref{fig:n7049_ellipse}) allows to constrain the different shape
and orientation of the nuclear ($\epsilon\approx0.1$,
PA$\approx85^\circ$) and ring ($\epsilon\approx0.3$,
PA$\approx55^\circ$) regions.
The gas velocity field of NGC 7049 is similar to that of NGC 2855,
displaying a S-shaped distortion in the nucleus and regular rotation
around the galaxy minor axis at large radii. The velocity dispersion
is systematically lower in \ha\ than in \niig . In the nuclear region
shows a central dip surrounded by a double peak where it reaches
$\approx$ 120 \kms\ in \ha\ and $\approx$ 170 \kms\ in \niig
.  Outwards it decreases smoothly with radius
(Fig. \ref{fig:n7049_vfield}).

In both galaxies the line-of-sight velocities measured from \ha\ and
\niig\ are consistent within the measurement errors. 
For this reason we modeled only the velocity field 
and surface-brightness distribution derived from the \niig\ line.
Typical errors on the line-of-sight velocity and velocity dispersions
were $\approx10$ and $20$ \kms , including the uncertainties due to
wavelength calibration.

\section{Analysis of the velocity field of the ionized gas}
\label{sec:model}

The field of view of VIMOS/IFU allows us to map the velocity field of
the kinematically-decoupled gaseous component, as well as that of the
gas residing in the main disk of the galaxy. As a first step, we fit
the whole velocity field by assuming that the gas is moving onto
circular and coplanar orbits (Sec. \ref{sec:maindisk}). To this aim we
masked the central region, where the S-shaped distorsion of the
isovelocity countours is observed.
We adopted the results of the circular velocity model to constrain the
orientation and inclination of the main disk of the galaxy. Then we
modeled the complex gas kinematics and surface-brightness distribution
of the inner regions. We assumed that the gaseous component was
distributed either on two distinct and orthogonally-rotating disks
(Sec. \ref{sec:orthodisk}) or onto a single and strongly warped disk
(Sec. \ref{sec:warpdisk}).

In our analysis we fitted iteratively a model to the observed velocity
field and surface-brightness distribution of the \niig\ line using a
non-linear least-squares minimization method. It was based on the
robust Levenberg-Marquardt method implemented by Mor\`e et
al. (1980). The actual computation was done using the {\tt MPFIT}
algorithm implemented by C. B. Markwardt under the {\tt IDL}
environment\footnote{The updated version of this code is available on
http://cow.physics.wisc.edu/~craigm/idl/idl.html}.
During each iteration the model was calculated on a subsampled pixel
grid with the bin size of $0\farcs335\times0\farcs335$ (i.e., a
subsampling factor of $2\times2$ relative to the VIMOS pixel
scale). This resulted the best compromise between good sampling and
reasonable computational time.  The model was rebinned on a spatial
scale of $0\farcs67\times0\farcs67$ and {\bf the seeing effects were
taken into account by convolving the model with a Gaussian kernel with
$FWHM=1''$.} Finally, the model was compared to the observed velocity
field and surface-brightness distribution.

\subsection{Single rotating disk}
\label{sec:maindisk}

\subsubsection{Model calculation}
\label{sec:maindisk_model}

The model of the gas velocity field is generated assuming that the
ionized-gas component is moving onto circular orbits in an
infinitesimally thin disk with a negligible velocity dispersion.
Hereafter this will be considered as the main disk of the galaxy.

Let $(X,Y,Z)$ be Cartesian coordinates with the origin in the center
of the gaseous disk, the $Y-$axis aligned along the line of nodes, and
the disk plane confined to the $(X,Y)$ plane. We can consider the gas
motion in this plane by introducing the polar coordinates $(R, \phi)$
with origin in the center of the gaseous disk. It is
$R=\sqrt{X^2+Y^2}$ (or $R=\sqrt{(X-X_c)^2+(Y-Y_c)^2}$ if the
coordinates of the kinematic and geometric center of the gaseous disk
are different) and $\cos \phi= Y/R$ with $\phi$ counted
counter-clockwise from $Y-$axis.
We assume that the gas circular velocity $V$ at a given radius $R$ is
\begin{equation}
V(R) = \frac{2}{\pi} V_{\it max} \arctan{\frac{R}{R_h}} 
\label{eqn:main_circ}  
\end{equation}
where $V_{\it max}$ and $R_h$ are the maximal velocity and a scale
radius, respectively. This empirical function has been proved to
reproduce fairly well the shape of optical rotation curves with the
smallest number of free parameters (Courteau 1997).

We now project the velocity field of the gaseous disk on the plane of
the sky. Let $(x,y,z)$ be Cartesian coordinates with the origin in the
center of the gaseous disk, the $x-$axis pointing eastwards, the
$y-$axis pointing northwards, and the $z-$axis along the line of sight
directed toward the observer. The sky plane is confined to the $(x,y)$
plane. 
In the reference frame of the sky $(x,y,z)$, the gaseous disk is
inclined by the zenithal and azimuthal angle $i$ and $\pa$,
respectively.  These angles correspond to the inclination of the disk
(with $i=0^\circ$ corresponding to the face-on case) and position
angle of its line of nodes (with $\pa$ counted counter-clockwise from
the $y-$axis and $\pa = 0^\circ$ corresponding to the case with the
line of nodes pointing northward).
The transformation of the coordinates of the main disk from its
reference frame $(Z=0)$ to the reference frame of the sky $(x,y,z)$ is
given by
\begin{equation}
\left( 
  \begin{array}{l}x \\ 
                  y \\
		  z
   \end{array}
\right) 
= 
\mathfrak{R}
\left( 
  \begin{array}{l}X \\ 
                  Y \\
		  0
  \end{array}
\right) 
\label{eqn:main1sky}
\end{equation}
where 
\begin{equation}
\mathfrak{R} =
\left( 
  \begin{array}{ccc}
    \cos \pa \cos i  & \sin \pa & \cos \pa \sin i \\ 
   -\sin \pa \cos i  &  \cos \pa & -\sin \pa \sin i \\ 
    -\sin i                & 0         & \cos i 
  \end{array}
\right). 
\label{eqn:r}
\end{equation}
which leads to the usual equations in the sky plane $(x,y)$:
\begin{eqnarray}
x &=&    X \cos \pa \cos i + Y \sin i  \nonumber \\  
y &=&  - X \sin \pa \cos i + Y \cos \pa \nonumber 
\end{eqnarray}

The ionized-gas velocity measured along the line of sight at a given
sky point $(x,y)$ is
\begin{equation}
v(x,y) = V(R) \sin i \cos \phi + V_{\it sys} 
\label{eqn:main_losv} 
\end{equation}
where the circular velocity $V$ is given in Eq. \ref{eqn:main_circ},
and $V_{\it sys}$ is the systemic velocity of the galaxy. 

The parameters of our model are the maximal velocity $V_{\it max}$,
scale radius $R_h$, inclination $i$ and position angle $\pa$ of the
gaseous disk, and the systemic velocity $V_{\it sys}$ of the galaxy.
The best-fitting model was obtained by masking the region of the
observed velocity field characterized by the S-shaped distorsion of
the isovelocity countours. The area of the masked region was
$9\times11$ pixels and $7\times8$ pixels in NGC 2855 and NGC 7049,
respectively.

\begin{figure}
\vbox{
\hbox{
 \psfig{file=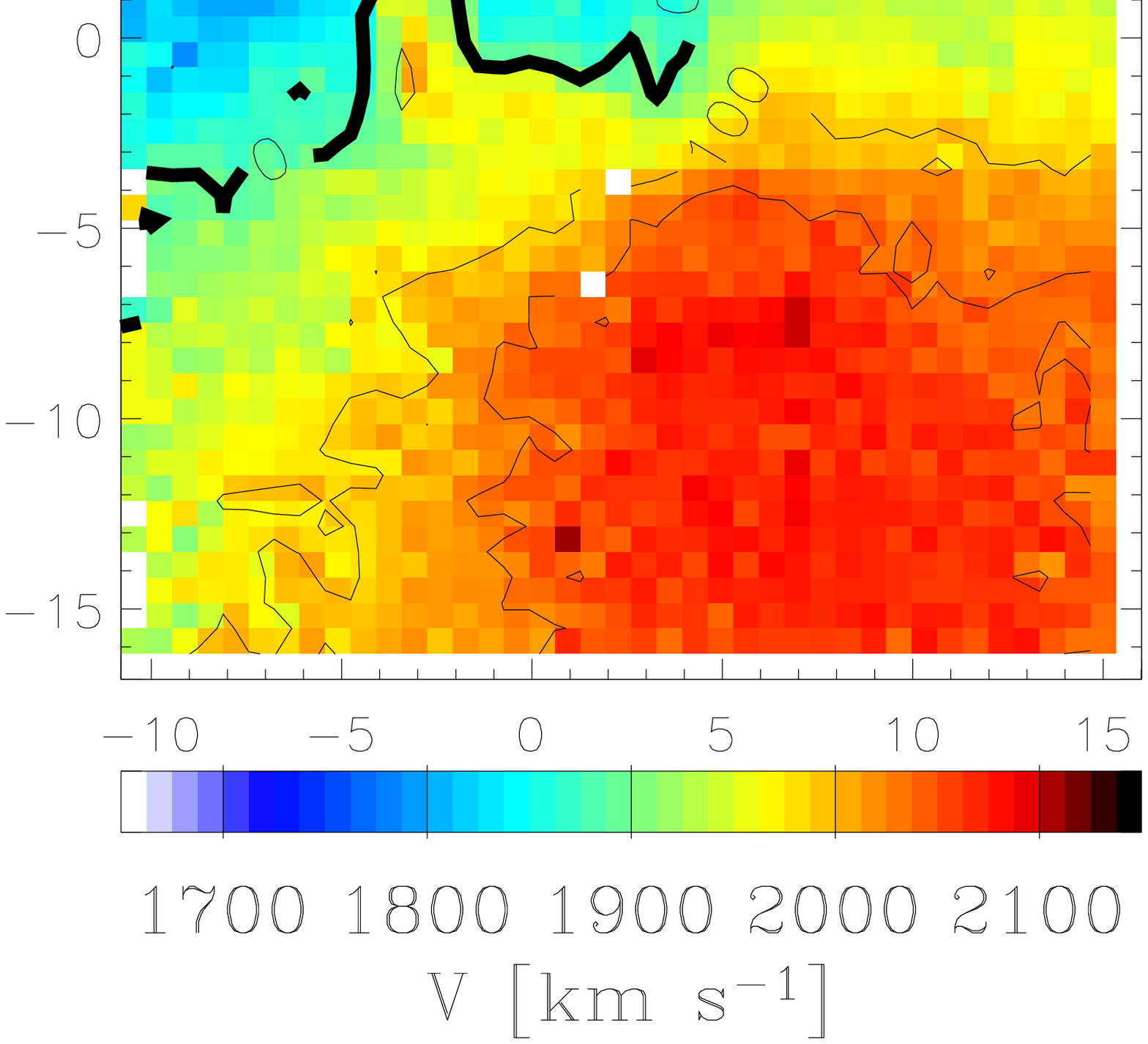,width=2.93cm,clip=}
 \psfig{file=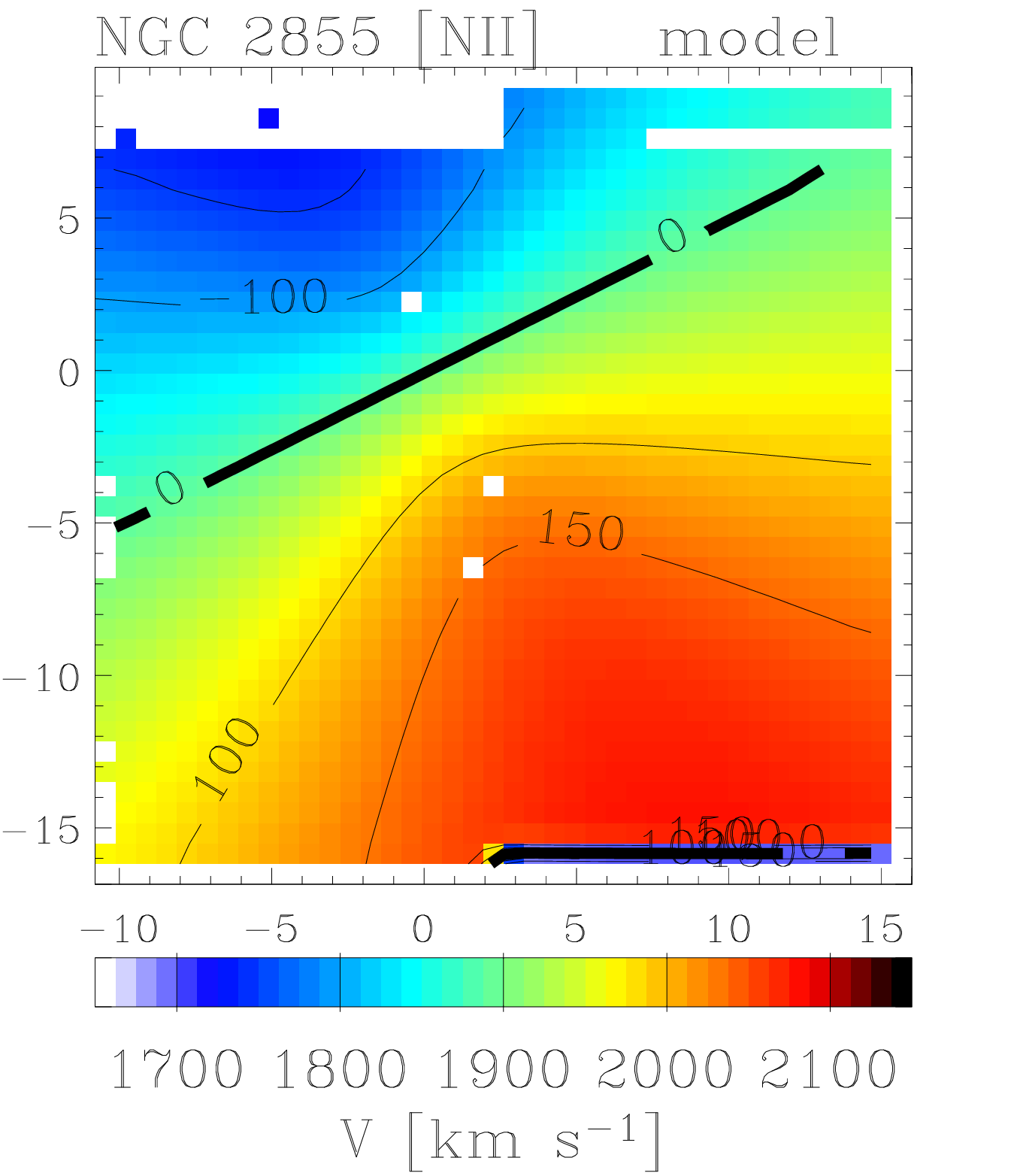,width=2.93cm,clip=} 
 \psfig{file=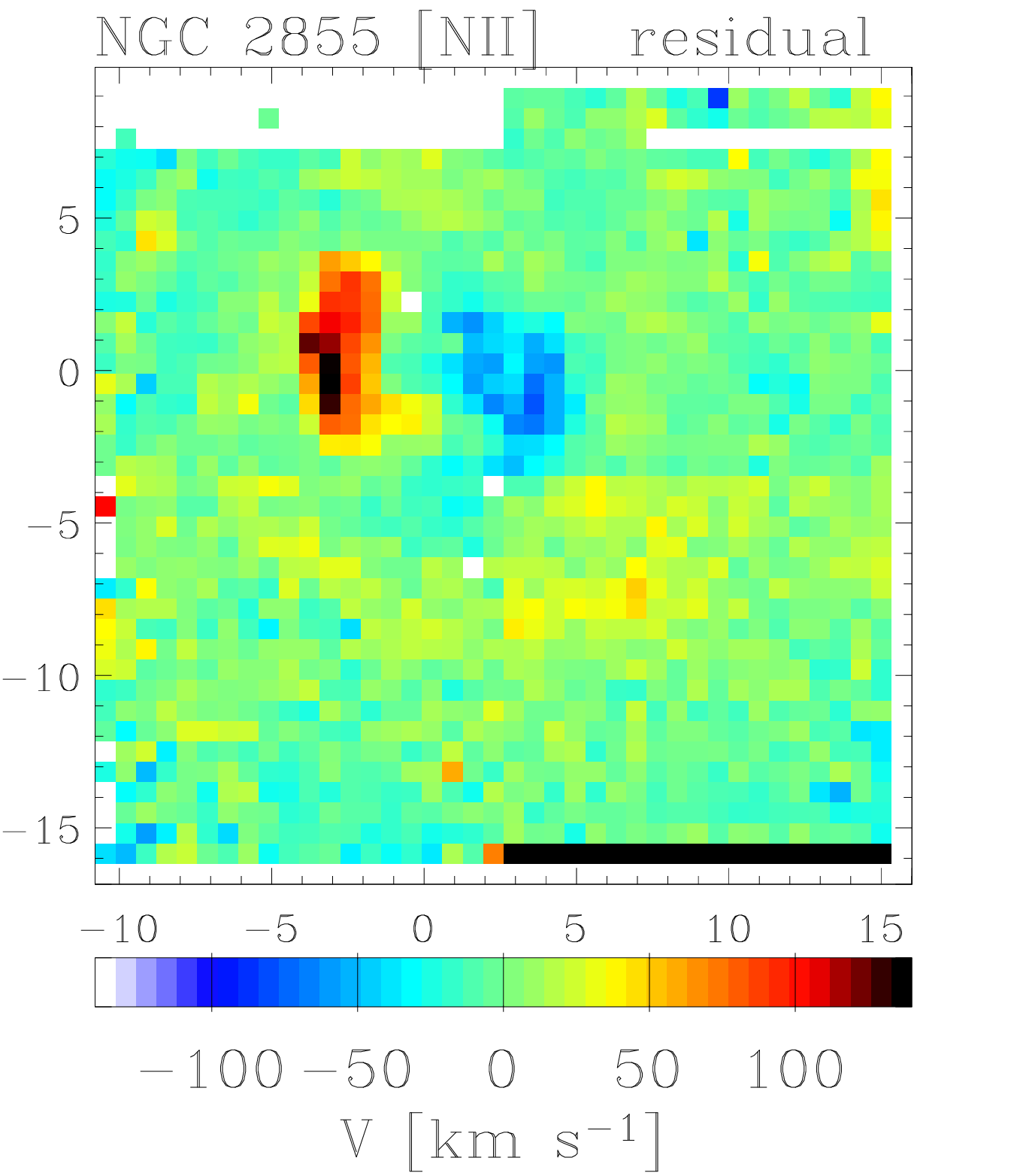,width=2.93cm,clip=}
}
\hbox{
 \psfig{file=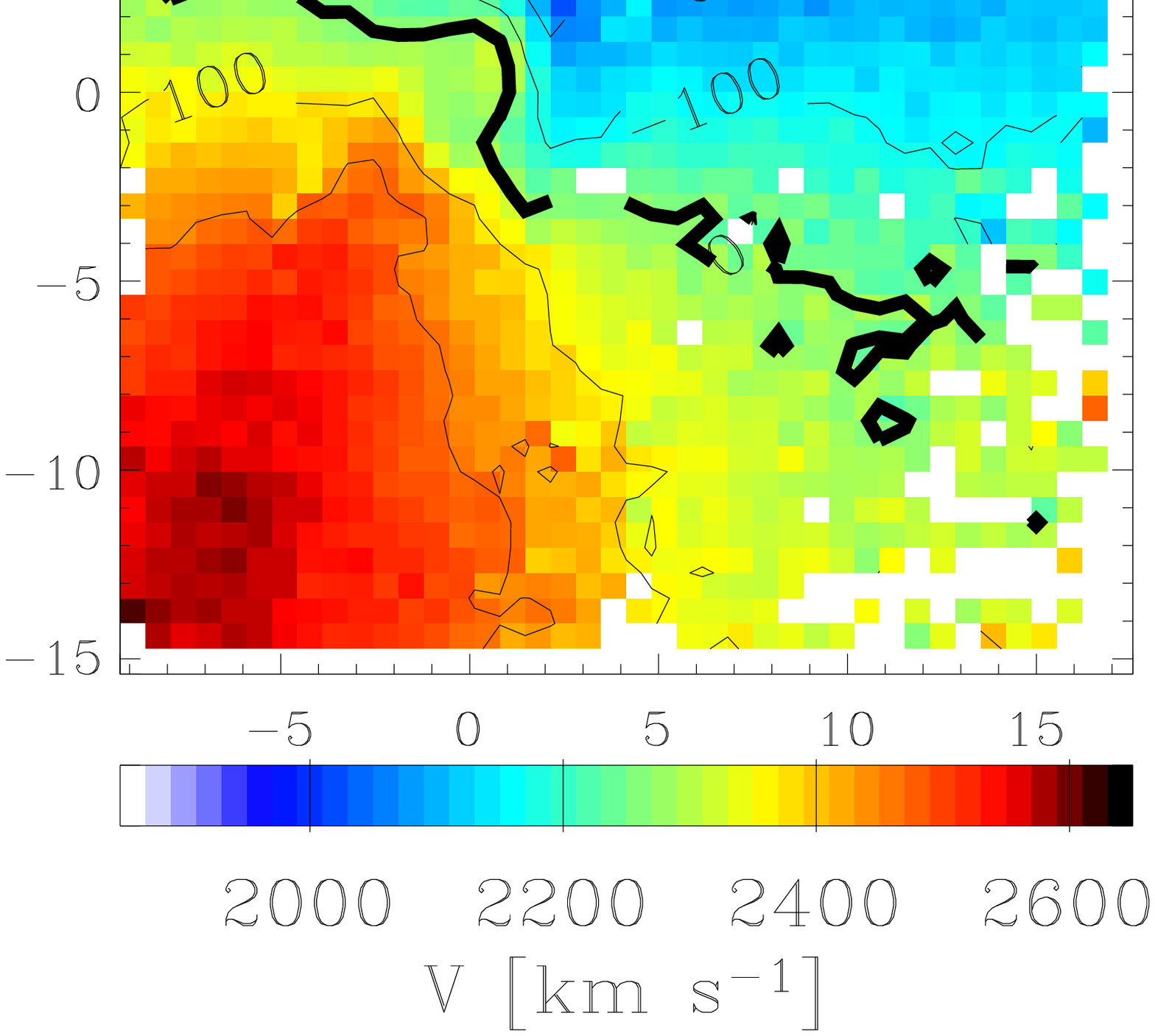,width=2.93cm,clip=}
 \psfig{file=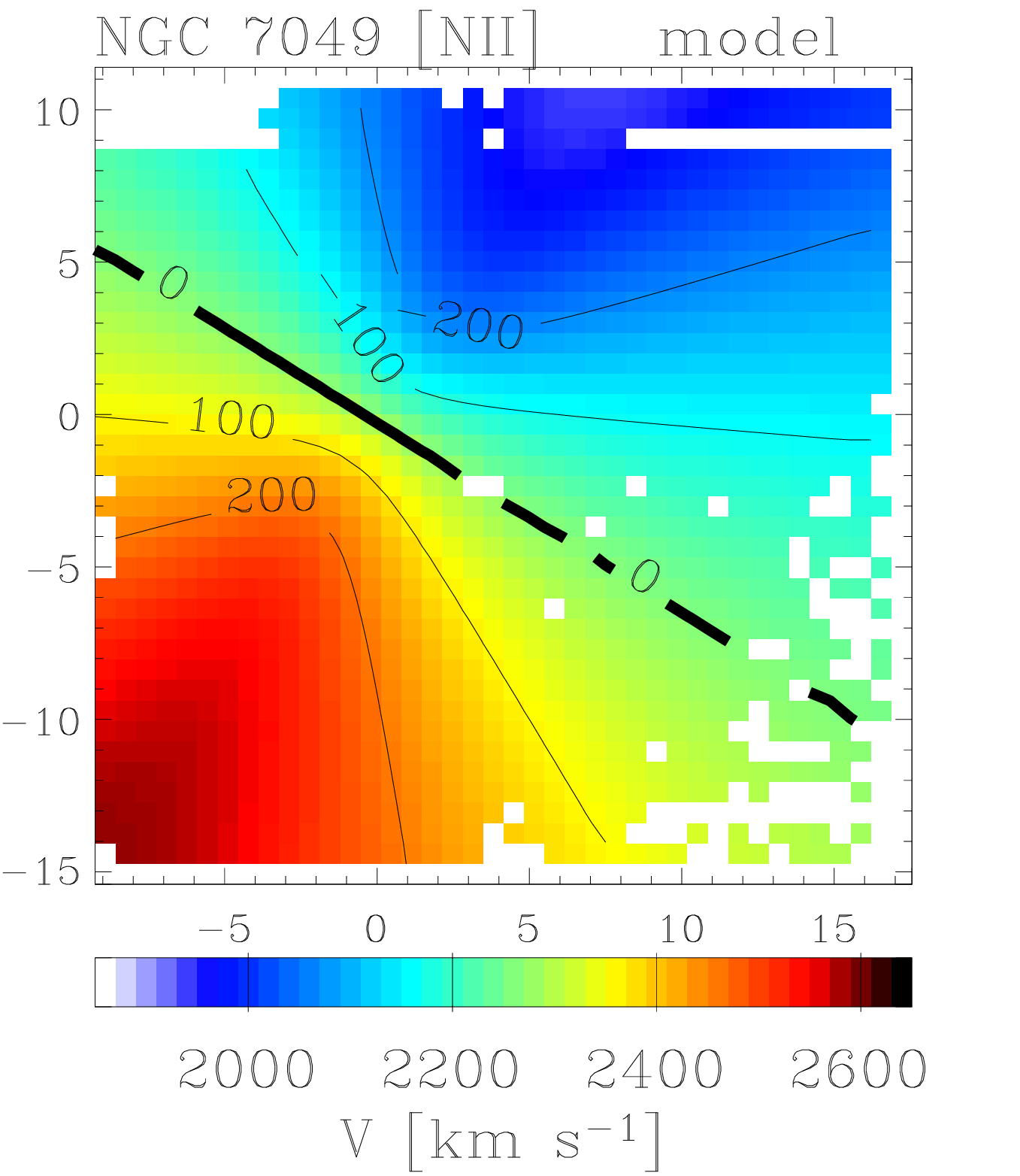,width=2.93cm,clip=}
 \psfig{file=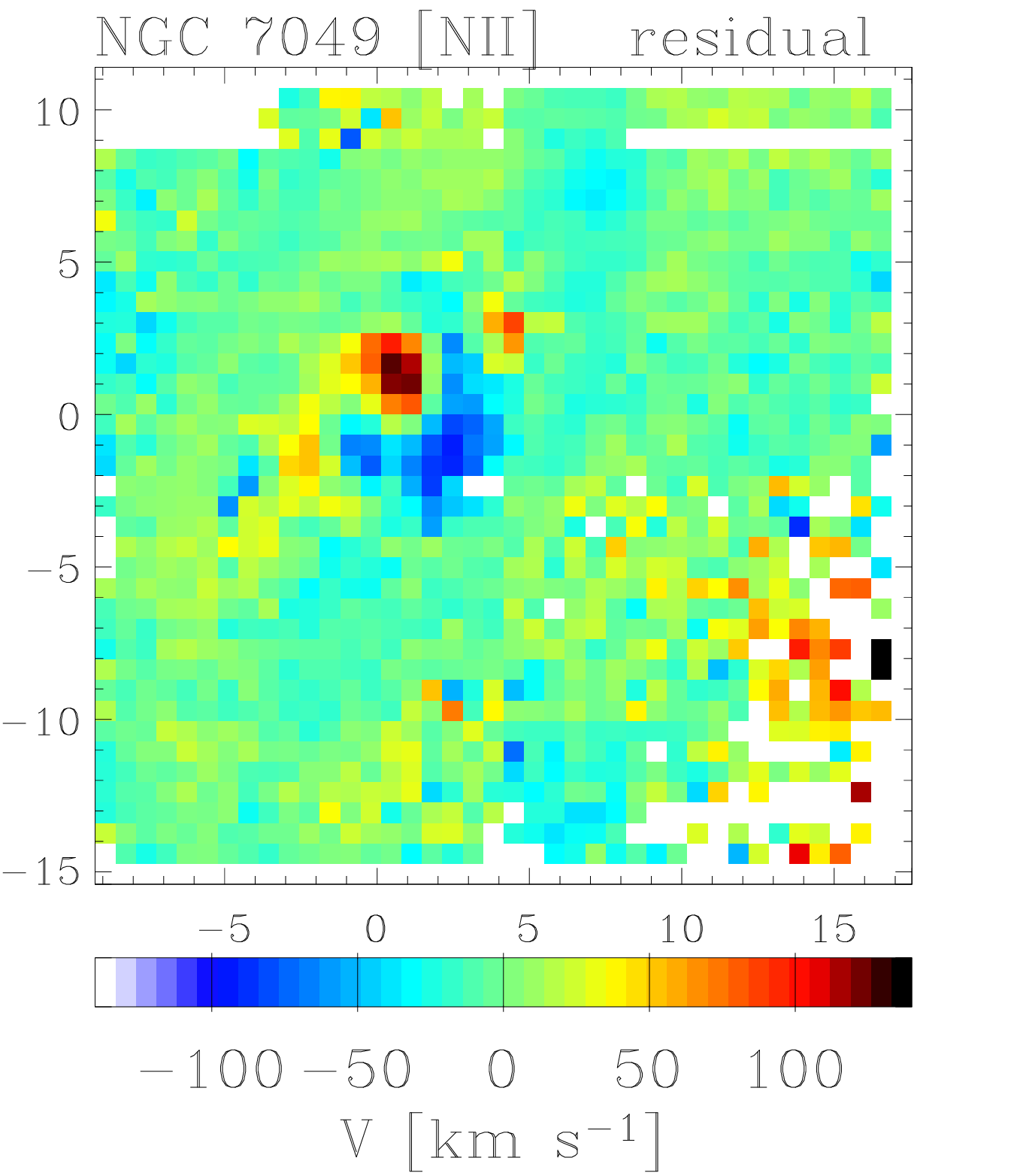,width=2.93cm,clip=}
}}
\caption{Model of the velocity field of NGC 2855 ({\em upper
panels}) and NGC 7049 ({\em lower panels}) with a single rotating
disk.  The field of view, orientation, ranges and isovelocity contours
are as in Fig. \ref{fig:n2855_vfield}.  {\em Left panel}: Observed
velocity field. {\em Central panel}: Model. {\em Right panel}:
Residuals.}
\label{fig:maindisk}
\end{figure}

\subsubsection{Results}
\label{sec:maindisk_results}

The velocity fields of the best-fitting disk models are compared to
observed ones in Fig. \ref{fig:maindisk}.  Best-fitting parameters are
given in Table \ref{tab:maindisk}.

\begin{table}
\centering
\caption{Model with a single rotating disk}
\begin{tabular}{l r r}
\noalign{\smallskip}
\hline
\hline
\noalign{\smallskip}
\multicolumn{1}{c}{Parameter}&
\multicolumn{1}{c}{NGC~2855}&
\multicolumn{1}{c}{NGC~7049}\\
\noalign{\smallskip}
\hline
\noalign{\smallskip}
$X_c$   [pixel]     &  $-$0.4$\pm$0.5 &      $-$0.2$\pm$0.3 \\
$Y_c$   [pixel]     &   0.4$\pm$0.6  &      $-$0.6$\pm$0.3 \\ 
$V_{sys}$ [\kms]    &  1893$\pm$6  &      2230$\pm$5 \\
$\theta$ [$^\circ$] &  {\bf 116.6}$\pm$0.8  &      58.1$\pm$0.7 \\
$i$ [$^\circ$]      &  42$\pm$4  &      {\bf 60}$\pm$2 \\
$V_{max}$ [\kms]    &  $-$344$\pm$22  &      $-$459$\pm$ 13 \\
$R_h$ [$''$]        &  3.9$\pm$0.7  &      3.5$\pm$0.4 \\
\noalign{\smallskip}
\hline
\noalign{\smallskip}
\end{tabular}
\label{tab:maindisk}
\end{table}

For the gaseous disk of NGC 2855, the position angle and inclination
measured at large radii with the isophotal analysis
(Sec. \ref{sec:analysis}) are not consistent with that derived from
the kinematic model (Tab. \ref{tab:maindisk}). We explain such a
discrepancy as due to the presence of the patchy spiral arm pattern,
which characterizes the surface brightness distribution of the \niig\
line at $r\gtrsim4''$.  This makes the photometric parameters not
reliable to describe the geometry of the disk. The coordinates of the
disk centers derived in the photometric and kinematic analysis
(Tab. \ref{tab:maindisk}) coincide within a pixel.

The position angle, inclination and center derived from the
photometric and kinematic analysis of the gaseous disk of NGC 7049 are
consistent within the errors.

In the rest of the paper we adopt as position angle and inclination of
the main gaseous disks of NGC 2855 and NGC 7049 the values we derived
from the kinematic model.

\subsection{Orthogonally-rotating disks}
\label{sec:orthodisk}

\subsubsection{Model calculation}
\label{sec:orthodisk_model}

The model of the gas velocity field is generated assuming that there
are two different ionized-gas components which are moving onto
circular orbits in two orthogonally-rotating disks. They are the main
galactic disk studied in Sect. \ref{sec:maindisk} and an inner
orthogonal disk. Both disks have the same center, are infinitesimally
thin, and have a negligible velocity dispersion.

\begin{figure}
\centering
\psfig{file=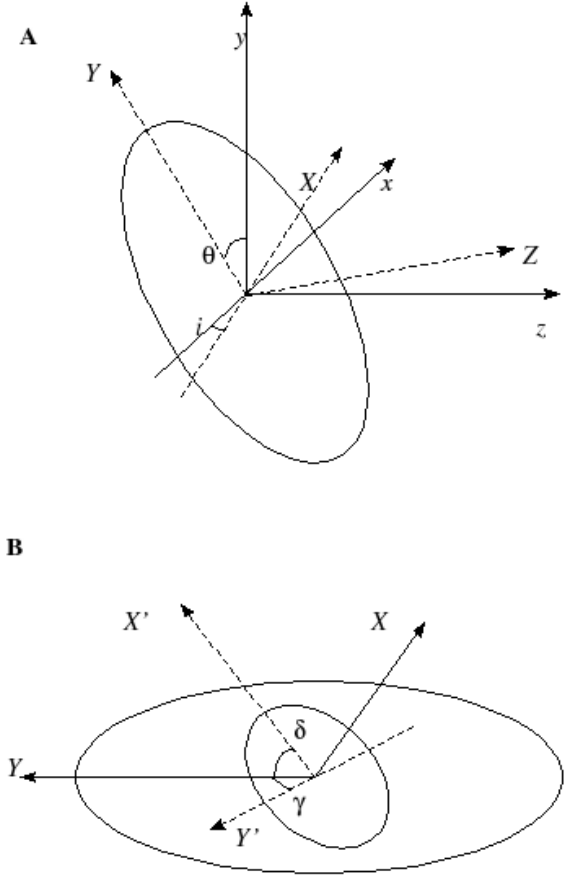,width=8cm,clip=0}
\caption{{\em Panel A}: Orientation of the plane ($X,Y$)
of the main disk of the galaxy with respect to the sky plane
($x,y$). {\em Panel B}: Orientation of the plane ($X',Y'$) of the
orthogonal disk with respect to the plane ($X,Y$) of the main disk. }
\label{fig:coord_sist}
\end{figure}

Let $(X',Y',Z')$ be Cartesian coordinates with the origin in the
center of the orthogonal disk, the $Y'-$axis aligned along the line of
nodes (defined as the intersection between the plane of the orthogonal
disk and that of the main disk), and the plane of the orthogonal disk
confined to the $(X',Y')$ plane.
In the reference frame of the main disk $(X,Y,Z)$, the orthogonal
disk is inclined by the zenithal and azimuthal angle $\delta$ and
$\gamma$, respectively.
These angles are the inclination of the inner disk (with
$\delta=90^\circ$ corresponding to the case of an orthogonal disk) and
position angle of its line of nodes (with $\gamma$ counted
counter-clockwise from the $Y-$axis and $\gamma = 0^\circ$
corresponding to the case with the line of nodes of the orthogonal
disk pointing in the same direction as the line of nodes of the main
disk). A sketch of the orthogonal disk in the reference frames
$(X,Y,Z)$ and $(X',Y',Z')$ is shown in Figure \ref{fig:coord_sist}.

We assumed that the circular velocity of the gaseous component in the
main disk and in the orthogonal disk are given by
Eq. \ref{eqn:main_circ} and
\begin{equation}
V_{\it OD}(R') = \frac{2}{\pi} V'_{\it max} \arctan{\frac{R'}{R'_h}}, 
\label{eqn:od_circ}  
\end{equation}
respectively. $V'_{\it max}$ and $R'_h$ are the maximal velocity and
scale radius of the orthogonal disk. It is $R'=\sqrt{X'^2+Y'^2}$ and
$\cos \phi'= Y'/R'$ with $\phi'$ counted counter-clockwise from
$Y'-$axis.

The flux radial profile of the main gaseous disk was assumed to be
exponential
\begin{equation}
F_{\it MD}(R) = F_0 + F_1 e^{-R/R_F}.
\label{eqn:main_flux}  
\end{equation}
For the flux radial profile of the orthogonal disk we assumed a
truncated exponential function
\begin{equation}
F_{\it OD}(R') = \left\{
\begin{array}{ll}
F'_0 + F'_1 e^{-R'/R'_F} & {\rm for}\ R'\leq R'_0 \\
0                        & {\rm for}\ R'> R'_0
\end{array}
\right.
\label{eqn:od_flux}  
\end{equation}
where $R'_0$ corresponds to the radial extension of the orthogonal
gaseous component.

We now project the velocity field of the two gaseous disks on the
plane of the sky. The transformation of the coordinates of the inner
polar disk from its reference frame $(X',Y',Z')$ to the reference
frame of the main disk $(X,Y,Z)$ is given by
\begin{equation}
\left( 
  \begin{array}{l}X \\ 
                  Y \\ 
                  Z
   \end{array}
\right) 
= 
\mathfrak{R}1
\left( 
  \begin{array}{l}X' \\ 
                  Y' \\ 
                  Z'
  \end{array}
\right) 
\label{eqn:od2main}
\end{equation}
where 
\begin{equation}
\mathfrak{R}1 =
\left( 
  \begin{array}{ccc}
    \cos \gamma \cos \delta  & \sin \gamma & \cos \gamma \sin \delta  \\ 
   -\sin \gamma \cos \delta  & \cos \gamma & -\sin \gamma \sin \delta \\ 
   -\sin \delta              & 0           & \cos \delta 
  \end{array}
\right). 
\label{eqn:r1}
\end{equation}

The transformation of the coordinates of the main disk from its
reference frame $(X,Y,Z)$ to the reference frame of the sky $(x,y,z)$
is given by
\begin{equation}
\left( 
  \begin{array}{l}x \\ 
                  y \\ 
                  z
   \end{array}
\right) 
= 
\mathfrak{R}2
\left( 
  \begin{array}{l}X \\ 
                  Y \\ 
                  Z
  \end{array}
\right) 
\label{eqn:main2sky}
\end{equation}
where 
\begin{equation}
\mathfrak{R}2 =
\left( 
  \begin{array}{ccc}
    \cos \pa \cos i  & \sin \pa &  \cos \pa \sin i  \\ 
   -\sin \pa \cos i  & \cos \pa & -\sin \pa \sin i \\ 
   -\sin i           & 0        & \cos i
  \end{array}
\right). 
\label{eqn:r2}
\end{equation}

The transformation of the coordinates of the orthogonal disk from
its reference frame $(X',Y',Z')$ to the reference frame of the sky 
$(x,y,z)$ is given by
\begin{equation}
\left( 
  \begin{array}{l}x \\ 
                  y \\ 
                  z
   \end{array}
\right) 
= 
\mathfrak{R}
\left( 
  \begin{array}{l}X' \\ 
                  Y' \\ 
                  Z'
  \end{array}
\right) 
\label{eqn:od2sky}
\end{equation}
where $\mathfrak{R} = \mathfrak{R}2 \mathfrak{R}1$. 

The inclination $i'$ of the orthogonal disk with respect to the
sky plane is given by the direction cosine between the $z-$axis of the
line of sight and the $Z'-$axis of the orthogonal disk
\begin{equation}
\cos i' = \mathfrak{R}_{3,3}= \cos \delta  \cos i 
  -\sin \delta  \cos \gamma \sin i. 
\label{eqn:odinc}
\end{equation}

The line of nodes of the orthogonal disk, defined as the intersection
between the disk plane and the sky plane, is
\begin{equation}
Z' = \mathfrak{R}'_{3,1} x + \mathfrak{R}'_{3,2} y 
  + \mathfrak{R}'_{3,3} z = 0 
\label{eqn:odlon1}
\end{equation}
where $\mathfrak{R}'$ denotes the inverse of $\mathfrak{R}$. In the
sky plane it is 
\begin{equation}
\mathfrak{R}'_{3,1} x + \mathfrak{R}'_{3,2} y = 0 
\label{eqn:odlon2}
\end{equation}
and therefore the position angle $\theta'$ of the apparent major axis of
the orthogonal disk with respect to the $y-$axis is given by $\tan
\theta' = -\mathfrak{R}'_{3,2}/\mathfrak{R}'_{3,1}$.  This is
\begin{equation}
\tan \theta' = \frac{\sin\pa (\sin\delta \cos\gamma \cos i+\cos\delta \sin i) + 
  \cos\pa \sin \delta \sin\gamma}
                 {\cos\pa (\sin\delta \cos\gamma \cos i+\cos\delta \sin i) -
  \sin\pa \sin\delta \sin\gamma}  
\label{eqn:odpa}
\end{equation}

The ionized-gas velocity measured along the line of sight at a given
sky point $(x,y)$ is
\begin{equation}
v(x,y) = \frac{v_{\it MD}(x,y) f_{\it MD}(x,y) + 
  v_{\it OD}(x,y) f_{\it OD} (x,y)}{f(x,y)} 
\label{eqn:total_losv}
\end{equation}
where 
\begin{eqnarray}
v_{\it MD}(x,y) &=& V_{\it MD}(R) \sin i \cos \phi + V_{\it sys} \\
f_{\it MD}(x,y) &=& F_{\it MD}(R) /\cos i 
\end{eqnarray}
are the line-of-sight velocity and surface brightness of the main disk,
\begin{eqnarray}
v_{\it OD}(x,y) &=& V_{\it OD}(R) \sin i' \cos \phi' + V_{\it sys} \\
f_{\it OD}(x,y) &=& F_{\it OD}(R) /\cos i' 
\end{eqnarray}
are the line-of-sight velocity and surface brightness of the
orthogonal disk, and $f(x,y)=f_{\it MD}(x,y) + f_{\it OD}(x,y)$ is the
observed surface brightness.

For the main disk we adopted the inclination $i$ and position angle
$\pa$ derived in Sect. \ref{sec:maindisk} and we assumed the inner
disk to be orthogonal with respect to the main one within
$\pm1^\circ$.
Therefore the parameters of our model are $V_{\it max}$, $R_h$, $F_0$,
$F_1$, and $R_F$ for the main disk, $V'_{\it max}$, $R'_h$, $F'_0$,
$F'_1$, $R'_F$, $R'_0$, $i'$ (with 
$89^\circ \leq |i'-i| \leq 91^\circ$)
and $\theta'$ for the orthogonal disk, and the systemic velocity
$V_{\it sys}$ of the galaxy.

We started fitting a model only to the observed \niig\
surface-brighteness distribution (Fit \#1 hereafter). Then we fitted a model
only to the observed velocity field adopting for the main disk the
maximal velocity $V_{\it max}$ and scale radius $R_h$ derived in
Sect. \ref{sec:maindisk} and for the orthogonal disk the inclination
$i'$ and position angle $\theta'$ derived in Fit \#1 (Fit
\#2 hereafter). Finally, we fitted simultaneously both the velocity field and
surface-brightness distribution (Fit \#3 hereafter).

\subsubsection{Results}
\label{sec:orthodisk_results}

\begin{figure}
\vbox{
\hbox{
 \psfig{file=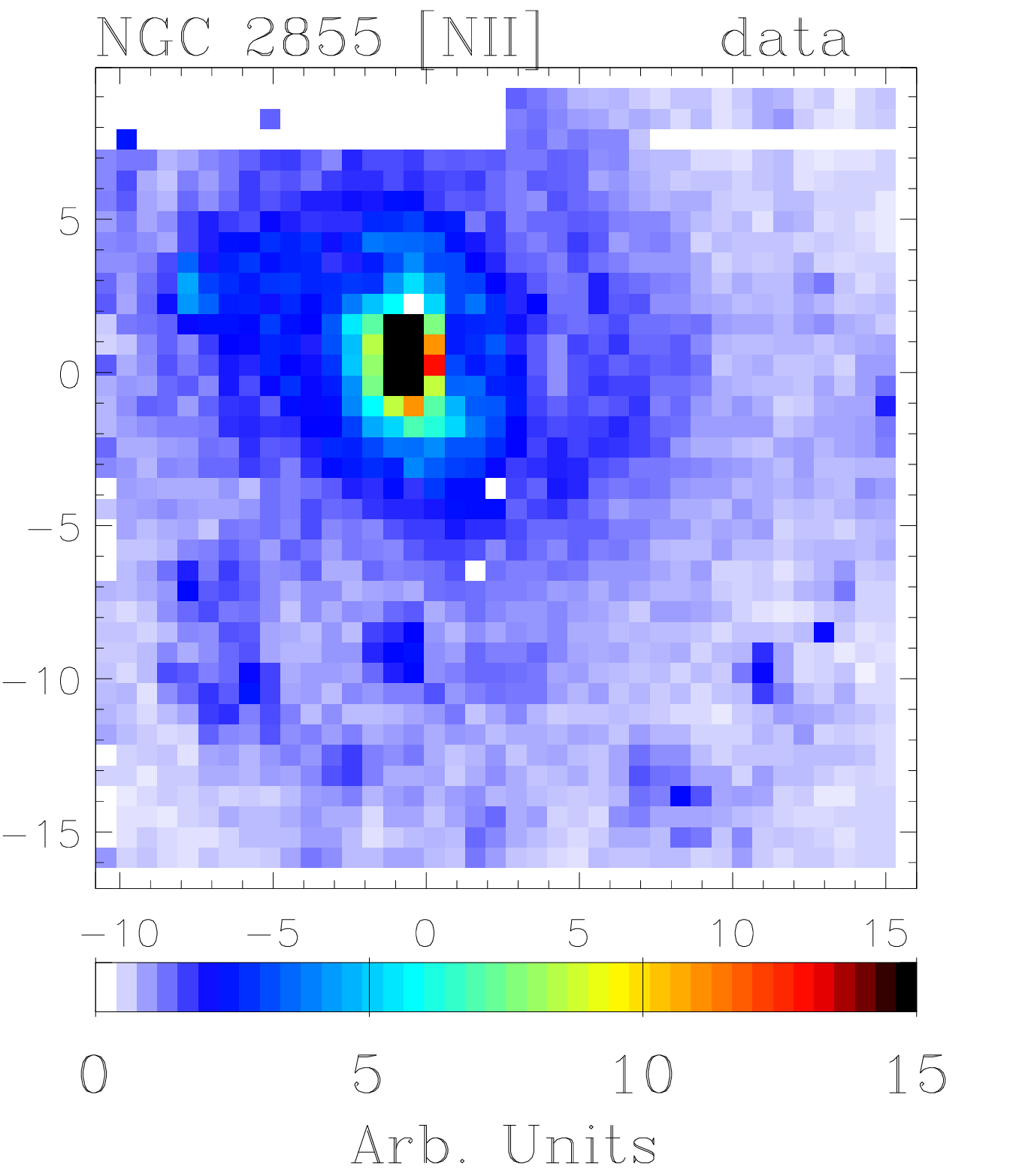,width=2.93cm,clip=}
 \psfig{file=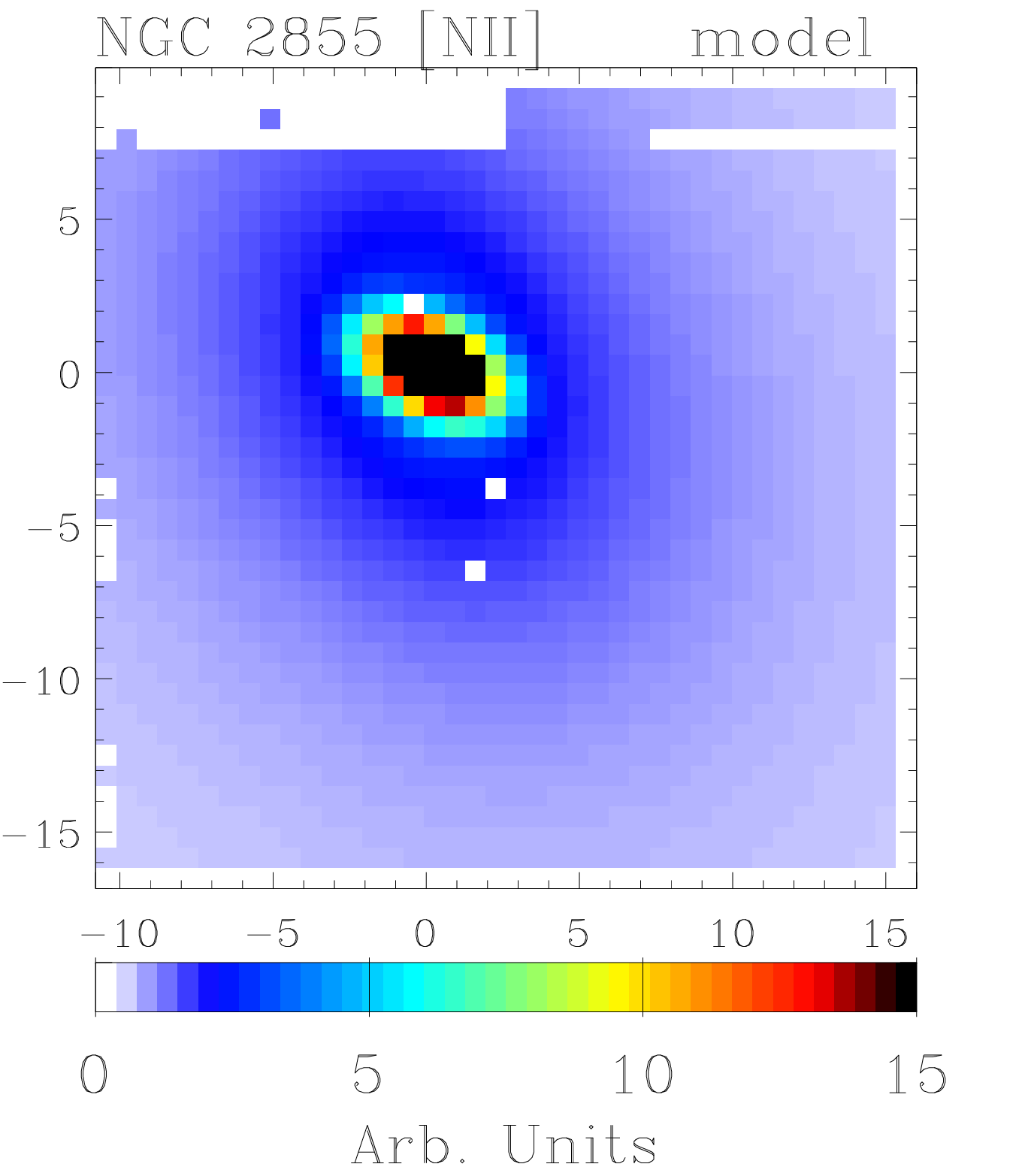,width=2.93cm,clip=}
 \psfig{file=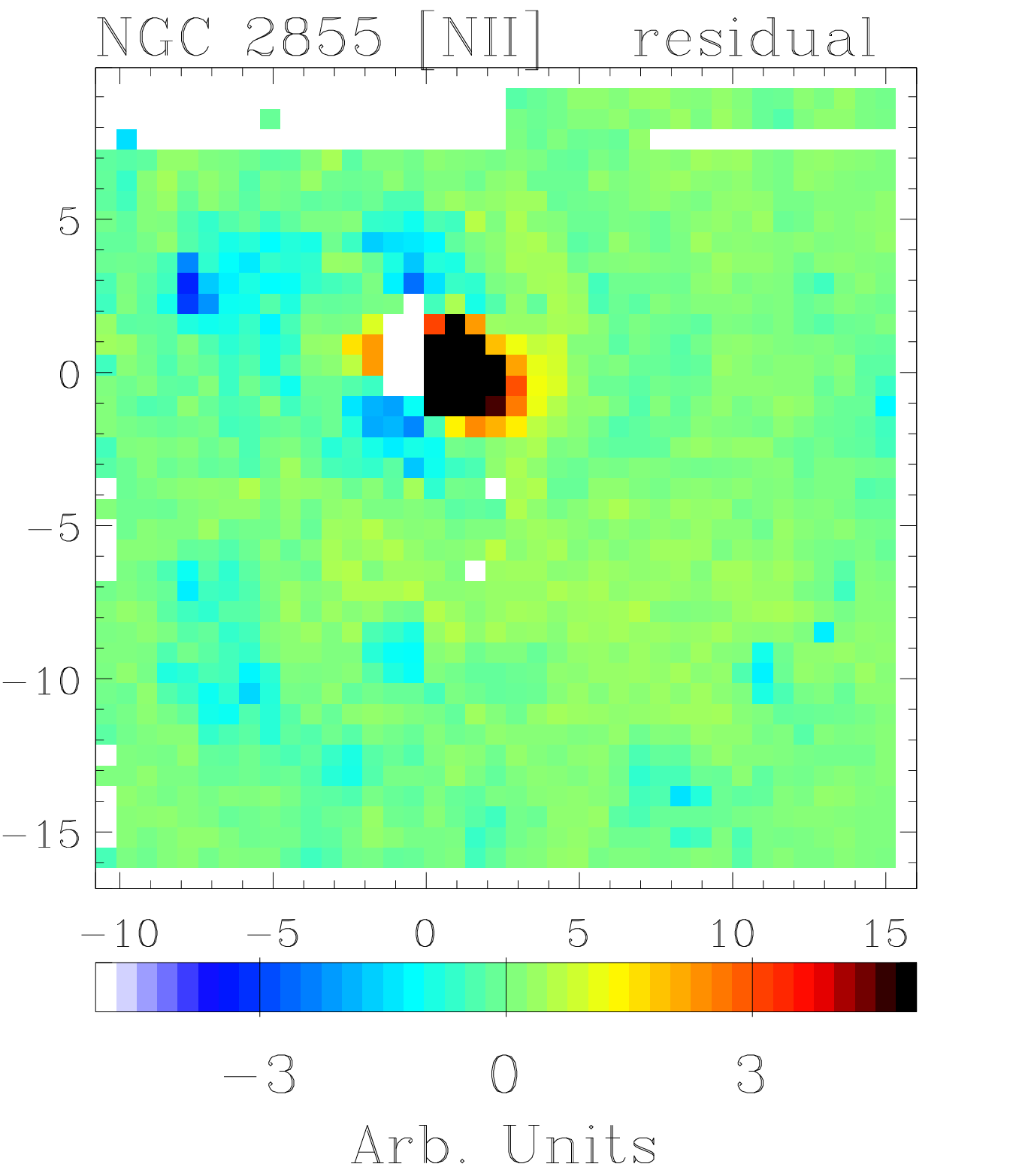,width=2.93cm,clip=} 
}
\hbox{
 \psfig{file=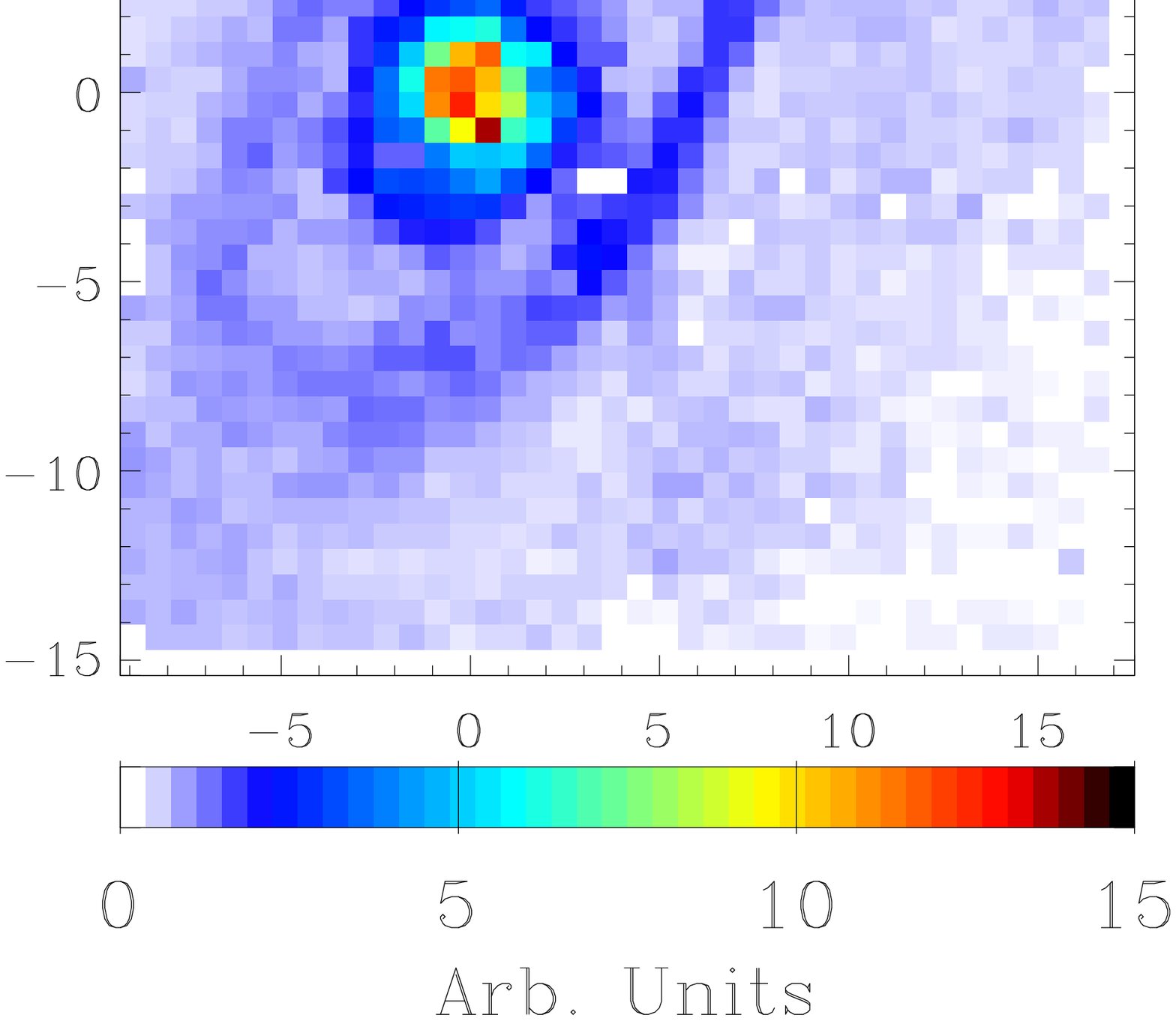,width=2.93cm,clip=}
 \psfig{file=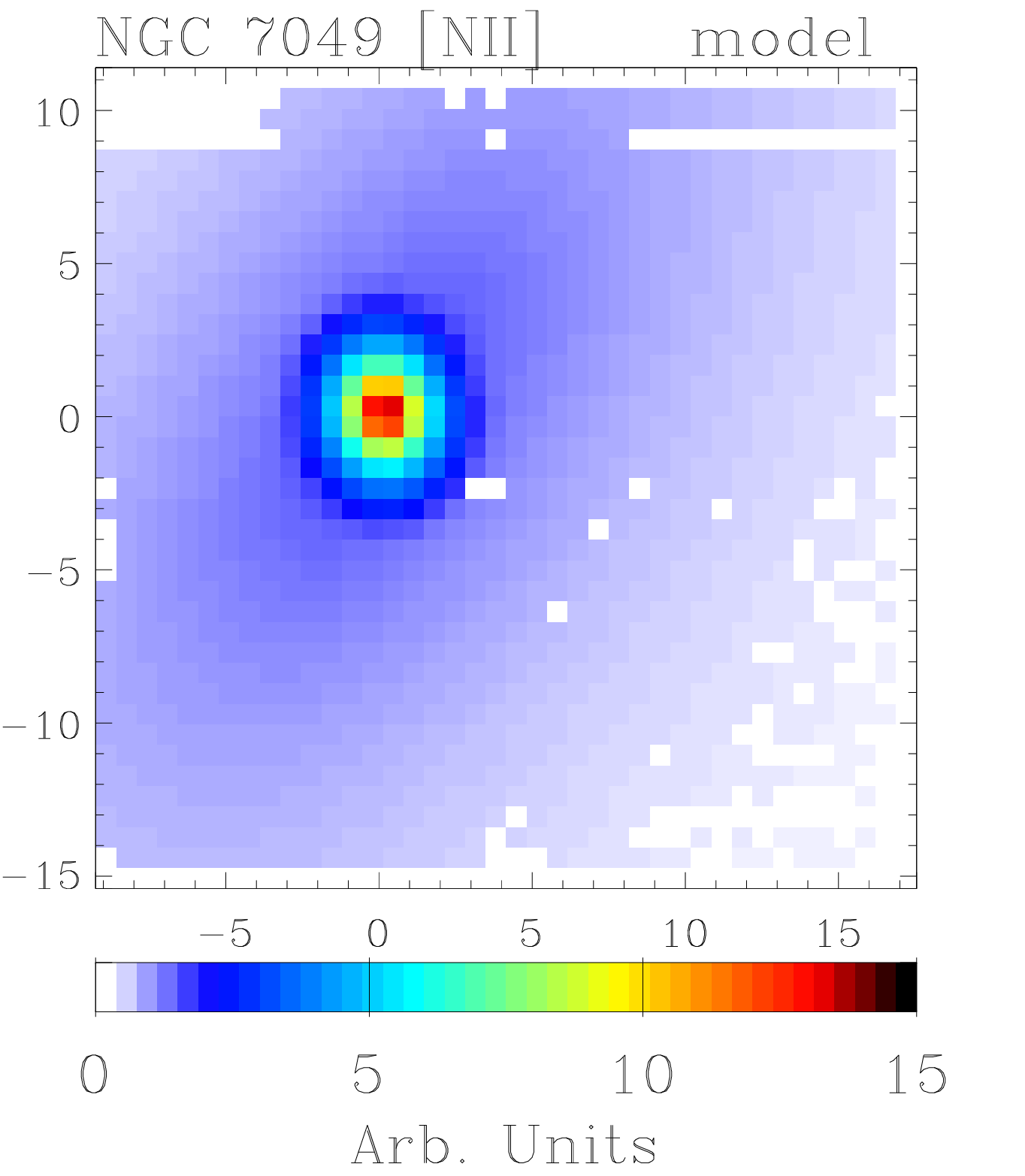,width=2.93cm,clip=}
 \psfig{file=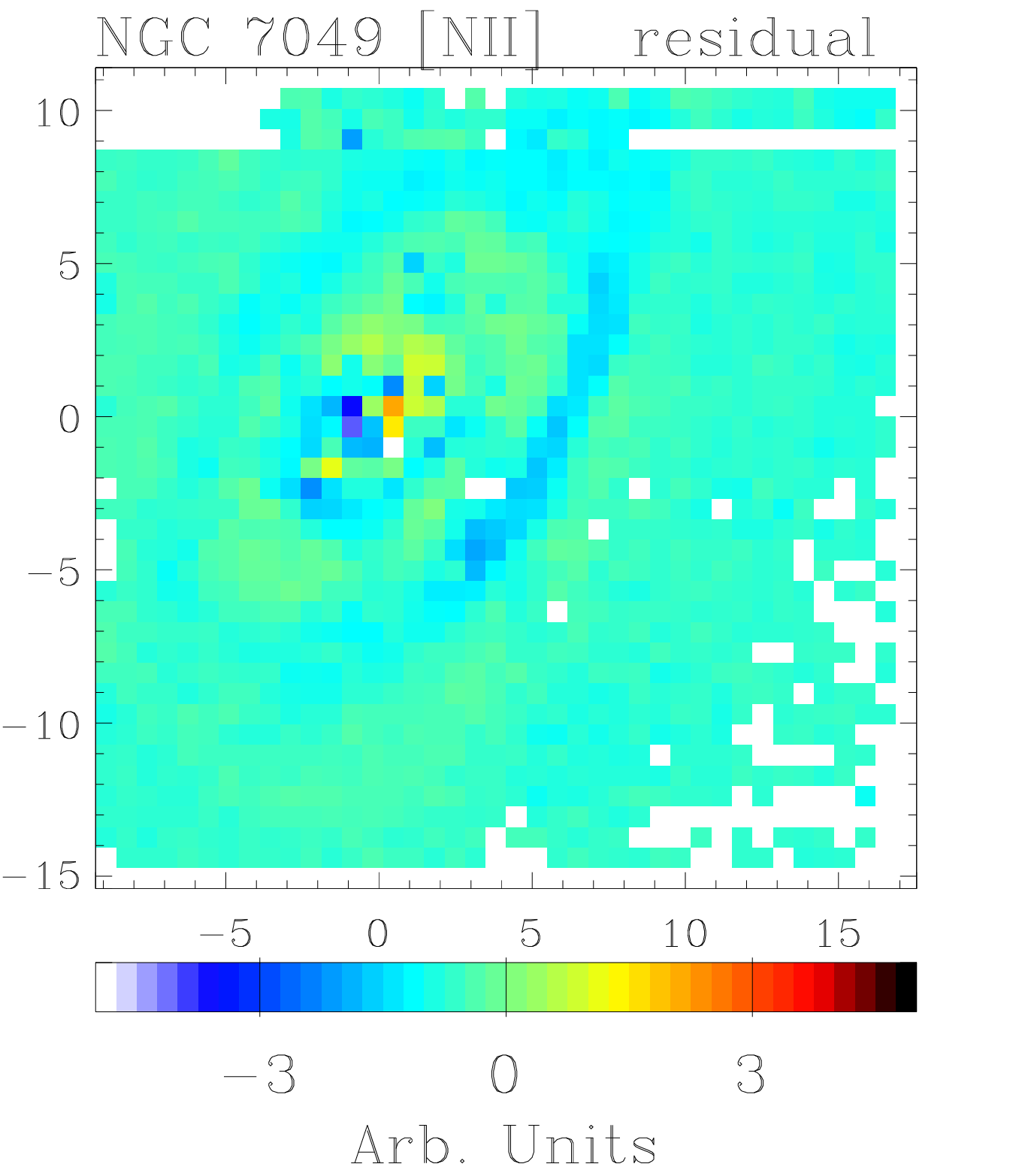,width=2.93cm,clip=} 
}}
\caption{Model of the surface-brightness distribution of NGC 2855
({\em upper panels}) and NGC 7049 ({\em lower panels}) with two
orthogonally-rotating disks (Fit \#1).  The field of view,
orientation, ranges and isovelocity contours are as in
Fig. \ref{fig:n2855_vfield}.  {\em Left panel}: Observed
surface-brightness distribution. {\em Central panel}: Model. {\em
Right panel}: Residuals.}
\label{fig:od_fit1}
\end{figure}

\begin{figure}
\vbox{
\hbox{
 \psfig{file=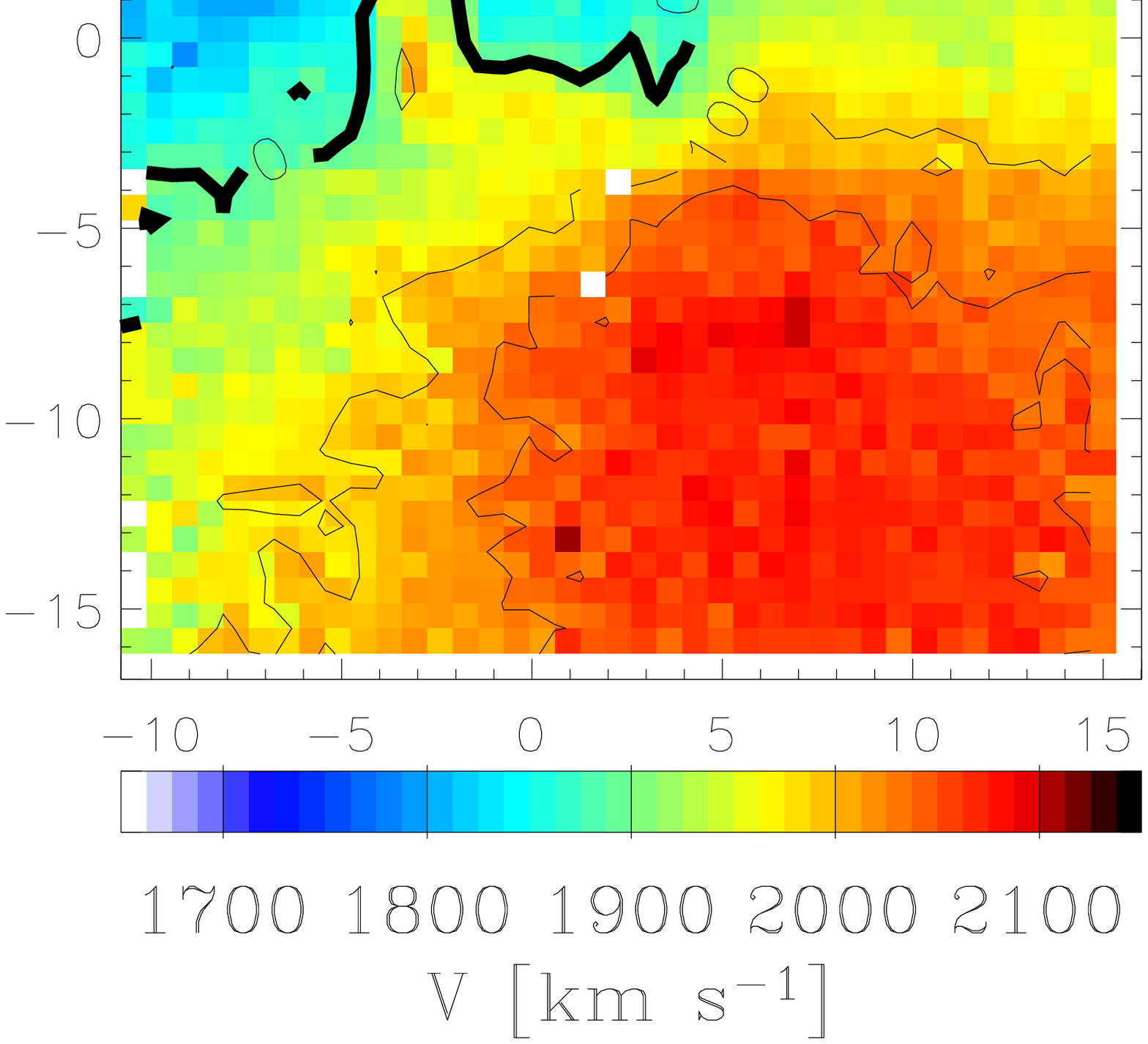,width=2.93cm,clip=}
 \psfig{file=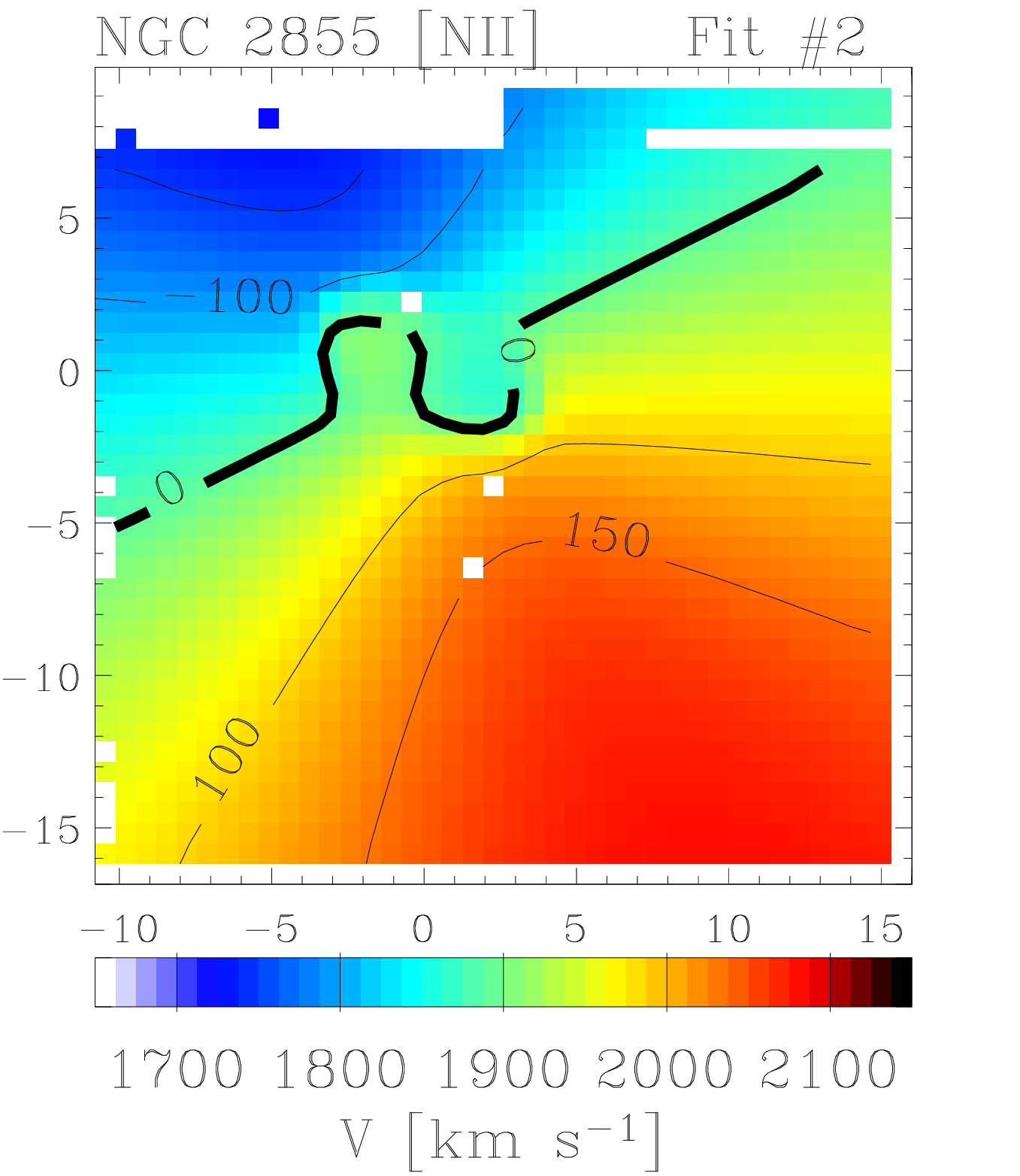,width=2.93cm,clip=} 
 \psfig{file=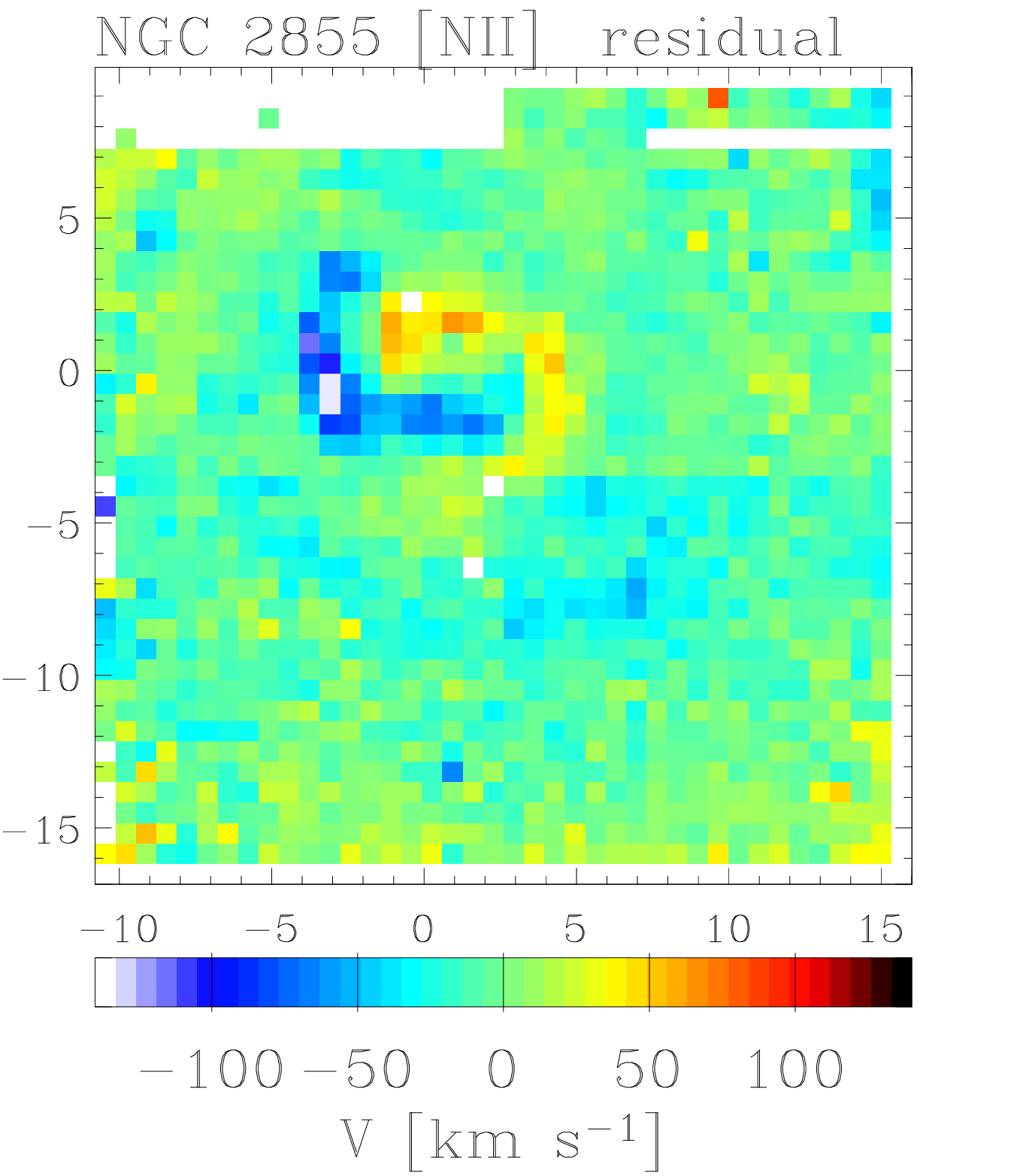,width=2.93cm,clip=}
}
\hbox{
 \psfig{file=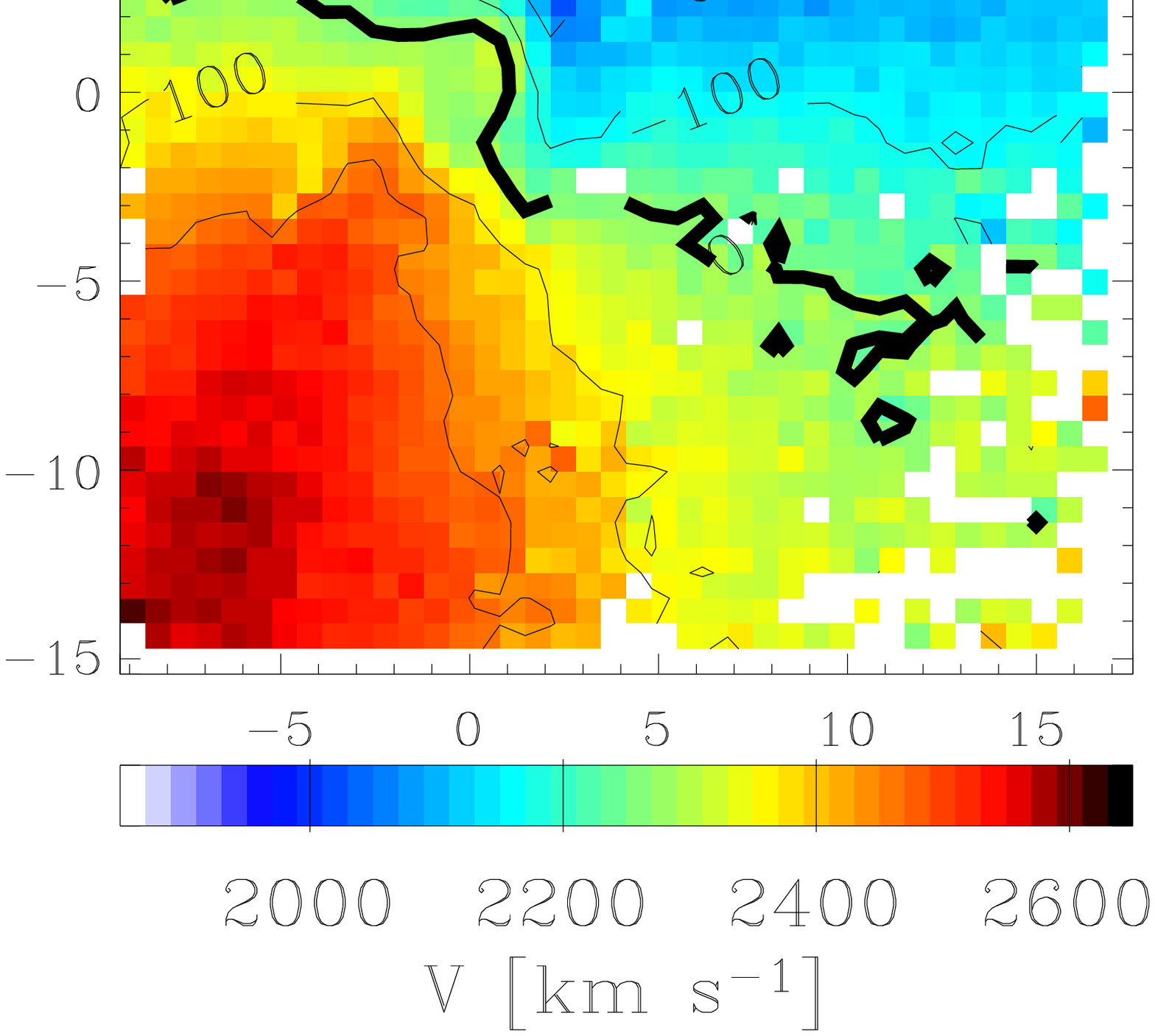,width=2.93cm,clip=}
 \psfig{file=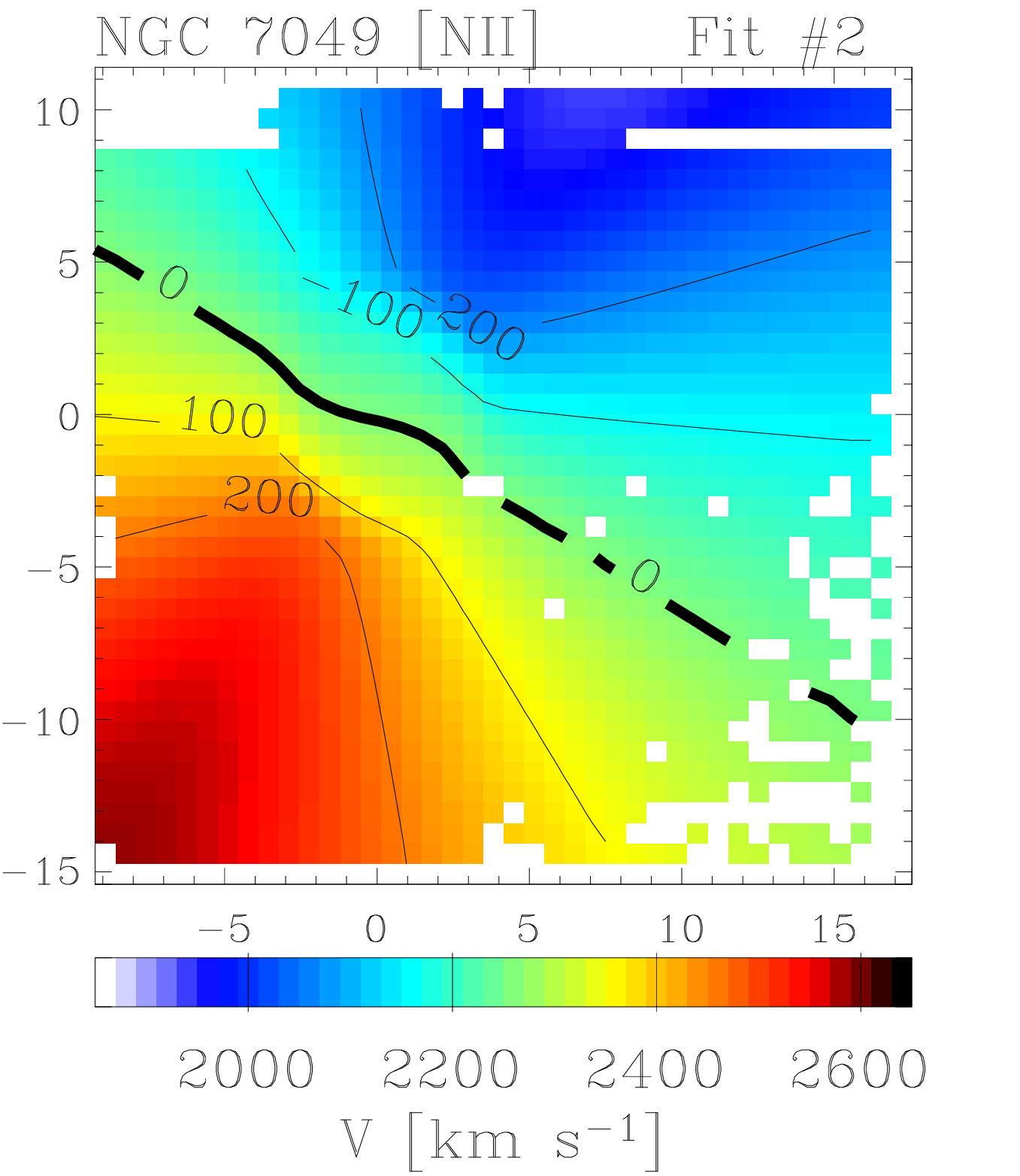,width=2.93cm,clip=} 
 \psfig{file=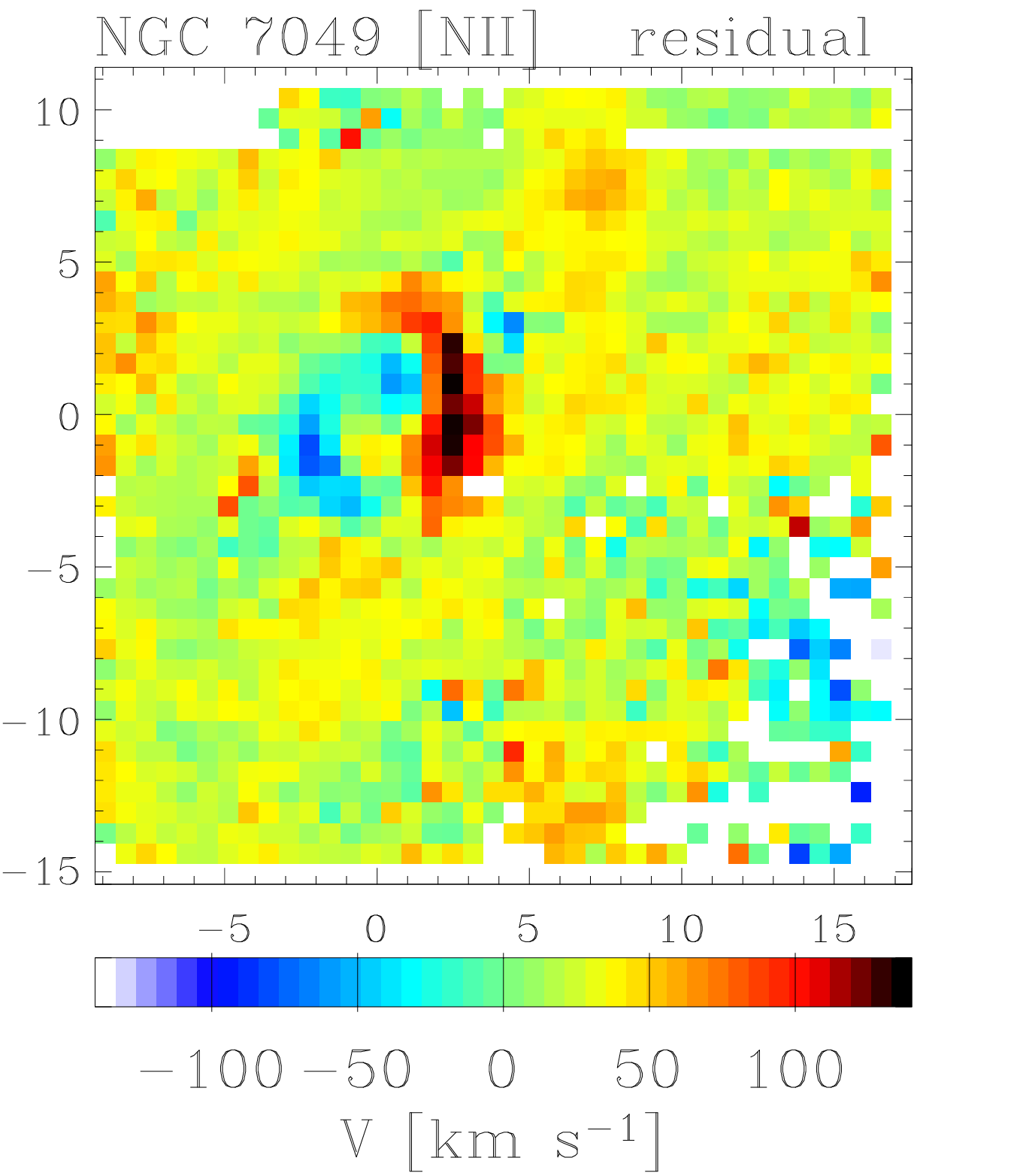,width=2.93cm,clip=}
}}
\caption{Model of the velocity field of NGC 2855
({\em upper panels}) and NGC 7049 ({\em lower panels}) with two
orthogonally-rotating disks (Fit \#2).  The field of view,
orientation, ranges and isovelocity contours are as in
Fig. \ref{fig:n2855_vfield}.  {\em Left panel}: Observed
velocity field. {\em Central panel}: Model. {\em
Right panel}: Residuals.}
\label{fig:od_fit2}
\end{figure}

\begin{figure}
\vbox{
\hbox{
 \psfig{file=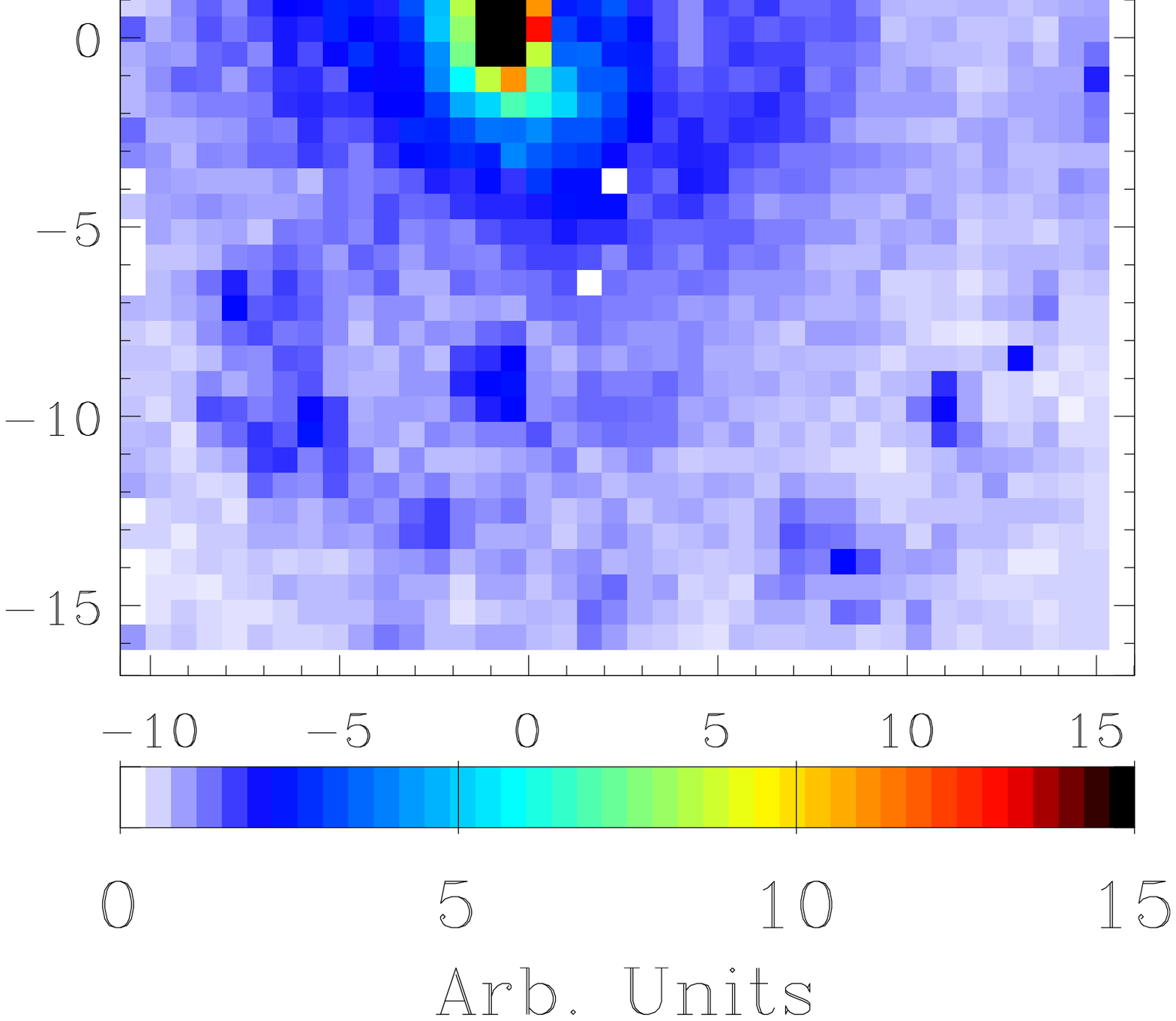,width=2.93cm,clip=}
 \psfig{file=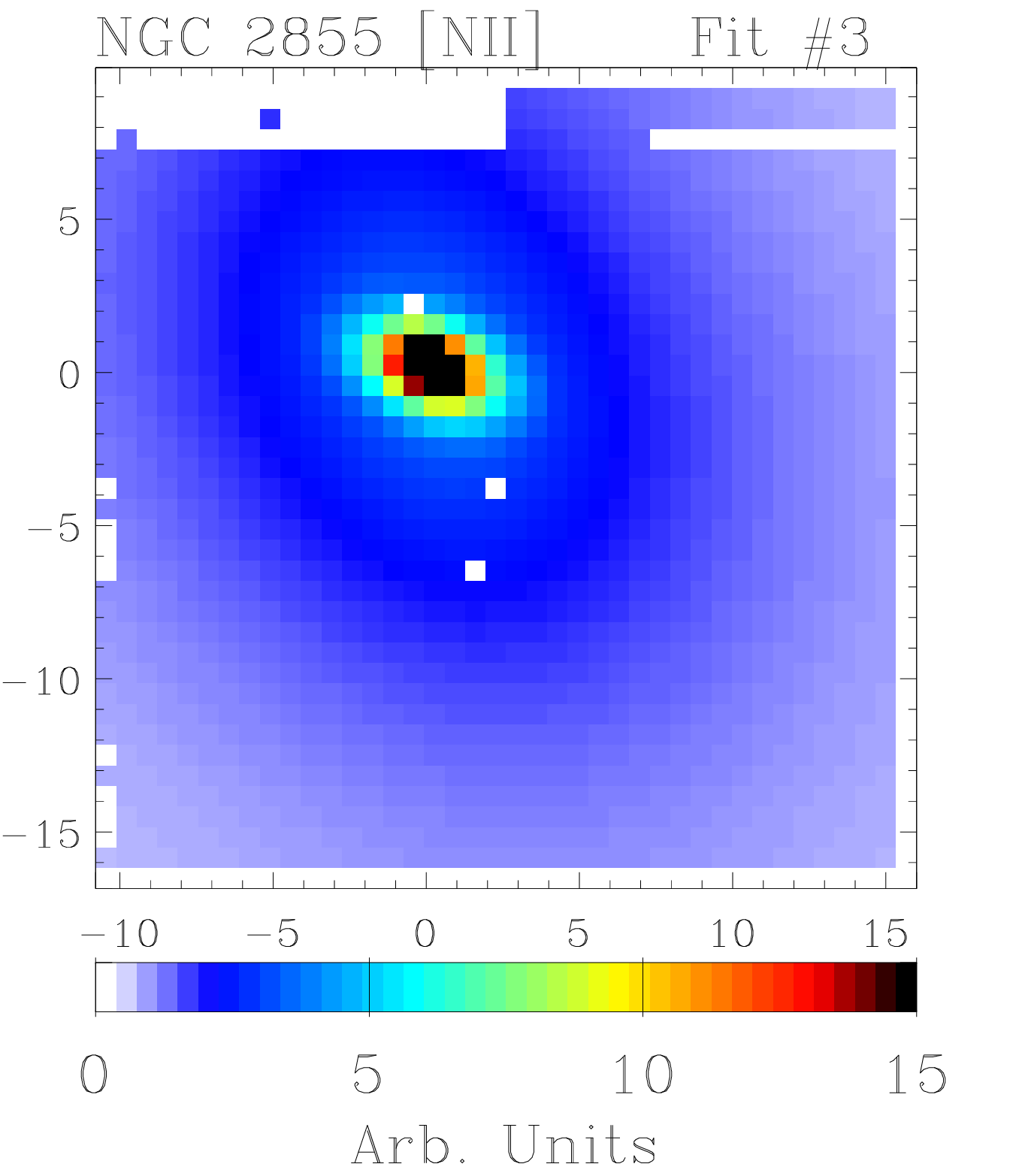,width=2.93cm,clip=} 
 \psfig{file=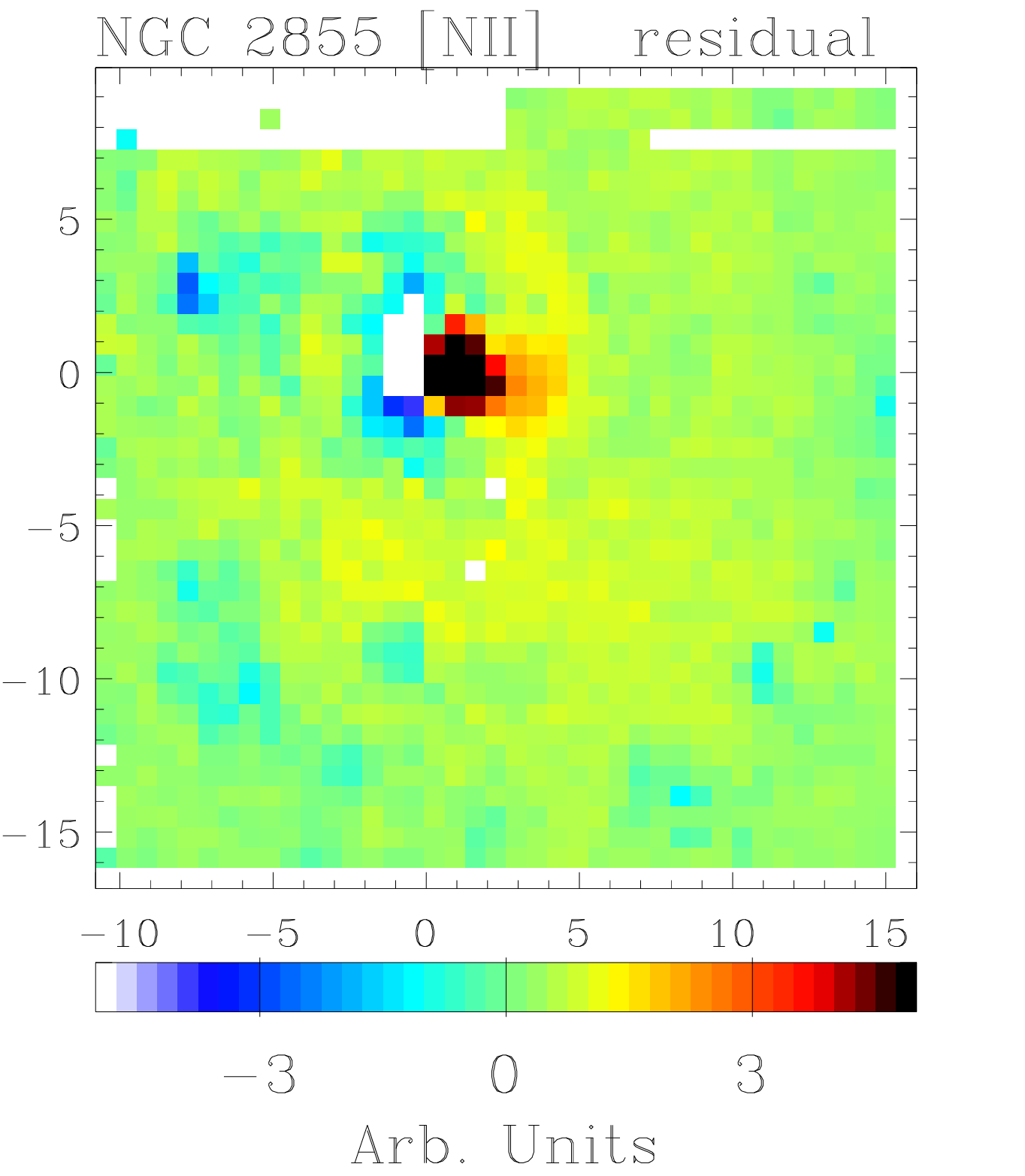,width=2.93cm,clip=}
}
\hbox{
 \psfig{file=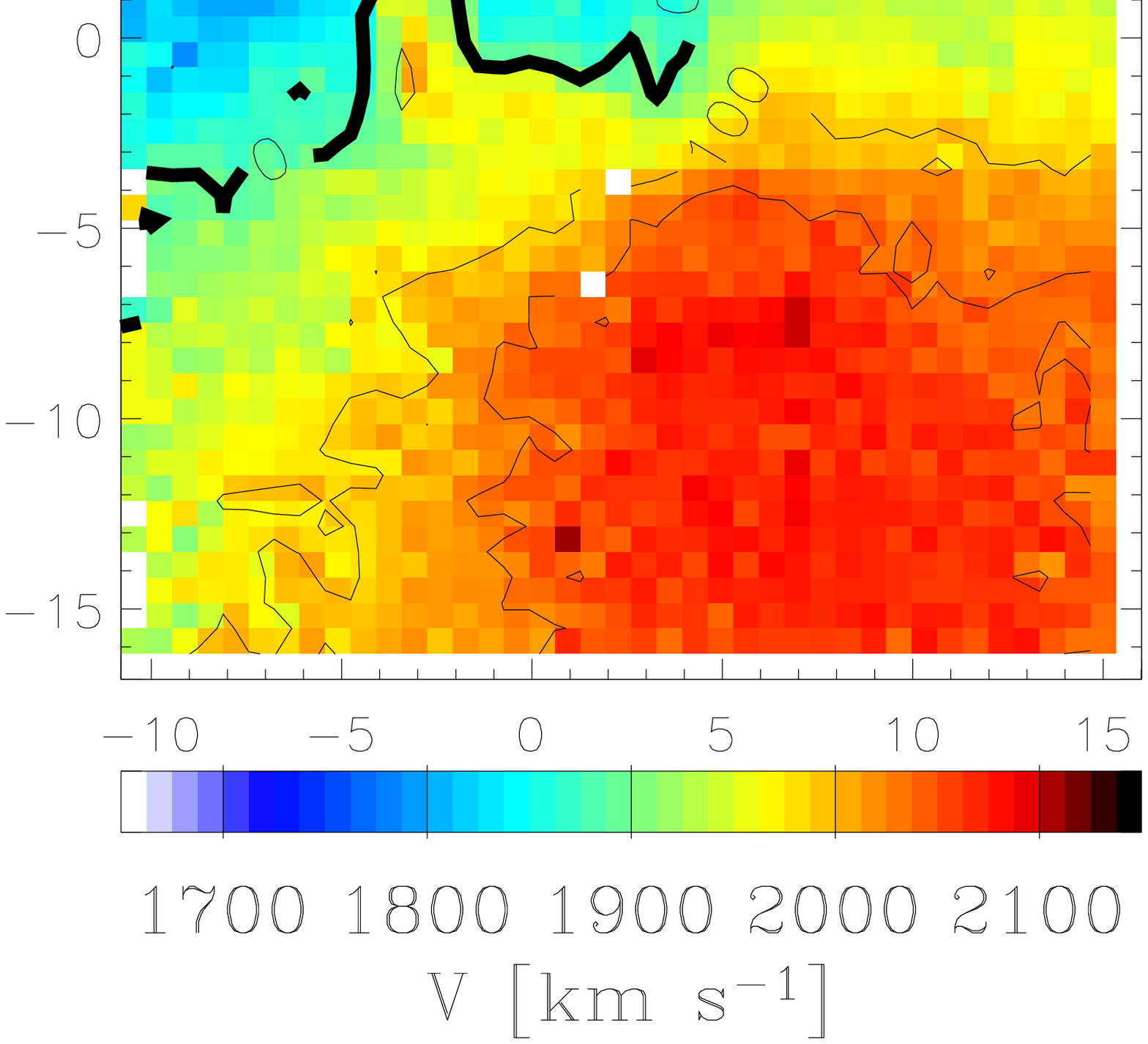,width=2.93cm,clip=}
 \psfig{file=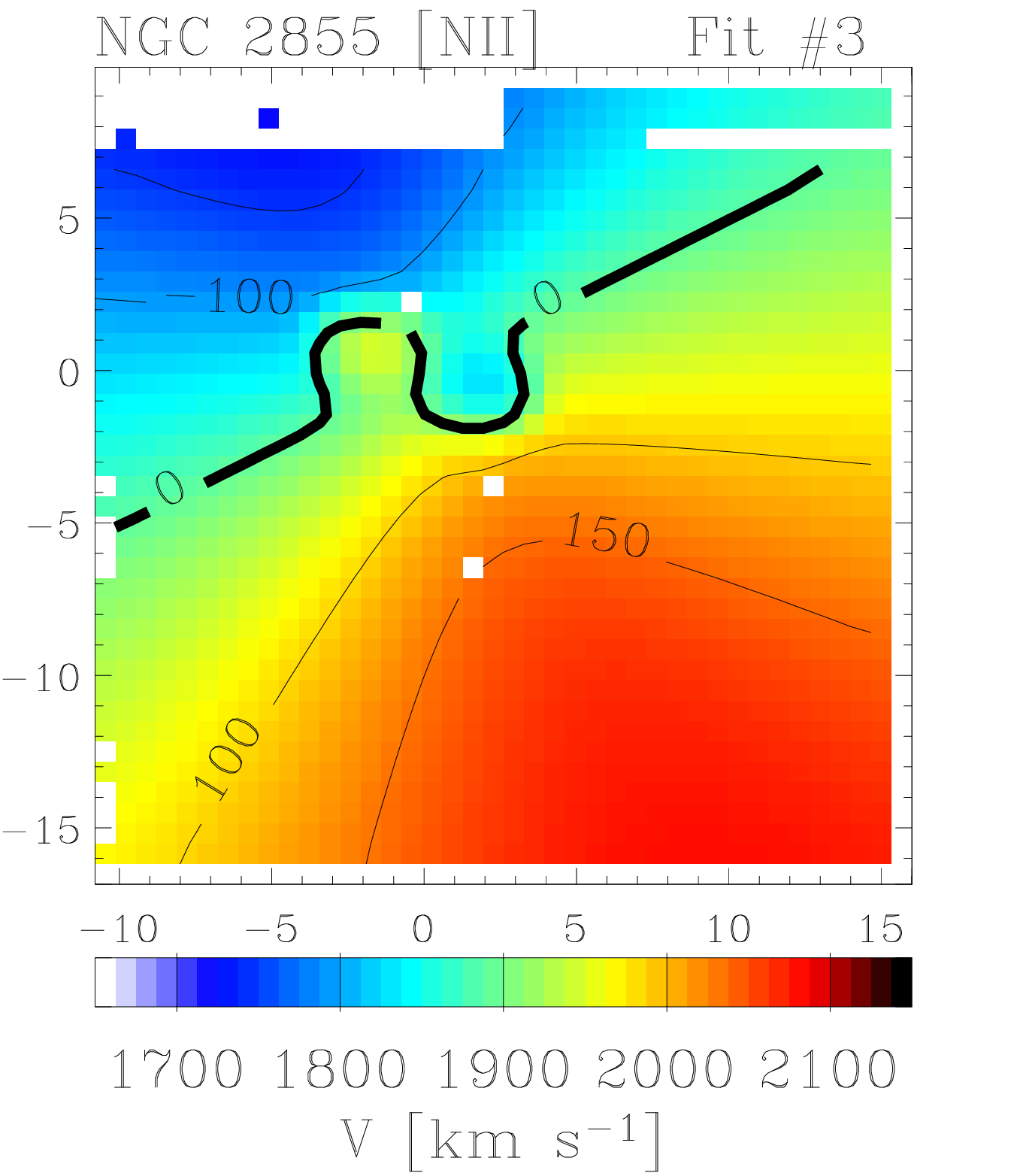,width=2.93cm,clip=} 
 \psfig{file=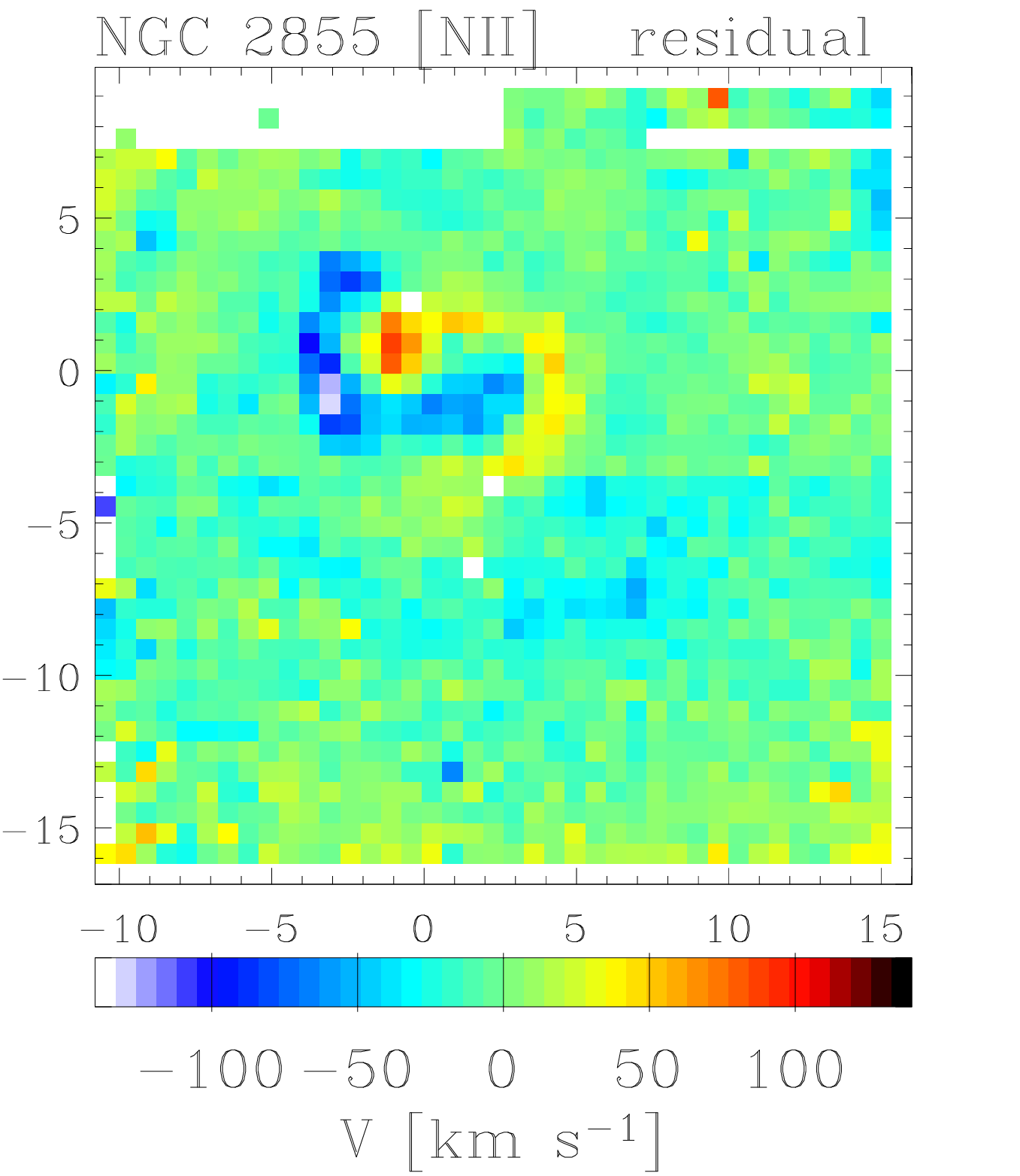,width=2.93cm,clip=}
}}
\caption{Model of the surface-brightness distribution ({\em upper panels})
and velocity field {\em lower panels}) of NGC 2855 with two
orthogonally-rotating disks (Fit \#3). The field of view, orientation,
ranges and isovelocity contours are as in Fig. \ref{fig:n2855_vfield}.
{\em Left panel}: Observed data. {\em Central panel}: Model. {\em
Right panel}: Residuals.}
\label{fig:od_fit3_n2855}
\end{figure}

\begin{figure}
\vbox{
\hbox{
 \psfig{file=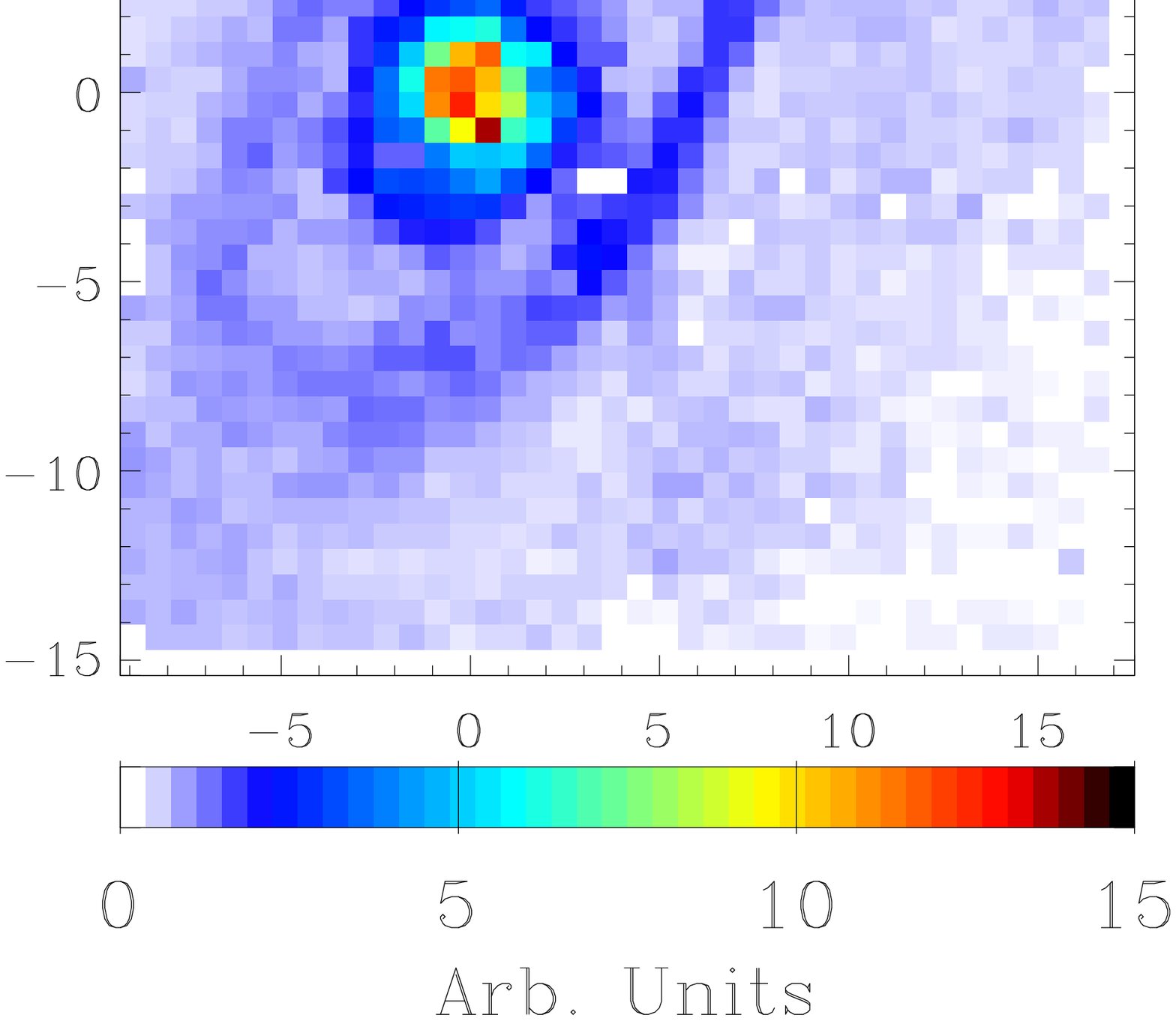,width=2.93cm,clip=}
 \psfig{file=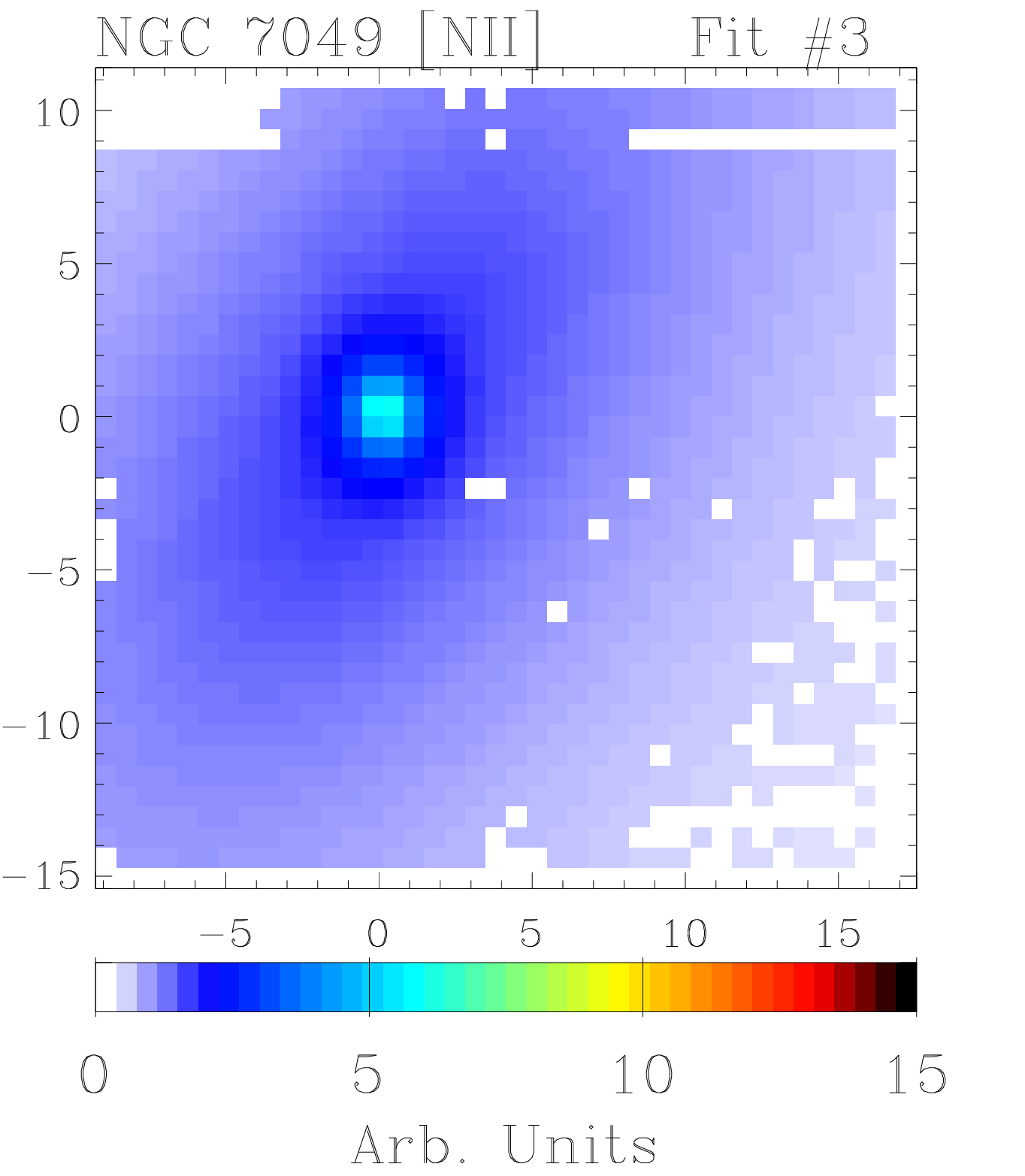,width=2.93cm,clip=} 
 \psfig{file=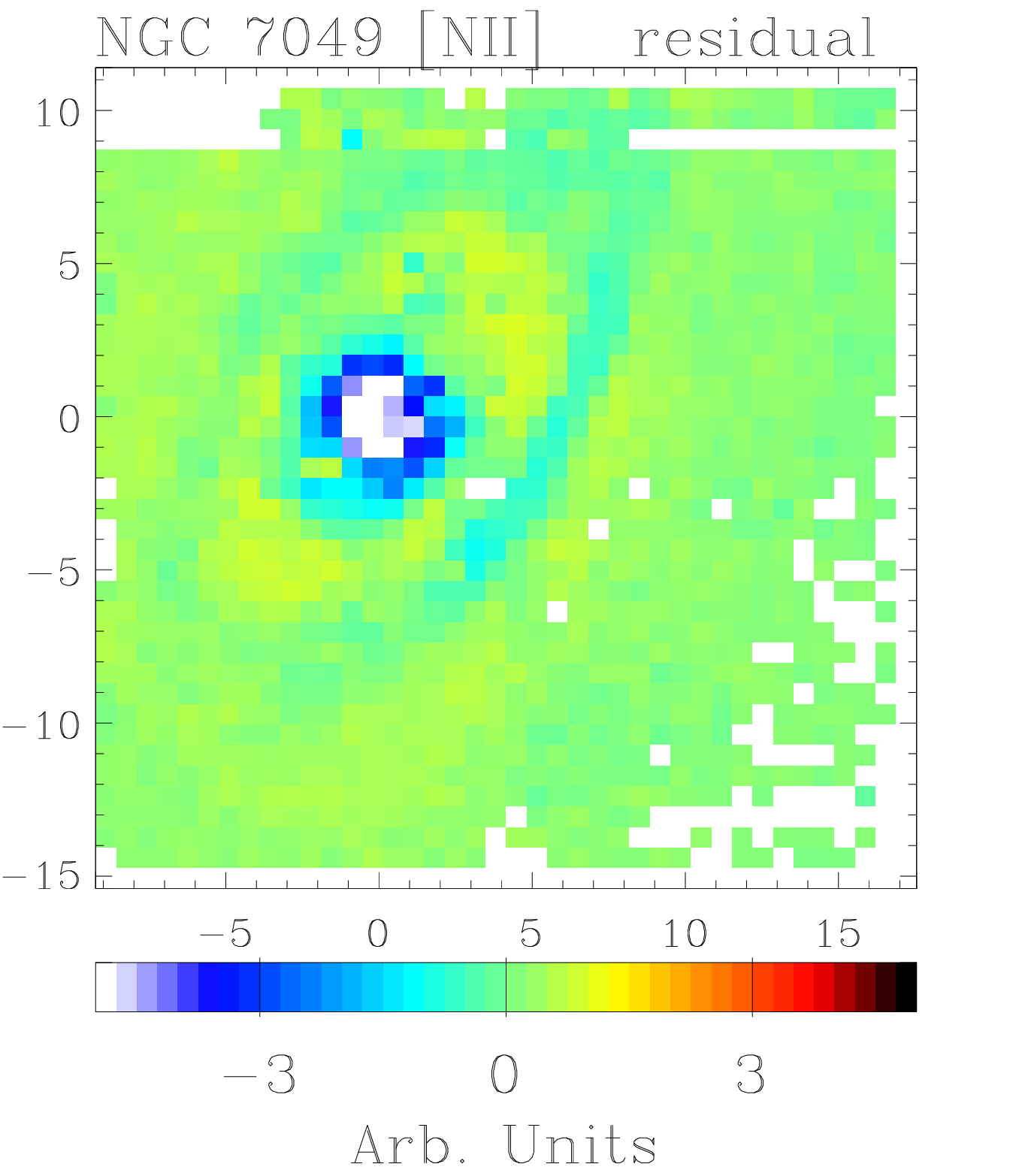,width=2.93cm,clip=}
}
\hbox{
 \psfig{file=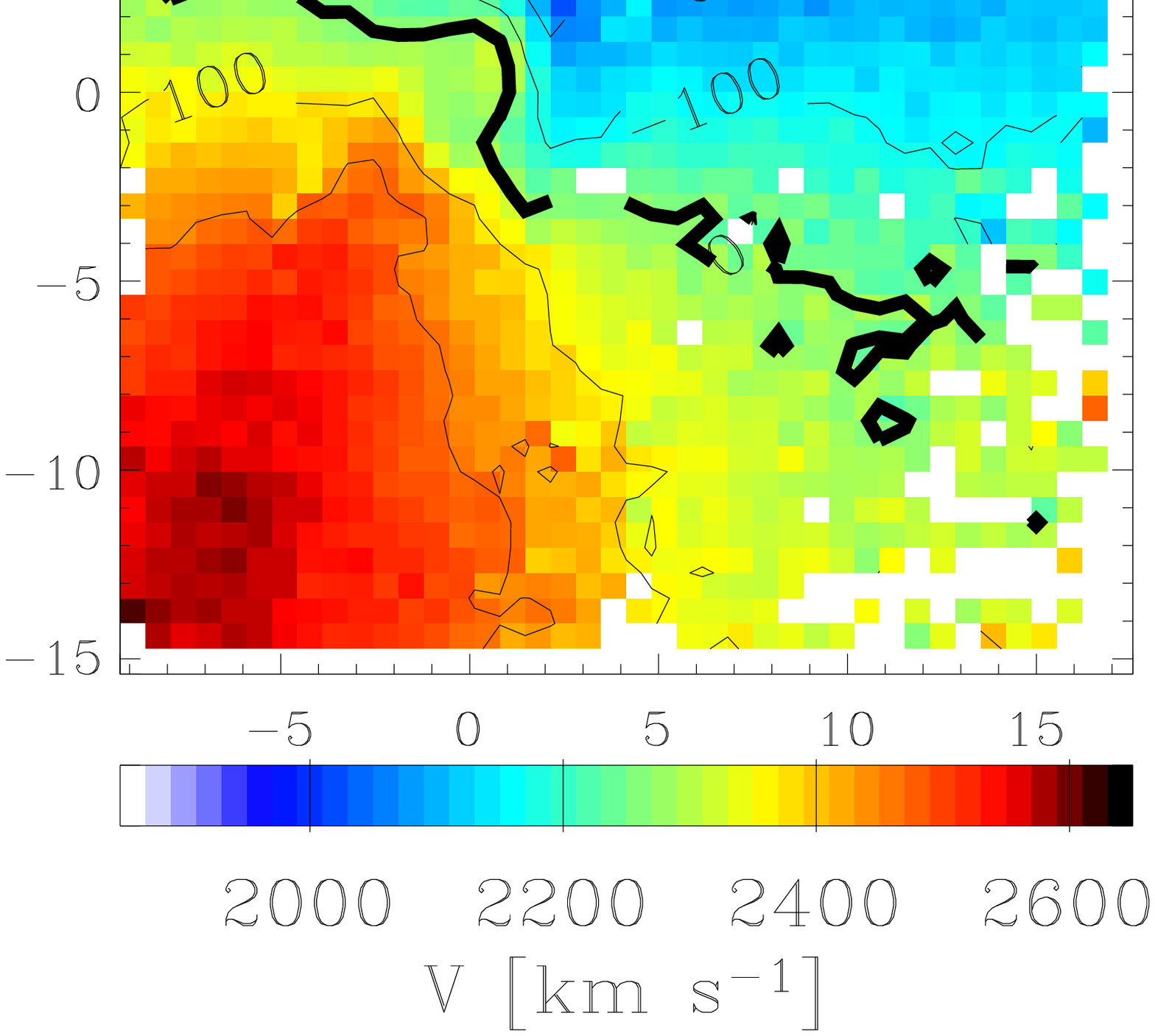,width=2.93cm,clip=}
 \psfig{file=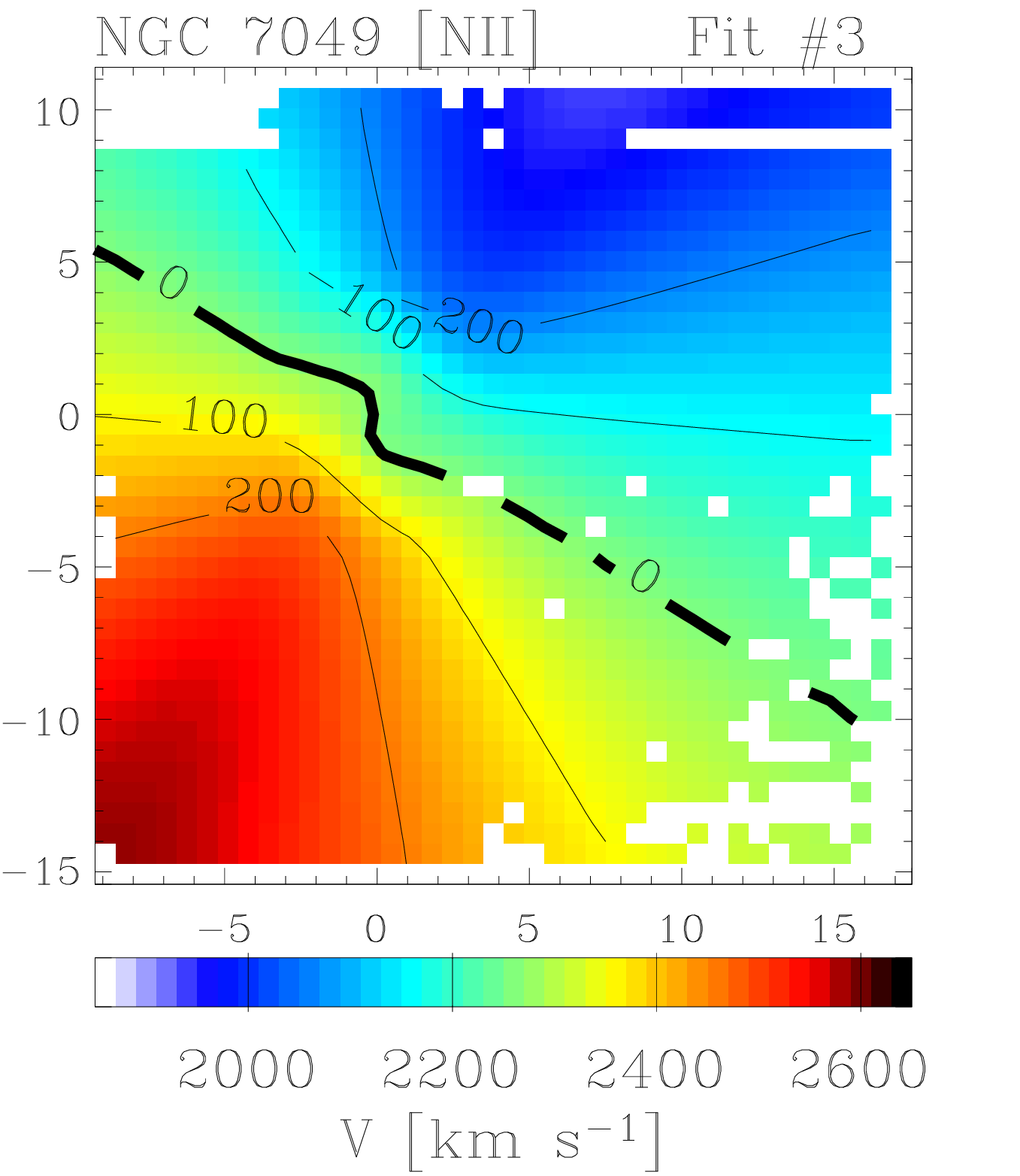,width=2.93cm,clip=} 
 \psfig{file=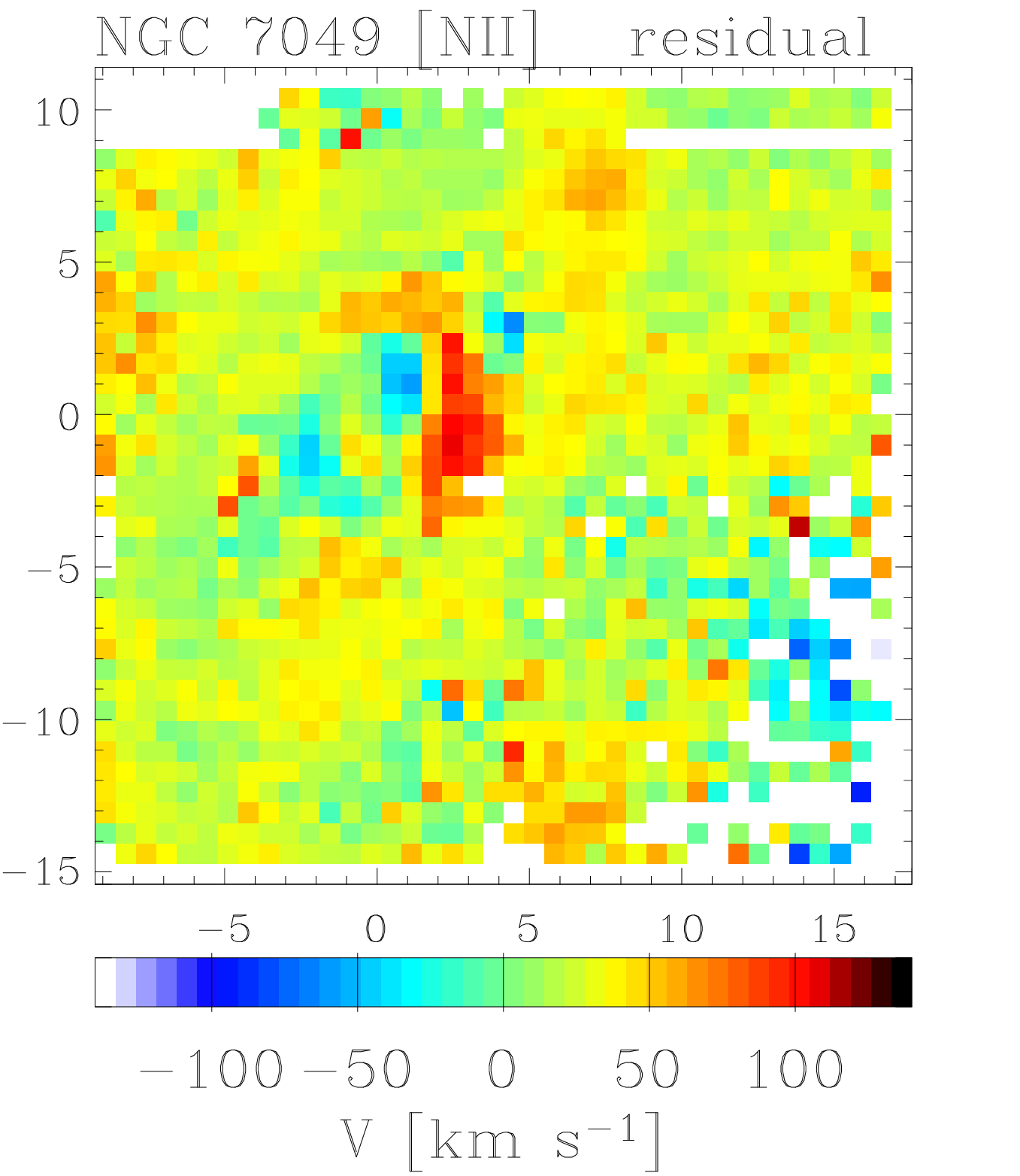,width=2.93cm,clip=}
}}
\caption{Same as in Fig. \ref{fig:od_fit3_n2855} but for NGC~7049.}
\label{fig:od_fit3_n7049}
\end{figure}

The surface-brightness distribution of the best-fitting models built
in Fit \#1 are shown in Fig. \ref{fig:od_fit1}. The radial profiles of
the surface brightness, ellipticity, and position angle derived from
the model are compared to those from the isophotal analysis in
Fig. \ref{fig:n2855_ellipse} and Fig. \ref{fig:n7049_ellipse} for NGC
2855 and NGC 7049, respectively.
The velocity field of the best-fitting models built in Fit \#2 are
shown in Fig. \ref{fig:od_fit2}.
The surface-brightness distribution and velocity field of the
best-fitting models built in Fit \#3 are compared to observations in
Fig. \ref{fig:od_fit3_n2855} and \ref{fig:od_fit3_n7049}.

The surface-brightness distribution and velocity field of NGC 2885 can
not be interpreted as due to presence of two different
orthogonally-rotating disks of gas. The largest deviations in the
residual maps are observed in the nuclear region.

\begin{figure}
\hbox{
\psfig{file=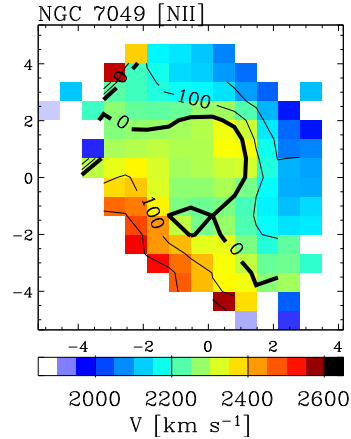,width=5cm,clip=}
}
\caption{The velocity field of the
orthogonal-rotating disk of NGC~7049, as obtained from
Eq. \ref{eqn:od_losvd} The field of view of the panel is
$9''\times9''$. East is up and North is right. The range is indicated
at the bottom of the panel. Isovelocity contours correspond to
velocities after the subtraction of the systemic velocity.}
\label{fig:od_losvd_n7049}
\end{figure}

For NGC 7079 the surface-brightness distribution of the model built in
Fit \#1 is consistent with observations suggesting the presence of an
inner gaseous component which is orthogonal with respect to the main
gaseous disk of the galaxy. However, as in the case of NGC 2855, the
observations are not reproduced by the models built in Fit \#2 and
\#3. We conclude that the circular velocity curve of the orthogonal
disk differs from Eq. \ref{eqn:od_circ}. We derive the velocity field
of the orthogonal disk as 
\begin{equation}
v_{\it OD} = \frac{v(x,y)f(x,y) - v_{\it MD}(x,y) 
  f_{\it MD}(x,y)}{f_{\it OD}}
\label{eqn:od_losvd}
\end{equation}
and show it in Fig. \ref{fig:od_losvd_n7049}.
We tested different parametric prescriptions (Brandt 1960; Freeman
1970; Bertola et al. 1991; Persic et al. 1996) to fit the velocity
field assuming the ionized gas is on circular orbits on a plane. But
none of them gave better results than that obtained adopting
Eq. 5. {\bf This suggests that probably the gas kinematics plotted in
Fig. \ref{fig:od_losvd_n7049} is not characterized by circular
motions. The limited radial extension of the region does not allow us
to adopt more complicated functions as well as a larger number of free
parameters to take into account for non-circular motions.}

\begin{table}
\centering
\caption{Model with two orthogonally-rotating disks (Fit \#1) for NGC~7049}
\begin{tabular}{l r}
\noalign{\smallskip}
\hline
\hline
\noalign{\smallskip}
\multicolumn{1}{c}{Parameter}&
\multicolumn{1}{c}{} \\
\noalign{\smallskip}
\hline
\noalign{\smallskip}
$F_0$          &   0.01$\pm$0.01    \\
$F_1$          &   1.6$\pm$0.2    \\
$R_F$ [$''$]   &   {\bf 15}$\pm$2   \\
$F'_0$         &   0.01$\pm$0.01      \\
$F'_1$         &   {\bf 22}$\pm$1     \\
$R'_F$ [$''$]  &   {\bf 1.13}$\pm$0.08   \\
$\gamma$ [deg] &   {\bf 23}$\pm$4  \\
$\delta$ [deg] &   91$\pm$1  \\
$R'_0$ [$''$]  &   {\bf 3.6}$\pm$0.1    \\
\hline
\noalign{\smallskip}
\end{tabular}
\label{tab:od_model}
\end{table}

\subsection{Warped disk}
\label{sec:warpdisk}

\subsubsection{Model calculation}
\label{sec:warpdisk_model}

The model of the gas velocity field is generated approximating the
warped gas distribution with a series of concentric and circular
wires.

Let $(X''_n,Y''_n,Z''_n)$ be Cartesian coordinates with the origin in
the center of the $n-$th wire, the $Y''_n-$axis aligned along the line
of nodes defined as the intersection between the plane of the wire
$(X''_n,Y''_n)$ and that of the main disk $(X,Y)$.
In the reference frame of the sky $(x,y,z)$, the wire is inclined by
the zenithal and azimuthal angle $\delta_n$ and $\gamma_n$,
respectively.

The transformation of the coordinates of the $n-$th wire from its
reference frame $(X''_n,Y''_n,Z''_n)$ to the reference frame of the
sky $(x,y,z)$ is given by Eq. \ref{eqn:od2sky}, where $\delta$ and
$\gamma$ are replaced by $\delta_n$ and $\gamma_n$,
respectively. Similarly, the inclination $i''_n$ and the position
angle $\theta''_n$ of the apparent major axis of the $n-$th wire are
given by
\begin{equation} 
\cos i''_n = \cos \delta_n \cos i -\sin \delta_n \cos \gamma_n \sin i. 
\label{eqn:wireinc} 
\end{equation}
and
\begin{equation}
\tan \theta''_n = \frac{\sin\pa (\sin\delta_n \cos\gamma_n \cos i+
  \cos\delta_n \sin i) + \cos\pa \sin \delta_n \sin\gamma_n}
  {\cos\pa (\sin\delta_n \cos\gamma_n \cos i+\cos\delta_n \sin i) -
  \sin\pa \sin\delta_n \sin\gamma_n}.  
\label{eqn:wirepa}
\end{equation}

We assumed that the circular velocity of the gaseous component in the
warped disk and is given by
\begin{equation}
V_{\it WD}(R''_n) = \frac{2}{\pi} V''_{\it max} 
  \arctan{\frac{R''_n}{R''_h}}, 
\label{eqn:warp_circ}  
\end{equation}
where $V''_{\it max}$ and $R''_h$ are the maximal velocity and scale
radius of the warped disk. It is $R''_n=\sqrt{X^{''2}_n + Y^{''2}_n}$
and $\cos \phi''_n= Y''_n/R''_n$ with $\phi''_n$ counted
counter-clockwise from $Y''_n-$axis.

The ionized-gas velocity and flux measured along the line of sight at
a given sky point $(x,y)$ where $M$ wires are observed are given
respectively by
\begin{equation}
v(x,y) = \frac{\sum_{n=1}^M v_n(x,y) f_n(x,y)}{\sum_{n=1}^M f_n(x,y)} 
\label{eqn:warp_losv}
\end{equation}
and
\begin{equation}
f(x,y) = M f_n(x,y), 
\label{eqn:warp_losf}
\end{equation}
where 
\begin{eqnarray}
v_n(x,y) &=& V_{\it WD}(R''_n) \sin i''_n \cos \phi''_n + V_{\it sys}
\end{eqnarray}
is the line-of-sight velocity and $f_n(x,y)$ is the line-of-sight flux
of the $n-$th wire. Eq. \ref{eqn:warp_losf} implies that all the wires
observed at a given position on the sky give the same contribution to
the observed surface brightness. This is supported by the fact that
all the wires observed at a given position are supposed to have
similar radii, warping and twisting. Moreover, this choice allowed to
better {\bf reproduce} the inhomogeneous distribution of surface brightness
due to the spiral arms of NGC 2855 and ring-like structure of NGC
7049.  The attempt to model the surface-brightness distribution
adopting an exponential radial profile was not successful. 

We evaluated the warping and twisting of the gaseous component from
the radial profiles of ellipticity and position angle obtained from
the isophotal analysis of the surface-brightness map of the \niig\
line in Sect. \ref{sec:analysis}.
We found that the following empirical functions
\begin{equation}
\delta_n(R''_n)=\frac{-k_2}{\pi}\arctan\left( 
  \frac{R''_n-k_0}{k_1}\right)+ \frac{k_2}{2}
\label{eqn:warp_delta}
\end{equation}
and
\begin{equation}
\gamma_n(R''_n) = c_0 + c_1 R''_n  
\label{eqn:warp_gamma}
\end{equation}
reproduced {\bf fairly} well the radial profiles of $\delta_n$ and $\gamma_n$
we derived from Eq. \ref{eqn:wireinc} and Eq. \ref{eqn:wirepa} by
adopting for $i$ and $\pa$ the values of the main disk fitted in
Sect. \ref{sec:maindisk}. This {\bf constraints} the warped structure to be
confined in the inner region of the galaxy.

Therefore the parameters of our model are $V''_{\it max}$ and $R''_h$
for the circular velocity, $k_0$, $k_1$, and $k_3$ for warping, $c_0$
and $c_1$ for twisting, and $V_{\it sys}$ for systemic velocity.
Forty wires were used to build the warped structure and cover the
whole velocity field with at least one wire for each pixel bin.

\subsubsection{Results}
\label{sec:warp_results}

\begin{figure}
\vbox{
\hbox{
  \psfig{file=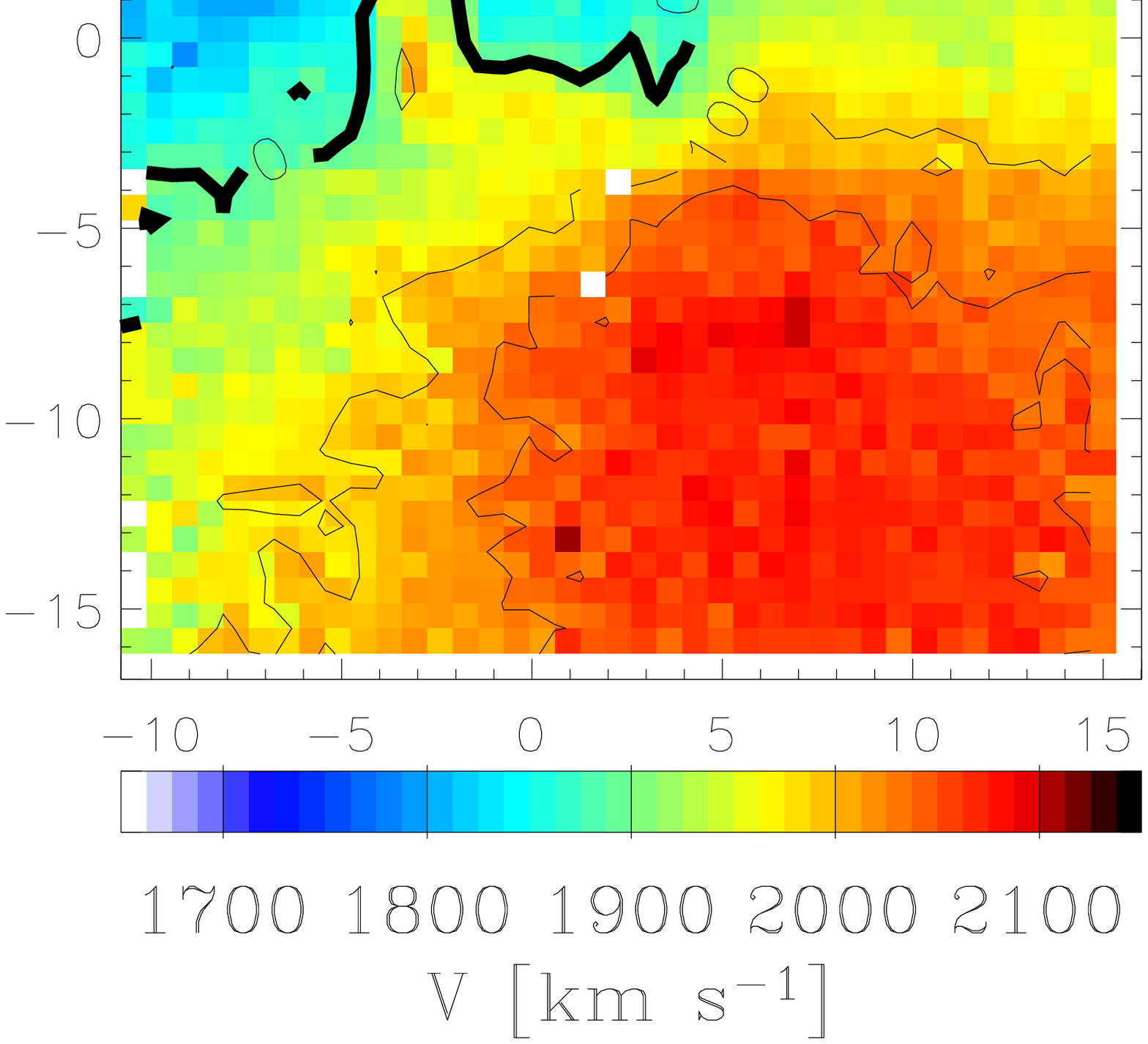,width=2.93cm,clip=}
  \psfig{file=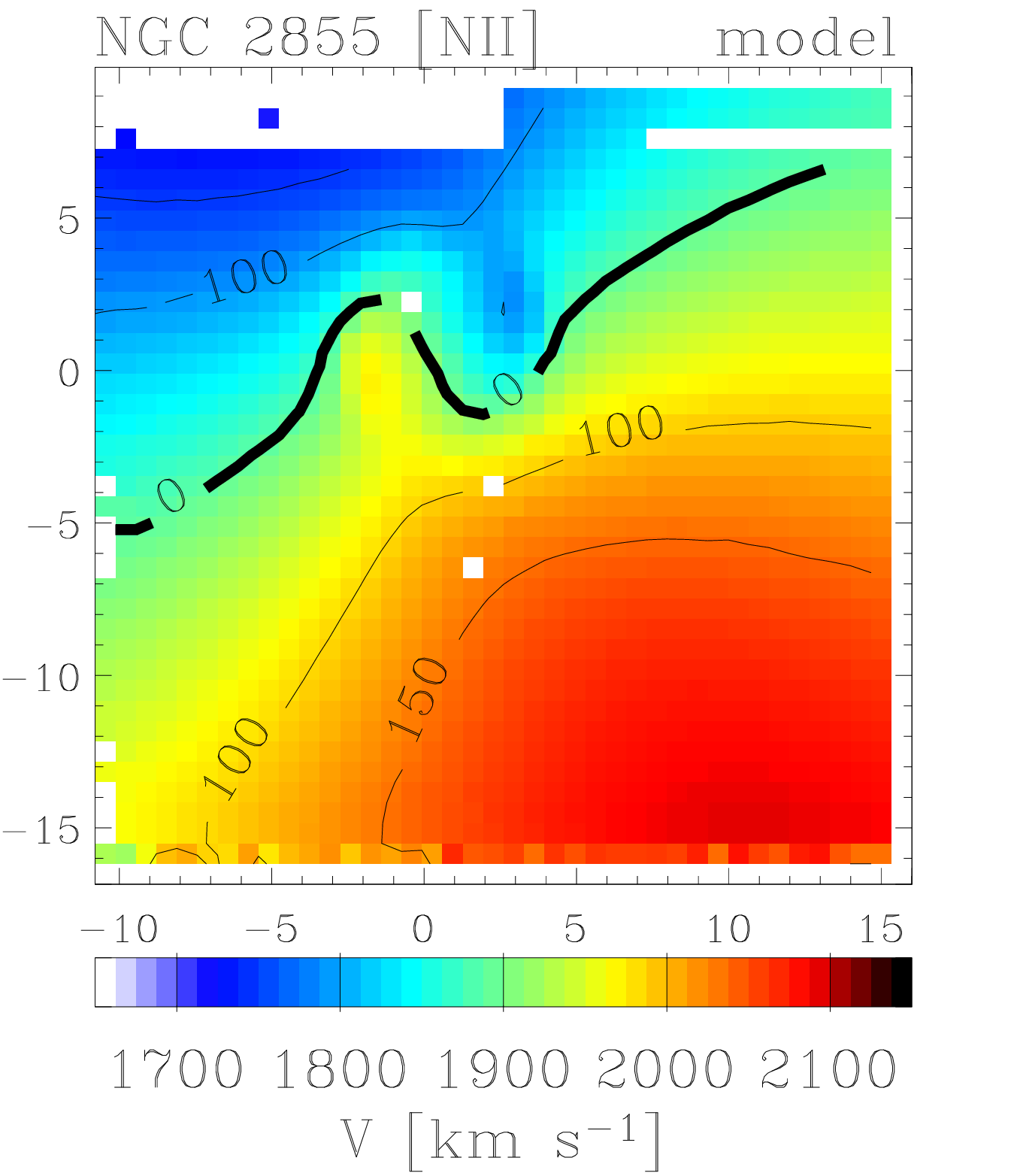,width=2.93cm,clip=}
  \psfig{file=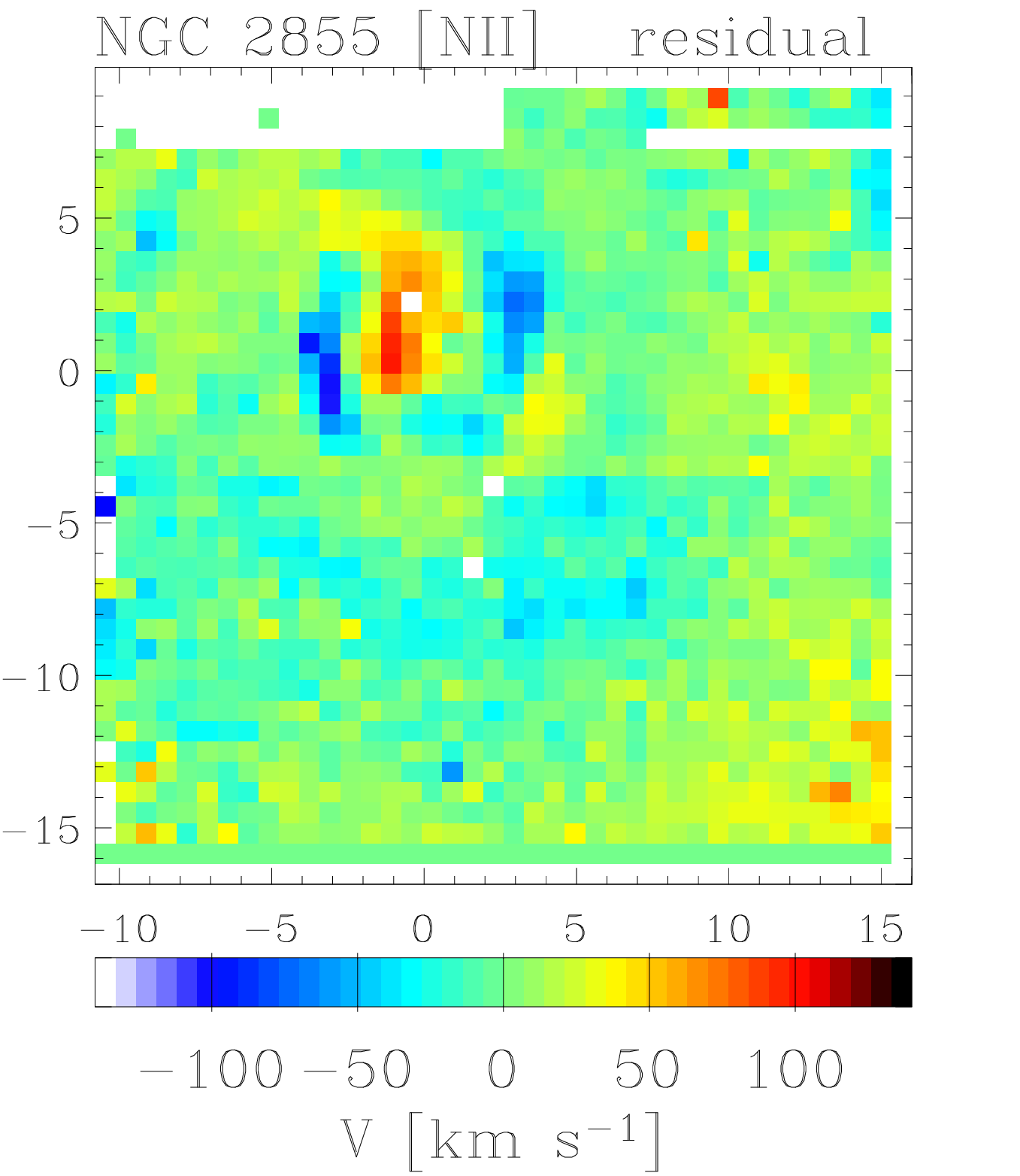,width=2.93cm,clip=}
}
\hbox{
  \psfig{file=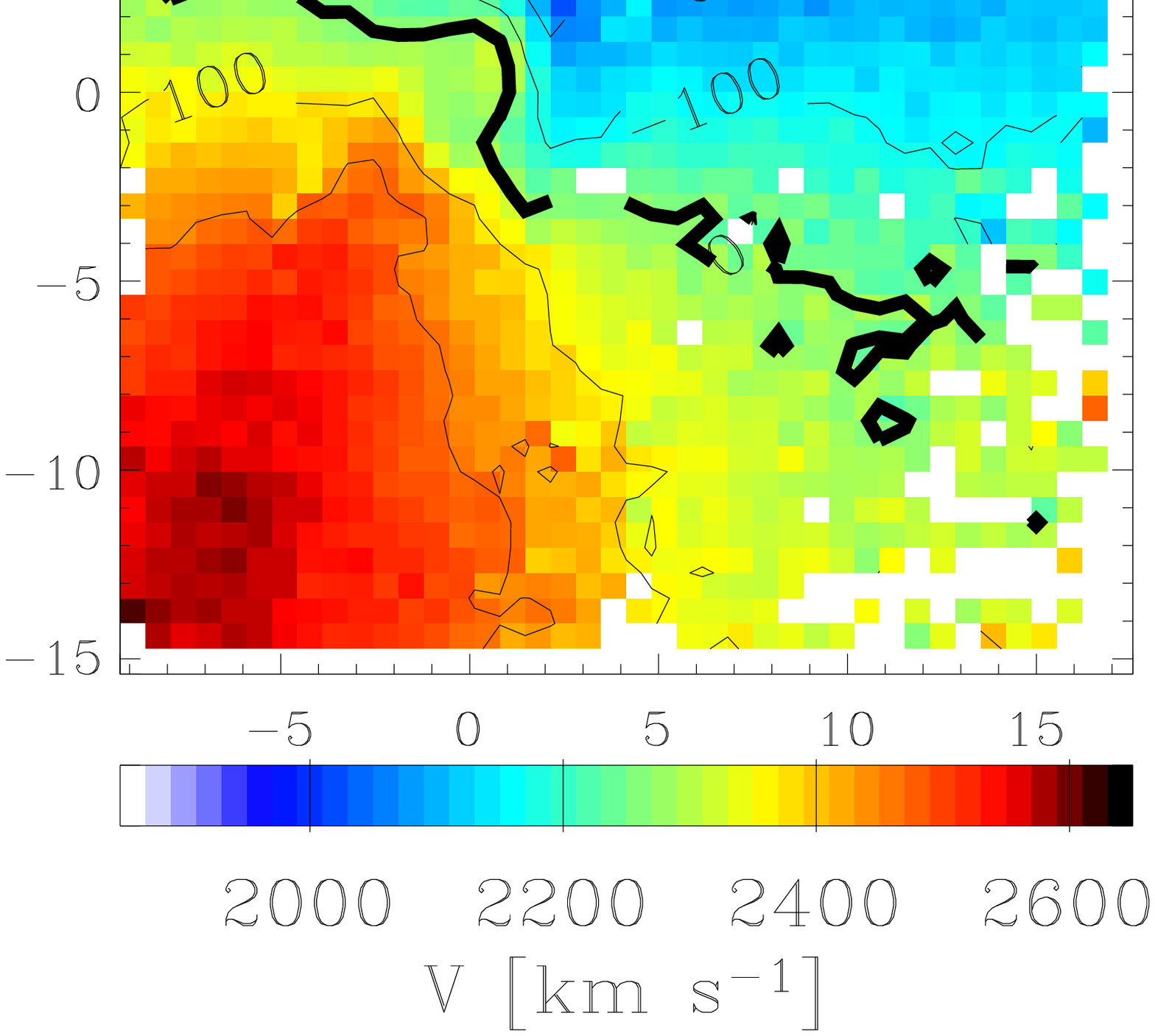,width=2.93cm,clip=}
  \psfig{file=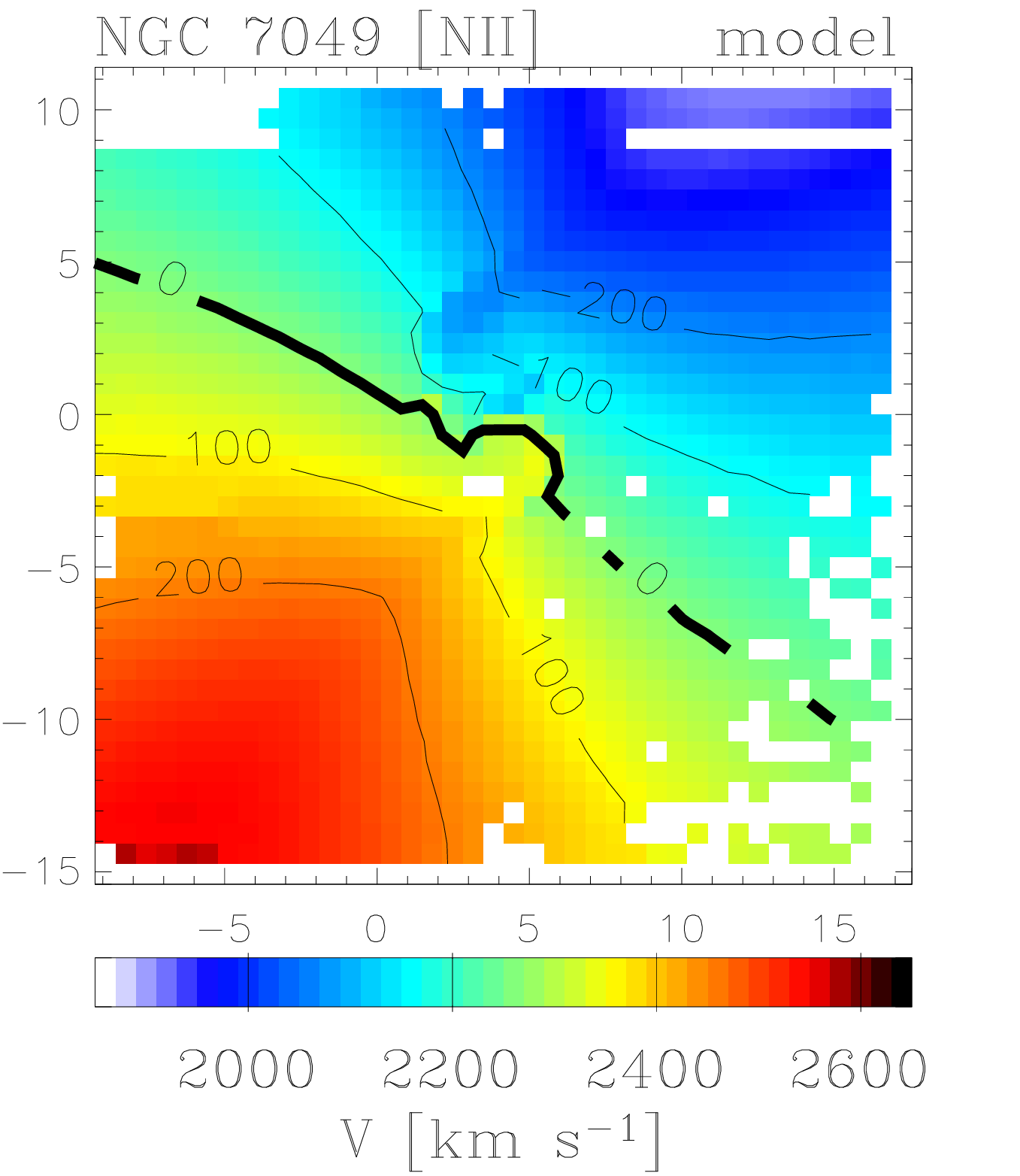,width=2.93cm,clip=}
  \psfig{file=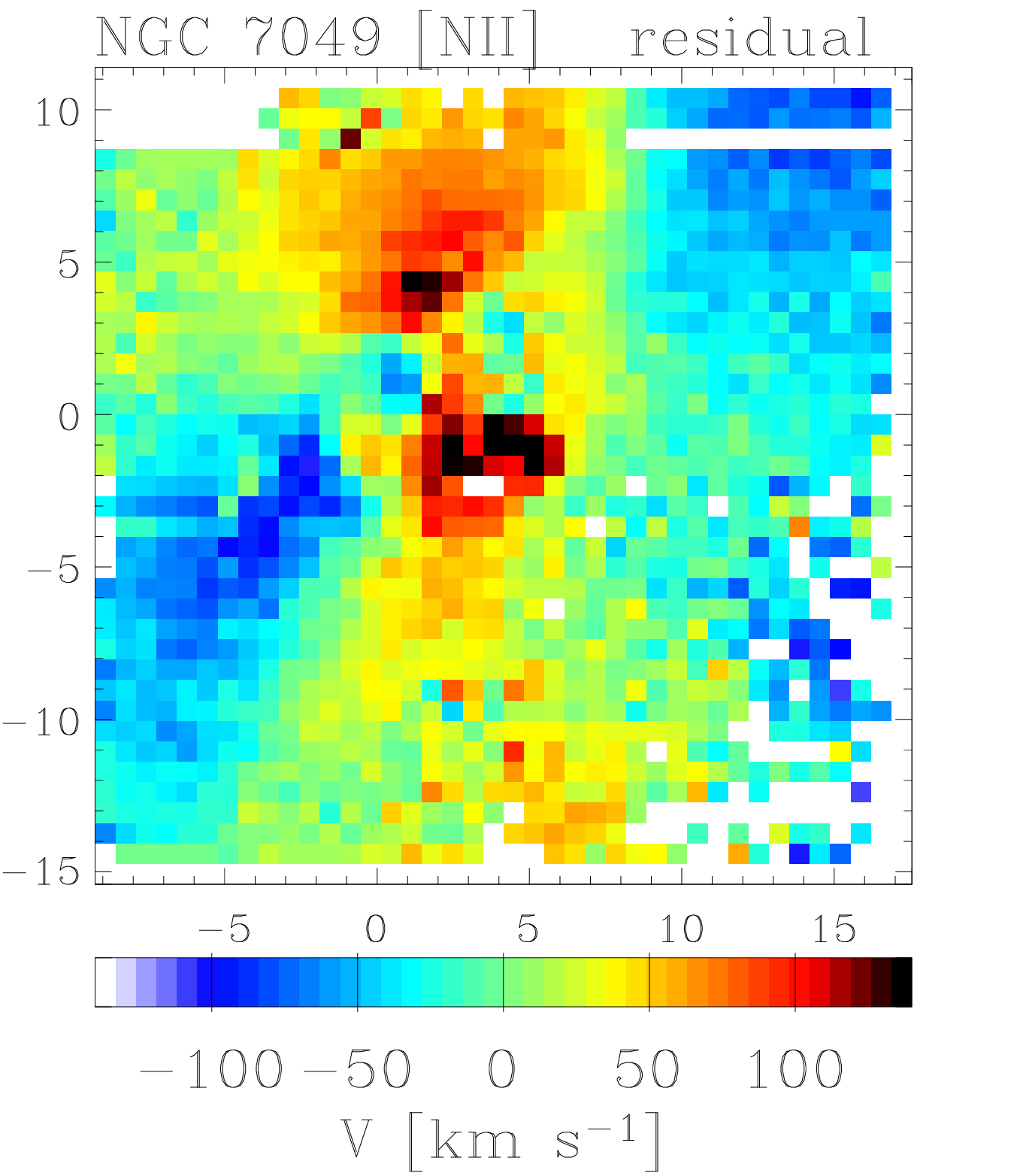,width=2.93cm,clip=}
}}
\caption{Model of the velocity field of NGC 2855
({\em upper panels}) and NGC 7049 ({\em lower panels}) with a warped
disk.  The field of view, orientation, ranges and isovelocity contours
are as in Fig. \ref{fig:n2855_vfield}.  {\em Left panel}: Observed
velocity field. {\em Central panel}: Model. {\em Right panel}:
Residuals.}
\label{fig:warp}
\end{figure}

The velocity fields of the best-fitting models and their residuals are
are shown in Fig. \ref{fig:warp}. Best-fitting parameters are given in
Table \ref{tab:warpdisk}.

\begin{figure}
\hbox{
  \psfig{file=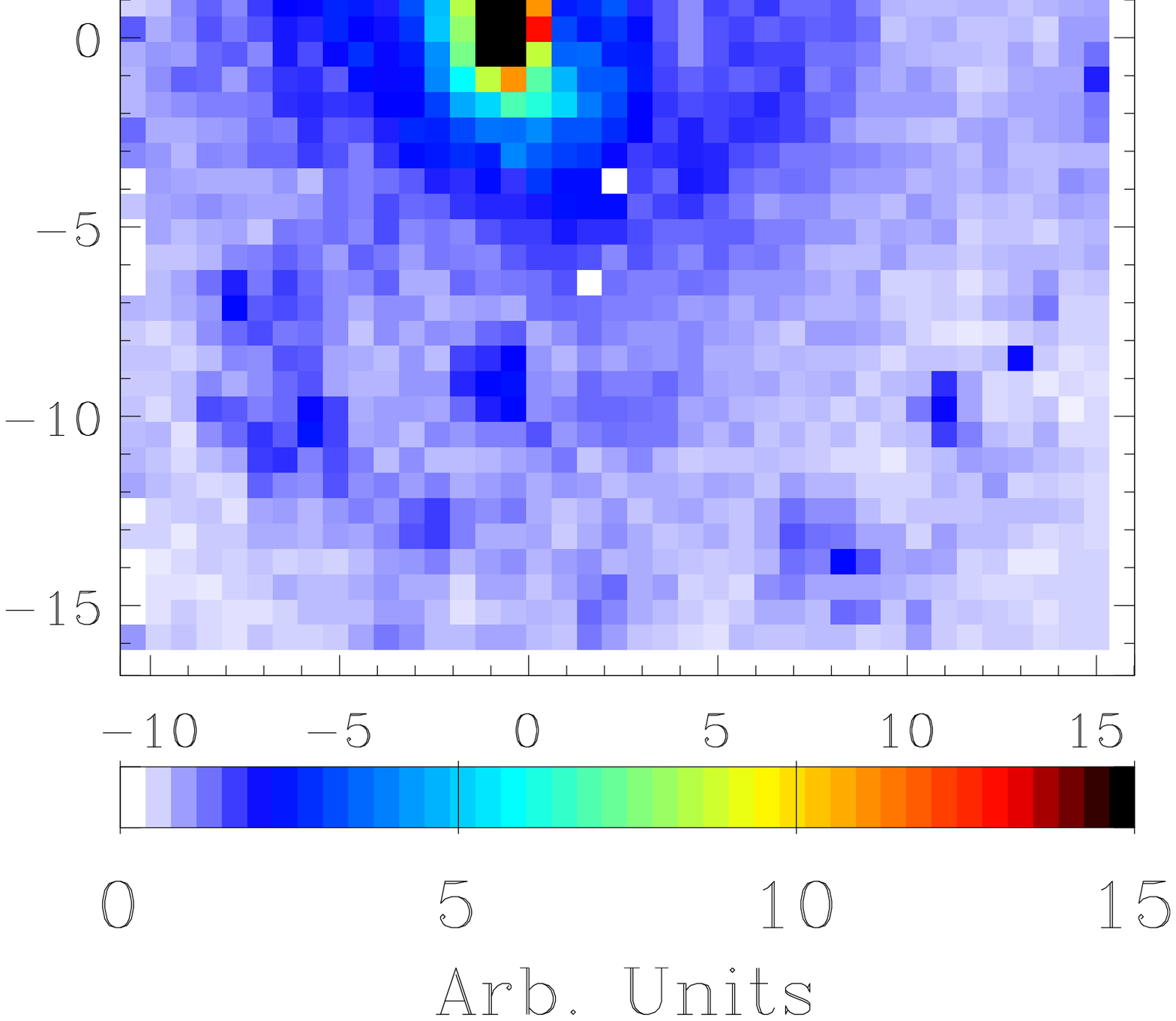,width=2.93cm,clip=}
  \psfig{file=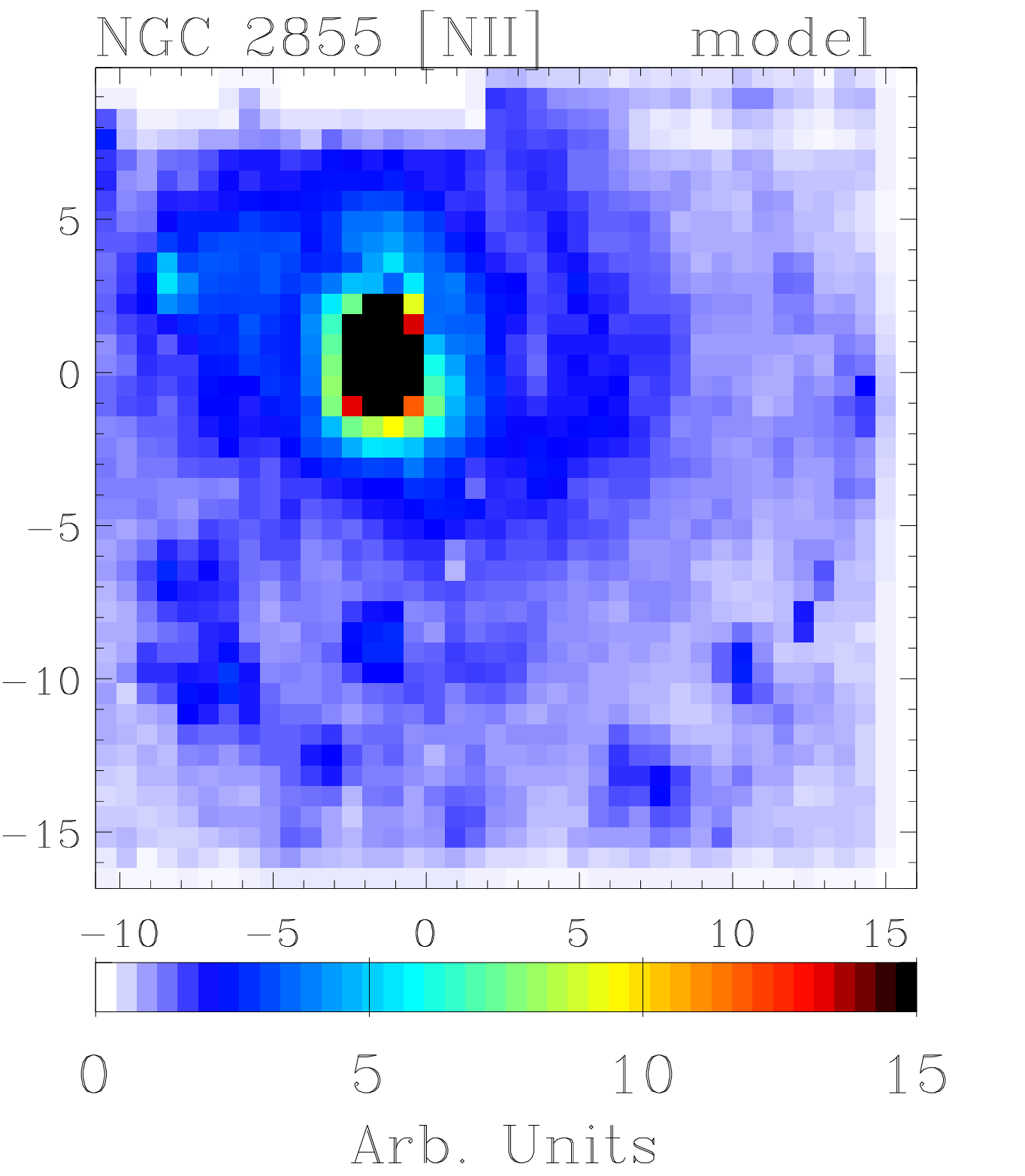,width=2.93cm,clip=}
  \psfig{file=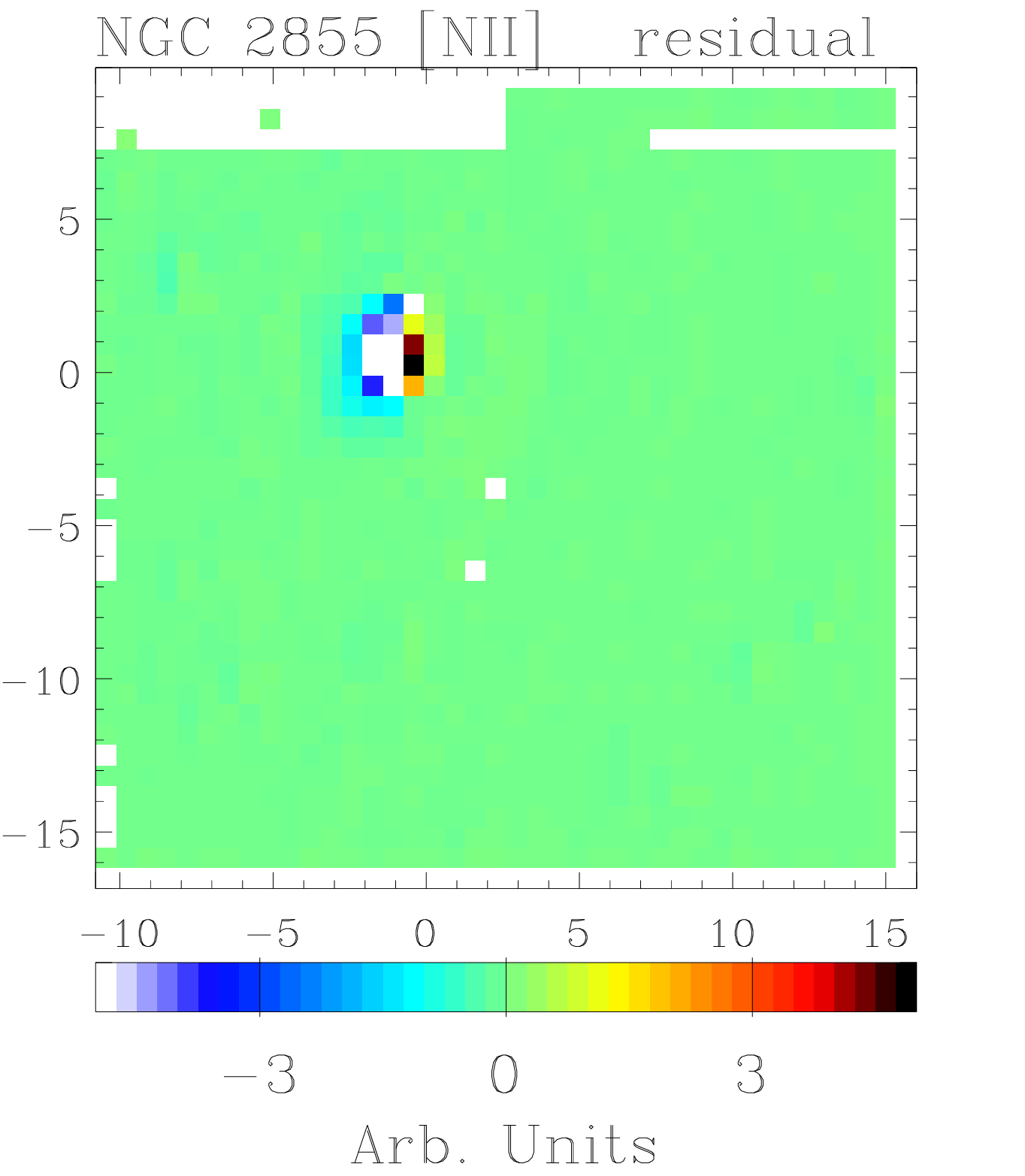,width=2.93cm,clip=}
}
\caption{Model of the surface-brightness distribution of NGC 2855 with a warped disk. 
The field of view, orientation, ranges and contours
are as in Fig. \ref{fig:n2855_vfield}.  {\em Left panel}: Observed
surface-brightness distribution. {\em Central panel}: Model. {\em Right panel}:
Residuals.} 
\label{fig:warp2}
\end{figure}

The warped model is able to reproduce the S-shaped zero velocity line
we observed in the nuclear region of NGC 2855 with velocity residuals
lower {\bf than} a few tens of \kms .
At large radii the position angle and inclination of the warped disk
are consistent with those of the main disk of the galaxy. At small
radii, the warped disk is almost orthogonal to the main disk
(Fig. \ref{fig:warp2}).

The velocity field of NGC 7049 can not be interpreted as due to
presence of a single gaseous component distributed in a warped disk.
In fact, the deviations of the residual map are larger than those we
found for the models with both a single disk and two
orthogonally-rotating disks. 

\begin{table}
\centering
\caption{Model with a warped disk for NGC~2885}
\begin{tabular}{l r}
\noalign{\smallskip}
\hline
\hline
\noalign{\smallskip}
\multicolumn{1}{c}{Parameter}&
\multicolumn{1}{c}{} \\
\noalign{\smallskip}
\hline
\noalign{\smallskip}
$V_{max}$ [\kms] &  {\bf $-373$}$\pm${\bf 16}     \\
$R_h$  [$''$]    &  {\bf 4.3}$\pm${\bf 0.6}   \\
$k_0$ [$''$]     &  3.4$\pm${\bf 0.1}  \\
$k_1$ [$''$]     &  {\bf 0.82}$\pm${\bf 0.03}  \\
$k_2$ [deg]      &  $-${\bf 79.0}$\pm${\bf 2}   \\
$c_0$ [deg]      &  {\bf 85.7}$\pm${\bf 0.7}   \\
$c_1$ [deg/$''$] &  $-${\bf 10.11}$\pm${\bf 0.07}  \\
$V_{sys}$ [\kms] &  {\bf 1885}$\pm${\bf 3}   \\
\noalign{\smallskip}
\hline
\noalign{\smallskip}
\end{tabular}
\label{tab:warpdisk}
\end{table}

\section{Conclusions}
\label{sec:conclusions}

We have measured with integral-field spectroscopy the
surface-brightness distribution and kinematics of the ionized gas in
NGC~2855 and NGC~7049. These two early-type spiral galaxies were
selected to possibly host an IPD on the basis of the analysis of their
long-slit spectra (Corsini et al. 2002, 2003).

In NGC~2855 the ionized-gas distribution peaks in the nucleus and
follows the patchy spiral pattern of the galaxy. In NGC~7049 the gas
is mostly concentrated in the nucleus and in a ring-like structure.
The ionized-gas velocity field of both galaxies is characterized by a
S-shaped line of zero velocity. It is aligned with the disk minor axis
in the inner regions ($r\lesssim8''$) and with the disk major axis at
larger radii. This remarkable misalignement between the kinematical
and photometrical major axes gives rise to the central velocity
gradient and zero-velocity plateau observed along the disk minor and
major axis, respectively. These kinematical features can not be
explained as due to non-circular gas motions in the principal
plane of the disk caused by the triaxial potential of the bulge or a
bar (de Zeeuw \& Franx 1989; Corsini et al. 2003).

As a first step we modeled the velocity field, except for the inner
region where the S-shaped distorsion of the isovelocity countours is
observed. We assumed that the gas is moving onto circular orbits in an
infinitesimally thin disk with a negligible velocity dispersion. This
allowed us to constrain the orientation and inclination of the main
disk of the galaxy.
Then we modeled the ionized-gas kinematics and distribution of both
the inner and outer regions. We assumed that the gaseous component is
distributed either on two orthogonally-rotating disks or in a single
and strongly warped disk.
In both galaxies the velocity field and distribution of the inner gas
are consistent with the presence of an IPD, which is in orthogonal
rotation with respect to the outer gas. In NGC~2855 the IPD correponds
to the innermost and strongly warped portion of the main disk. In
NGC~7049 it is a central and geometrically-decoupled disk, which is
nested in the main disk.

We exclude that warped gaseous disk of NGC~2855 is due moving onto
anomalous orbits in the triaxial bulge or in bar that is tumbling
around the minor axis. In fact, we do not observe any bar structure
neither in optical (Corsini et al. 2002) nor in near-infrared images
(Peletier et al. 1999; M\"ollenhoff \& Heidt 2001) of the
galaxy. Moreover, there is no evidence of gas in retrograde motion
relative to the stars at large radii from the galaxy center, where the
gas orbits are expected to be highly inclined with respect to the
figure rotation (van Albada et al. 1982; Friedli \& Benz 1993).
On the other hand, the galaxy is acquiring external material in a
direction close to the disk minor axis (Malin \& Hadley 1997).  We
suggest that the central portion of the warped disk is composed by
accreted gas, which is settling in the plane perpendicular to the long
axis of the bulge. It is in orthogonal rotation with respect to the
outer gaseous component, which lies on the plane perpendicular to the
long axis of the bulge. This is the one containing the main disk of
the galaxy. 
In NGC~7049 the two orthogonally-decoupled gaseous components are
already settled onto the principal planes of the triaxial bulge.

To date the presence of an inner polar disk has been confirmed by
means of integral-field spectroscopy in more than 20 early-type disk
galaxies (Sil'chenko 2006 and this paper). In
spite of the growing statistics, the question about their formation is
still open. Different qualitative arguments support either an external
or internal origin. They are the acquisition of external gas via
accretion on nearly polar orbits by the triaxial bulge of a
pre-existing galaxy, and the transfer of gas onto highly-inclined
anomalous orbits of a triaxial tumbling bulge or bar,
respectively. The comparison between the observed kinematics and
distribution of gas in bulges and bars hosting an inner polar disk
with gas dynamical models including external acquisition and internal
evolution in a variety of triaxial potentials are now required to
address this issue.

\begin{acknowledgements} 

We kindly acknowledge Michele Cappellari, Katia Ganda, Marc Sarzi, and
Reynier Peletier for useful comments to paper. {\bf We thank the
anonymous referee for suggestions that improved the paper.} LC thanks
Sandro D'Odorico and Carlo Izzo for their hints during the early
stages of data reduction. EMC acknowledges the Kapteyn Astronomical
Institute for the hospitality while this paper was in progress. This
work was made possible through grant PRIN 2005/32 from Istituto
Nazionale di Astrofisica (INAF).

\end{acknowledgements}

\label{lastpage}

\begin{thebibliography}{}


\bibitem[]{ber91} Bertola, F., Bettoni, D., Danziger, J. et al. 
1991, ApJ, 373, 369 

\bibitem[]{ber99} Bertola, F., Corsini, E. M., {\bf Vega Beltr{\'a}n}, J. C. et 
al. 1999, ApJ, 519, 127

\bibitem[]{bra60} Brandt, J.~C.\ 1960, ApJ, 131, 293 

\bibitem[]{coc04} Coccato, L., Corsini, E. M., Pizzella, A., et
al. 2004, A\&A, 416, 507

\bibitem[]{coc05} Coccato, L., Corsini, E. M., Pizzella, A., \&
Bertola. F. 2005, A\&A 2005, A\&A, 440, 107

\bibitem[]{cor02} Corsini, E.M., Pizzella, A., \& Bertola, F. 2002,
A\&A, 382, 488

\bibitem[]{cor03} Corsini, E. M., Pizzella, A., Coccato, L., \& Bertola,
F. 2003, A\&A, 408, 873

\bibitem[]{cou97} Courteau, S. 1997, AJ, 114, 2402

\bibitem[xxx]{dez89} de Zeeuw, T., \& Franx, M.\ 1989, ApJ, 343, 617

\bibitem[]{fal06}  {\bf Falc{\'o}n}-Barroso, J., Bacon, R., Bureau, M., et al. 2006, MNRAS, 369, 529

\bibitem[]{fre70} Freeman, K.~C.\ 1970, ApJ, 160, 811 

\bibitem[]{fri93} Friedli, D., \& Benz, W. 1993, A\&A, 268, 65

\bibitem[]{gan06} Ganda, K., {\bf Falc{\'o}n}-Barroso, J., Peletier, R. F., et al. 2006, MNRAS, 367, 46

\bibitem[]{mal97} Malin, D. \& Hadley, B. 1997, PASA, 14, 52

\bibitem[]{mer83} Merritt, D. \& de Zeeuw, T. 1983, ApJ 267, L19

\bibitem[]{mol01}M\"ollenhoff, C., \& Heidt, J. 2001, A\&A, 368, 16

\bibitem[]{ost96} Osterbrock, D. E., Fulbright, J. P., Martel, A. R., et al. 1996, PASP, 108, 227

\bibitem[]{pel99} Peletier, R. F., Balcells, M., Davies, R. L., et al. 1999, MNRAS, 310, 70

\bibitem[]{per96} Persic, M., Salucci, P., \& Stel, F.\ 1996, MNRAS, 281, 27 

\bibitem[]{san94} Sandage, A. \& Bedke, J. 1994, The Carnegie Atlas of Galaxies (Washington: Carnegie Institution of Washington)

\bibitem[]{sar00} Sarzi, M., Corsini, E. M., Pizzella, A., et
al. 2000, A\&A, 360, 439

\bibitem[]{sar06} Sarzi, M., Falc\'on-Barroso, J., Davies, R. L., et al. 2006, MNRAS, 366, 1151


\bibitem[]{sha04} Shalyapina, L. V., Moiseev, A. V., Yakovleva, V. A., Hagen-Thorn, V. A. \& Barsunova, O. Yu. 2004, AstL, 30, 583

\bibitem[]{sil06A}  Sil'chenko O., 2006, In: Progress in Study of Astrophysical
Disks, Ed, A. M. Fridman, M. Ya. Marov \& I. G. Kovalenko, ASSL,
Vol. 337, (Dordrecth: Springer) p. 275

\bibitem[]{sil04} Sil'chenko, O. K.\& Afanasiev, V. L. 2004,  AJ, 127, 2641

\bibitem[]{sil06B} Sil'chenko, O. K.\& Moiseev, A. V. 2006, AJ, astro-ph/0512431

\bibitem[]{van82} van Albada, T. S., Kotanyi, C. G.\& Schwarzschild,
M. 1982, MNRAS, 198, 303

\end{thebibliography}
\end{document}